\documentclass[pra,twocolumn,showpacs,amsmath,amssymb,superscriptaddress,floatfix]{revtex4-2}
\usepackage{braket}
\usepackage{cases}
\usepackage{color}
\usepackage[toc,page]{appendix}
\usepackage{graphicx,subfigure}
\usepackage{float}
\usepackage{bm}
\usepackage[export]{adjustbox}
\usepackage{commath}
\usepackage{color}
\usepackage[normalem]{ulem}
\usepackage{qcircuit}
\usepackage{hyperref}
\hypersetup{
    colorlinks=true, 
    linkcolor=cyan,
    citecolor=magenta, 
    filecolor=magenta, 
    urlcolor=cyan,
    runcolor=cyan
}
\usepackage[capitalise]{cleveref} 

\newcommand{\ignore}[1]{}

\newcommand{\bes} {\begin{subequations}}
\newcommand{\ees} {\end{subequations}}
\newcommand{\beq}{\begin{equation}}
\newcommand{\eeq}{\end{equation}}

\def\>{\rangle}
\def\<{\langle}

\newcommand{\mc}[1]{\mathcal{#1}}

\newcommand{\qopt}{q_{\text{opt}}}

\usepackage[bottom]{footmisc}
\begin{document}
\title{Better-than-classical Grover search via quantum error detection and suppression}

\author{Bibek Pokharel}
\email{pokharel@usc.edu}
\affiliation{Department of Physics \& Astronomy}
\affiliation{Center for Quantum Information Science \& Technology}

\author{Daniel A. Lidar}
\email{lidar@usc.edu}
\affiliation{Department of Physics \& Astronomy}
\affiliation{Center for Quantum Information Science \& Technology}
\affiliation{Department of Electrical \& Computer Engineering}
\affiliation{Department of Chemistry\\
University of Southern California, Los Angeles, CA 90089, USA}

\date{\today}

\begin{abstract}
	Grover's search algorithm is one of the first quantum algorithms to exhibit a provable quantum advantage. It forms the backbone of numerous quantum applications and is widely used in benchmarking efforts. Here, we report better-than-classical success probabilities for a complete Grover search algorithm on the largest scale demonstrated to date, of up to five qubits, using two different IBM superconducting transmon qubit platforms. This is enabled, on the four and five-qubit scale, by error suppression via robust dynamical decoupling pulse sequences, without which we do not observe better-than-classical results. Further improvements arise after the use of measurement error mitigation, but the latter is insufficient by itself for achieving better-than-classical performance. For two qubits, we demonstrate a success probability of $99.5\%$ via the use of the \([[4,2,2]]\) quantum error-detection (QED) code. This constitutes a demonstration of quantum algorithmic breakeven via QED. Along the way, we introduce \emph{algorithmic error tomography,} a method of independent interest that provides a holistic view of the errors accumulated throughout an entire quantum algorithm, filtered via the errors detected by the QED code used to encode the circuit. We demonstrate that algorithmic error tomography provides a stringent test of an error model based on a combination of amplitude damping, dephasing, and depolarization. 
\end{abstract}
\maketitle

\section{Introduction}
\label{sec:introduction}

The best possible classical strategy for finding a particular ``marked'' element in an unsorted list of length $N$ requires querying half of the elements in the list on average; a quantum computer (QC) can do this in quadratically fewer queries using Grover's search algorithm~\cite{Grover:97a}. This algorithm is optimal and provably better than all classical strategies~\cite{Bennett:1997lh}. As one of the first algorithms with a provable quantum speedup, Grover search is often used as a subroutine for other quantum algorithms~\cite{durrQuantumQueryComplexity2006,magniezQuantumAlgorithmsTriangle2007}. Over the last two decades, Grover search has been implemented on various quantum computing platforms~\cite{lubinskiApplicationOrientedPerformanceBenchmarks2021,royProgrammableSuperconductingProcessor2020,figgattComplete3QubitGrover2017,Zhang_2021}, albeit for relatively
small $N$.

Encoding a list of length $N$ requires $n=\lceil \log_2(N) \rceil$ qubits. The list can be queried classically or using quantum queries; in both cases, one finds the marked element with some probability, which we refer to as the classical or quantum success probability. The largest implementation of Grover's algorithm to date is for $n=8$ qubits, but without demonstrating a better-than-classical quantum success probability~\cite{lubinskiApplicationOrientedPerformanceBenchmarks2021}. Such better-than-classical performance has been achieved for $n=3$~\cite{royProgrammableSuperconductingProcessor2020,figgattComplete3QubitGrover2017} and $n=4$~\cite{Zhang_2021} qubits. Here, employing two seven-qubit IBM Quantum Experience (IBMQE) transmon qubit platforms ibm\_nairobi (Nairobi) and ibmq\_jakarta (Jakarta), we demonstrate higher success probabilities than all previous implementations, for $n\le 5$. 

Key to our demonstrations is the use of error suppression and mitigation strategies. In particular, we use the $[[4,2,2]]$ quantum error-detecting code~\cite{Vaidman:1996vs,gottesman}, which encodes $k=2$ logical qubits into $n=4$ physical qubits and detects arbitrary single-qubit errors, to demonstrate a significant success probability enhancement relative to using two copies of $n=2$ physical qubits. These success probabilities are further improved by combining error detection with measurement error mitigation~\cite{kandalaErrorMitigationExtends2019,nachman2020unfolding}. 

We use the quantum error detection results to perform what we call \emph{algorithmic error tomography}: for each algorithm execution we compute the probability of an output $X$, $Y$, or $Z$ error (corresponding to the three Pauli matrices) on one of the four physical qubits, or a logical error. This allows us to compute a detailed map of the errors that arise after executing the entire algorithm. In this sense, algorithmic error tomography provides a holistic and complementary perspective to techniques such as gate set tomography~\cite{blume2013robust,Merkel:2013aa}, which instead focus on individual gates.  

We demonstrate better-than-classical performance for three or more physical qubits by employing error suppression via dynamical decoupling (DD)~\cite{Viola:98,Viola:99,Zanardi:1999fk,Vitali:99}. Toward this end, we consider three robust DD families: universally robust (UR)~\cite{Genov:2017aa}, concatenated DD (CDD)~\cite{Khodjasteh:2005xu}, and robust genetic algorithm (RGA)~\cite{Quiroz:2013fv} sequences. We find that robust sequences with few pulses are vital in achieving better-than-classical algorithmic performance. 

We compare the experimentally obtained results for Grover's algorithm with an error model based on the concatenation of amplitude damping, phase damping, and depolarization maps. Each map is parameterized by the calibration metrics provided by the IBM Quantum Experience (IBMQE) backend~\cite{IBMQuantum2022}. We test this model using the observed success probabilities and the algorithmic error tomography results; the latter provides a much more stringent test. We find good agreement with the model, but only after using DD. We interpret this in terms of the suppression of crosstalk by DD~\cite{tripathi2021suppression,Zeyuan:22}, which is unaccounted for by the error model.

In summary, we demonstrate a better-than-classical Grover search for up to 5 qubits, enabled by quantum error detection and dynamical decoupling. That is, we demonstrate algorithmic performance that is enhanced beyond the break-even point -- where protected operations outperform their unprotected counterparts -- and the capabilities of the best possible classical algorithm executing the same task. Along the way, we introduce algorithmic error tomography -- a characterization of errors afflicting an entire quantum algorithm based on the syndromes of a quantum error detecting code.

The structure of this paper is as follows. In \cref{sec:Grover-background}, we summarize Grover's algorithm's salient aspects and discuss its implementation. In \cref{sec:sim}, we describe the open system model we use to compute the theoretically expected algorithmic performance. Details about our dynamical decoupling implementation are in \cref{sec:dd-description}. \cref{sec:2q_results} focuses on the performance of Grover's algorithm on $n=2$ qubits with and without error detection. Algorithmic error tomography is introduced in \cref{sec:2q_results} as well. The results for $2 < n \leq 5$, where DD plays a crucial role in achieving better-than-classical performance, are given in \cref{sec:3q_and_higher_results}. We conclude with observations and the implications of our results in \cref{sec:conclusion}. 

\section{Grover's Algorithm: background and implementation} 
\label{sec:Grover-background}

\begin{figure*}[t]
\includegraphics[width=\textwidth]{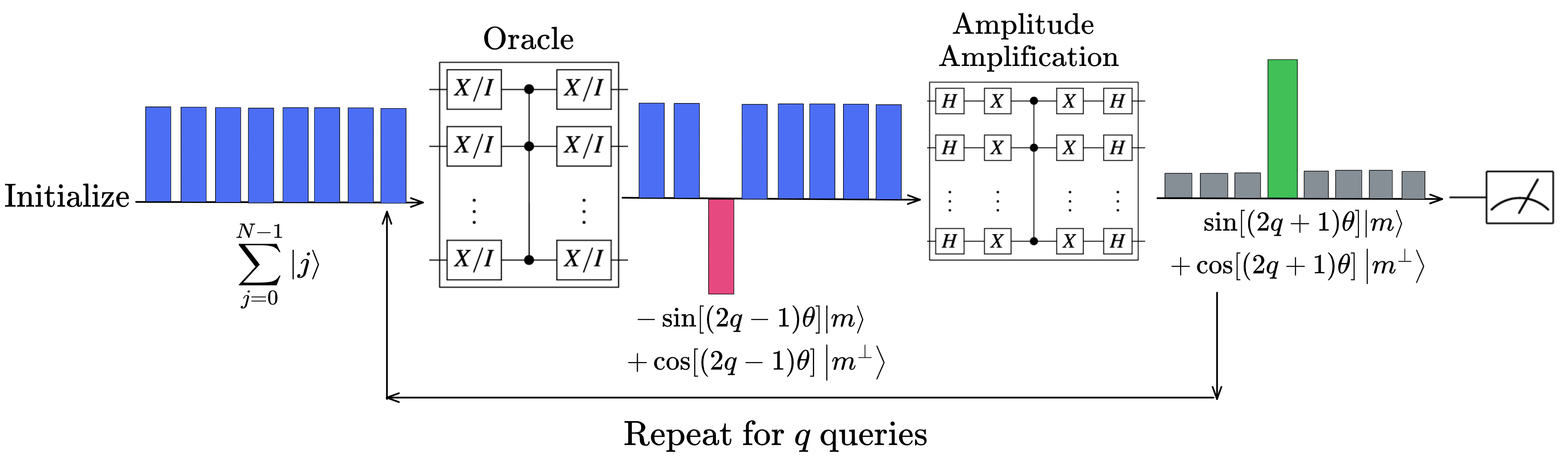}
\caption{Circuit description for Grover's algorithm. The relative amplitudes of all the states at each stage of the algorithm are shown. Starting with an equal superposition state, the oracle assigns a relative phase difference of $\pi$ to the marked state. The amplitude amplification step then performs an inversion about the mean, allowing $| m \rangle$ to have a larger probability amplitude than all other states. This round of querying and amplifying is repeated $q$ times. The optimal number of rounds for the $n$-qubit Grover problem is $\qopt = \lfloor \frac{\pi}{4} 2^{n/2}  \rfloor$. The only multi-qubit operation required to implement both the oracle and the amplitude amplification step is C$_{n-1}Z$ (vertical line in the Oracle and Amplitude Amplification boxes). 
}
 \label{fig:grover-setup}
\end{figure*}

\subsection{Problem Description}

Informally, the Grover problem is to search an unsorted list with $N=2^n$ elements for a marked element. Formally, the goal is to find the marked $n$-bit bitstring $m$ using the smallest number of queries of an oracle that implements a function $f_m:\{0,1\}^n\mapsto \{0,1\}$ defined as $f_m(x)=\delta_{x,m}$. Classically, after $q$ queries, the probability of correctly identifying the marked element, which hereafter we refer to as the \emph{success probability}, is $p_s^{\text{C}}(q,N) = (q+1)/N$ (see \cref{app:classical-p_s}). Consequently, the classical algorithm requires $O(N)$ queries.

Grover's algorithm provides a quadratic quantum speedup, requiring only $ O(\sqrt{N}) $ queries~\cite{Grover:97a}. This scaling remains valid with more than one marked element~\cite{Boyer:96}, or even for an arbitrary initial amplitude distribution over the list elements~\cite{Biham:1999ye}. In the original setting of a single marked element, the state after $q$ queries to the oracle is
\begin{equation}
    \left|\psi_{q}\right\rangle=\sin [(2 q+1) \theta]\ket{m}+\cos [(2 q+1) \theta]\ket{m^{\perp} },
\end{equation}
where $\ket{m^{\perp}} = \frac{1}{\sqrt{N-1}} \sum_{x \neq m}|x\rangle$ and $\theta=\arcsin \left( \frac{1}{\sqrt{N}} \right)$. Thus, the quantum success probability is $p_s^{\text{Q}}(q,N)= \sin^2 \left[ (2q +1) \theta \right]$, and the theoretically optimal number of queries is $\qopt = \lfloor \frac{\pi}{4}\sqrt{N} \rfloor $. Note that $p_s^{\text{C}}(q,N) < p_s^{\text{Q}}(q,N)$ for all $q < \qopt$.  However, the theoretically optimal $q$ is often not experimentally optimal. As circuit depth increases with the number of queries and the problem size, there is a trade-off between the added decoherence and the increase in the success probability. Most experimental implementations of Grover's algorithm have focused on a single query~\cite{royProgrammableSuperconductingProcessor2020,Zhang_2021, lubinskiApplicationOrientedPerformanceBenchmarks2021,figgattComplete3QubitGrover2017}, but this strategy does not scale well, as both $p_s^{\text{C}}(1,N)$ and  $p_s^{\text{Q}}(1,N)$ decrease exponentially with $n$. We adopt an empirical approach to identify the optimal number of queries such that $p_s$ is maximized. We set $q=2$ for all problem sizes other than $n=2$ where $\qopt=1$. We justify our choice of the number of queries in \cref{app:DD-survey}.  

\subsection{Implementation}

A schematic illustrating the implementation of the $n$-qubit Grover algorithm is shown in~\cref{fig:grover-setup}. The only multi-qubit operation is the $n$-qubit controlled-phase gate C$_{n-1}Z$, which needs to be implemented twice for each oracle query: once for the oracle and again for the amplitude amplification step. Different marked elements are represented by sandwiching the C$_{n-1}Z$ gate with $X_i$ or $I_i$ depending on whether the corresponding bit $b_i$ in the marked bitstring $m$ is 0 or 1. I.e., letting $m=b_1b_2 \dots  b_n$, then C$_{n-1}Z$ in the oracle layer is preceded and followed by $ X^{1-b_1} \otimes X^{1-b_2} \cdots \otimes  X^{1-b_n}$. Likewise, amplitude amplification is implemented as $H^{\otimes n}  X^{\otimes n}  (\text{C}_{n-1}Z)  X^{\otimes n}  H^{\otimes n}$.

For all problem sizes and oracles, we repeated each circuit for the maximum number of shots allowed on the QPU: $20000$ and $32000$ for Nairobi and Jakarta, respectively. The reported success probabilities were extracted by bootstrapping over these trials and all $N$ possible marked states. All error bars reflect 95\% confidence intervals obtained after bootstrapping unless specified otherwise.

\section{Open system model}
\label{sec:sim}

The QPUs used here are calibrated daily, and the following calibration metrics are recorded: the gate error $e_g$ and gate duration $\tau_g$, the qubit damping timescale $T_{1}$ and dephasing timescale $T_{2}$, and the response matrix $M$ for readout errors (see \cref{app:device,app:MEM}). In this section, we describe how we estimate the theoretical performance of Grover's algorithm using these metrics. The model described here is mathematically equivalent to the one used in Qiskit's Aer API (see the supplementary information of Ref.~\cite{blankQuantumClassifierTailored2020}). 

In a closed system described by a state $\rho$, a unitary gate $U$ acts as $\mathcal{U}(\rho) = U \rho U^{\dagger}$. In reality, the system is open, so we model gate $U$ as a CPTP map $\mathcal{E} = \mathcal{D} \circ \Phi \circ \mathcal{A}  \circ \mathcal{U}$, where $\mathcal{D}, \mathcal{A}, \Phi$  are depolarizing, amplitude damping and phase damping maps respectively~\cite{rivas_open_2012}. The amplitude damping and phase damping maps account for thermal relaxation, which we represent as $\mathcal{R} = \Phi \circ \mathcal{A}$. The single-qubit Kraus operators for $\mathcal{A}=\{A_0,A_1\}$ and $\Phi = \{F_0,F_1\}$ are

\bes
\begin{align}
    A_{0}&=\left(\begin{array}{cc}
        1 & 0 \\
        0 & \sqrt{1-p_{A}}
    \end{array}\right),\quad  A_{1}=\sqrt{p_{A}}|0\rangle\langle 1| \\
    F_{0} &= \sqrt{ p_{\Phi} } I, \quad F_{1} = \sqrt{ 1-p_{\Phi} } Z.
\end{align}
\ees
The $n$-qubit depolarizing map is
\begin{align}
    \mathcal{D}: \rho\mapsto (1-p_{D})\rho + p_{D} \frac{\mathbb{I}}{ 2^n } =  \sum_{j} K_{j} \rho K_{j}^{\dagger} ,   
	\label{eq:depolarizing}  
\end{align}
where $\mathcal{D}$ has $4^n$ Kraus operators: 
\begin{subequations}
\label{eq:D}
	\begin{align}
		K_{0} &= \sqrt{ 1 - \frac{4^n-1}{4^n} p_D }\  \mathbb{ I }^{\otimes n}, \\
		K_{j} &= \sqrt{ \frac{p_D}{4^n} } {P}_{j}, \quad {P}_{j} \in \{ \mathbb{ I }, X, Y, Z \}^{\otimes n} \smallsetminus \mathbb{ I }^{\otimes n}. 
	\end{align}
\end{subequations}
We parameterize these maps by their respective error probabilities $p_A$, $p_\Phi$ and $p_D$, which in turn depend on the calibration metrics $e_g$, $\tau_g$, $T_1$, and $T_2$. In particular,
\begin{subequations}
    \begin{align}
        p_{A} & = 1-e^{-\tau_g/T_{1}} \\
        p_{\Phi} & = \frac{1}{2} (1+e^{-\tau_g/T_{\Phi}}) \\
        p_{D} & = \frac{d(F(\mathcal{R}) - 1 + e_g)}{d F(\mathcal{R}) - 1}, 
    \label{eq:cptp-params}
    \end{align}
\end{subequations}
where 
\beq
\frac{1}{T_\Phi} = \frac{1}{2 T_1} - \frac{1}{T_2},
\eeq 
and 
\bes
\begin{align}
F(\mathcal{E}) &= \int d \psi\bra{\psi}U^{\dagger}\mathcal{E}(\ket{\psi}\!\bra{\psi}) U \ket{\psi} \\
& =\frac{d F_{\mathrm{pro}}(\mathcal{E})+1}{d+1}
\end{align}
\ees 
is the average gate fidelity for a CPTP map $\mathcal{E}$. Here  $F_{\text{pro}}(\mathcal{E})$ is the process fidelity of the map $\mathcal{E}$ with the target map $\mathcal{U}$, and $d$ is the dimension of the map~\cite{nielsenSimpleFormulaAverage2002}. 

For each gate, we know the total gate error $e_g$, and so we compute $p_D$ by setting $1-e_g=F(\mathcal{D}\circ\mathcal{R})$, which gives us \cref{eq:cptp-params} (see \cref{app:depolarizing} for more details). 
In other words, we assign any gate error not accounted for by relaxation to depolarization.   
Since $p_D\ge 0$, we must have $F(\mathcal{R}) \geq F(\mathcal{D} \circ \mathcal{R})$, i.e., the error due to relaxation alone cannot exceed the error due to relaxation followed by depolarization. If this condition is not met, then we assume that the error is entirely due to depolarization, so that $F(\mathcal{R})=1$ and hence we set $p_D = e_g d/(d-1) $ and $\mathcal{E} = \mathcal{D} \circ \mathcal{U}$. For idle intervals in the circuit, no gate error $e_g$ is reported by IBMQE~\cite{IBMQuantum2022}, and therefore we model idle intervals with duration $\tau$ as identity operations where only the relaxation $\mathcal{R}$ matters. This is equivalent to setting the gate error for idle intervals to $e_{\text{idle}} = 1 - F(\mathcal{R})$.

In summary, we model single-qubit gates $U_{1Q}$ with gate duration $\tau_g$ as 
\begin{equation}
	\mathcal{D}(p_{D}) \circ \mathcal{R}(T_{1}, T_{2},\tau_g) \circ \mathcal{U}_{1Q},
\end{equation}
and two-qubit gates $U_{2Q}$ with duration $\tau_g$ acting on qubits $j,k$ are modeled as
\begin{equation}
	\mathcal{D}(p_{D}) \circ (\mathcal{R}(T_{1}^j, T_{2}^j, \tau_g) \otimes \mathcal{I}_{k}) \circ (\mathcal{I}_{j} \otimes \mathcal{R}(T_{1}^k, T_{2}^k, \tau_g) ) \circ \mathcal{U}_{2Q}
\end{equation}
where for $n=2$ $\mathcal{D}(p_{D})$ has $16$ Kraus operators [see \cref{eq:D}].
Idle intervals with duration $\tau$ are described by 
\begin{equation}
    \mathcal{R}(T_1, T_2, \tau).
\end{equation}

The quantum circuit for each experiment is first compiled into the QPU's native gate set and then scheduled using IBMQE's API~\cite{IBMQuantum2022}. We use this circuit to determine the order of operations and then replace each unitary map $\mathcal{U}$ with the corresponding CPTP channel $\mathcal{E}$. In the end, we acquire a probability distribution corresponding to the theoretical estimate of the circuit's output as measured in the computational basis.

\section{Dynamical Decoupling}
\label{sec:dd-description}

\begin{figure}[t]
	\includegraphics[width=\columnwidth, valign=c]{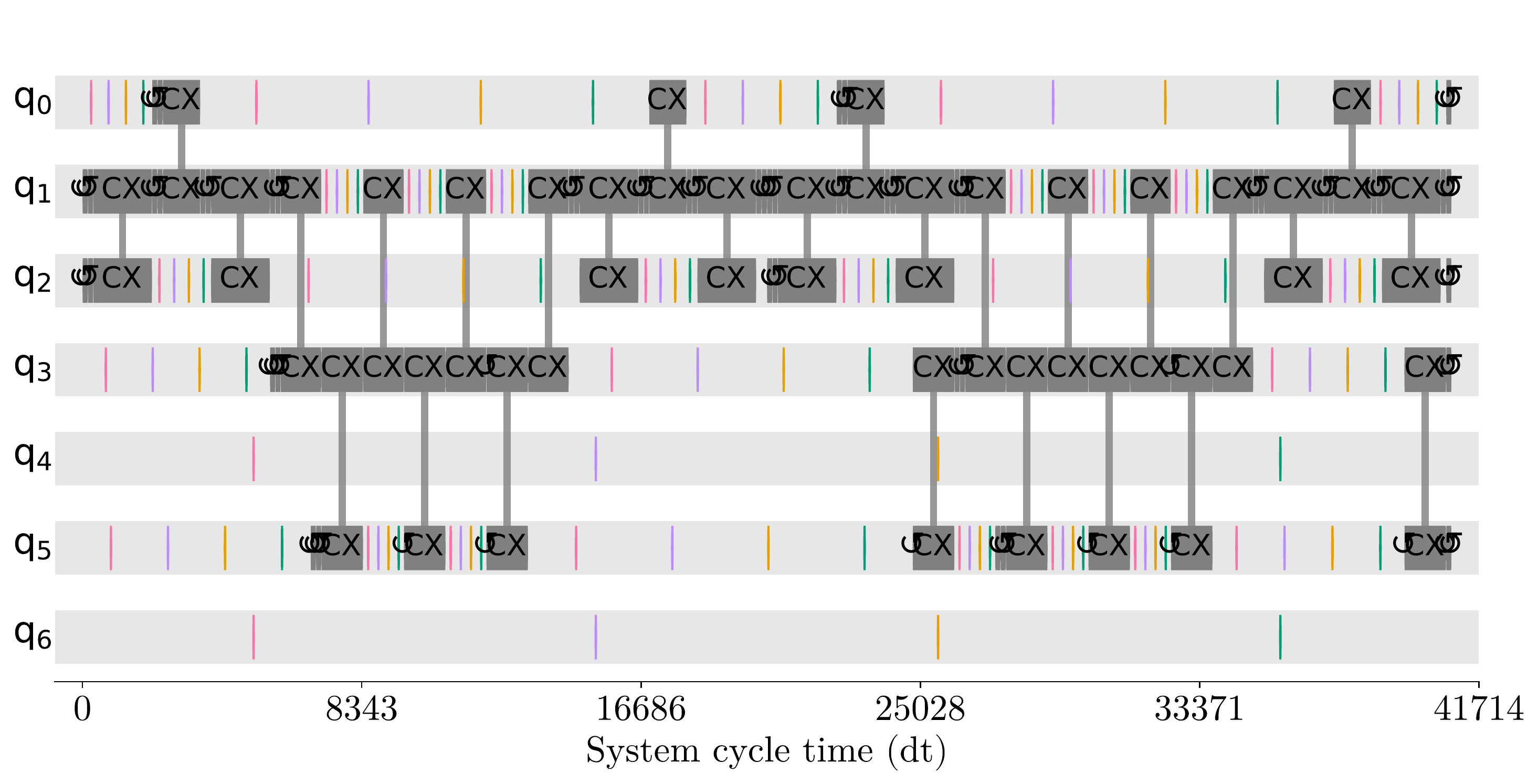} 
	\caption{Grover and DD. 
    The timeline for one oracle query for 4-qubit Grover with the marked state $\ket{1111}$ is shown. 
    Qubits q$_4$ and q$_6$ are spectators in this example. Recall that each oracle query for 4-qubit Grover requires two C$_3Z$ gates. 
    C$_3Z$ requires 14 CNOTs, and the entire circuit uses 28 CNOTs; see \cref{app:circuit-constr} for circuit compilation details. 
    The pre-DD circuit elements are grayed out, and the colored lines represent the DD pulses. The DD sequence exemplified here uses four pulses for illustration purposes; in reality, we used longer sequences. 
    The scheme demonstrated highlights four primary features of our implementation: (1) all idle intervals, including the ones on inactive qubits, are filled, (2) only one repetition of each sequence is performed, and the pulse interval is adjusted accordingly, (3) each pulse in the sequence can be unique, (4) a single qubit can experience multiple DD repetitions if there are multiple idle intervals.
	}
	 \label{fig:dd-on-grover}
	\end{figure}

DD is an open-loop quantum control technique wherein a sequence of pulses is strategically inserted between gates to suppress unwanted system-bath interactions~\cite{Viola:98,Viola:99,Zanardi:1999fk,Vitali:99}. While DD is fully compatible with quantum error correction~\cite{Paz-Silva:2013tt}, its most economical form requires no encoding,  measurements, or post-processing. It is, therefore, perhaps the least resource-intensive error suppression strategy. Error suppression via DD has a long history of experimental demonstrations on various quantum devices (see Ref.~\cite{Suter:2016aa} for a review). Here, we employ a ``decouple then compute'' strategy~\cite{West:10,Ng:2011dn}, whereby control pulses constituting short but complete DD sequences are interleaved with the quantum circuit, exploiting intervals when individual qubits in the corresponding quantum circuits are idle. A scheme demonstrating our strategy is shown in \cref{fig:dd-on-grover}. This interleaving strategy has been used to improve quantum volume~\cite{jurcevicDemonstrationQuantumVolume2021}, variational quantum algorithms~\cite{raviVAQEMVariationalApproach2021}, and most recently to demonstrate an algorithmic quantum speedup~\cite{pokharel2022demonstration}. 

In addition to the popular basic DD sequences -- CPMG~\cite{Maudsley:1986ty} and XY4~\cite{Viola:99} -- we consider three robust sequence families: universally robust (UR) DD~\cite{Genov:2017aa}, concatenated DD (CDD)~\cite{Khodjasteh:2005xu}, and robust genetic algorithm (RGA) DD~\cite{Quiroz:2013fv} (see \cref{app:DD-survey} for more details). Other than CPMG, these are all high-order, multi-axis sequences that are universal for single qubits, i.e., they suppress arbitrary single-qubit errors beyond first order in the Magnus or Dyson expansion~\cite{Lidar-Brun:book}. Robustness refers to the mitigation of axis-angle and over/under-rotation errors. In addition, these sequences can cancel crosstalk errors~\cite{tripathi2021suppression,Zeyuan:22}. Our sequence choice is informed by the results of Ref.~\cite{DD-survey}, which reported on a significantly more comprehensive survey of sequences using superconducting qubits and concluded that robust sequences are preferred default choices. Here we do not utilize the OpenPulse functionality of the IBMQE platforms, nor do we implement Uhrig-type~\cite{Uhrig:2007qf} non-uniform pulse interval DD sequences such as quadratic DD (QDD)~\cite{West:2010:130501}, which were also found to perform well in the survey~\cite{DD-survey}. Both have the potential to enhance our results and are attractive options for future studies.

\section{Two-qubit encoded Grover algorithm protected by quantum error detection}
\label{sec:2q_results}

\begin{figure*}
	\begin{equation*}
\Qcircuit @C=1em @R=0.8em @!R {
	& & & & \textsc{Initialization} & & & & & & \ \ \ \ \  \textsc{Oracle} & &  & & & & & \ \ \ \ \ \ \    \textsc{Amplitude Amplification} & & & & & & & &  \textsc{Measurement} & & & & & &  \\
	& & & \ \ \ \   U_{\text{encoding}} & & & & & & &  \ \ \ \ \ \ \ \text{ Oracle}_{01} & &  & & & & & \ \ \ \ \    \text{Oracle}_{00} & & & & & & & \ \ \ \ \ \ \    U_{\text{decoding}}  & & & & & & &\\
	&\qw& \gate{H} & \ctrl{1} & \qw     & \qw     & \gate{H} &\qw&\qw& \gate{X} & \gate{P} & \gate{I} & \gate{X} &\qw&\qw&\gate{H} & \gate{I} & \gate{P} & \gate{I} & \gate{I} &\gate{H}&\qw&\qw& \qw      & \qw     & \ctrl{1} & \gate{H} & \meter\\
	&\qw& \qw      & \targ    & \ctrl{1}& \qw     & \gate{H} &\qw&\qw& \gate{I} & \gate{P} & \gate{Z} & \gate{I} &\qw&\qw&\gate{H} & \gate{X} & \gate{P} & \gate{Z} & \gate{X} &\gate{H}&\qw&\qw& \qw      & \ctrl{1}& \targ    & \qw      & \meter\\
	&\qw& \qw      & \qw      & \targ   & \ctrl{1}& \gate{H} &\qw&\qw& \gate{X} & \gate{P} & \gate{Z} & \gate{X} &\qw&\qw&\gate{H} & \gate{X} & \gate{P} & \gate{Z} & \gate{X} &\gate{H}&\qw&\qw& \ctrl{1} & \targ   & \qw      & \qw      & \meter \\&\qw& \qw      & \qw      & \qw     & \targ   & \gate{H} &\qw&\qw& \gate{I} & \gate{P} & \gate{I} & \gate{I} &\qw&\qw&\gate{H} & \gate{I} & \gate{P} & \gate{I} & \gate{I} &\gate{H}&\qw&\qw& \targ    & \qw 	 & \qw      & \qw      & \meter \\ 
	& & & & & & & & & & & & & & & &  & & & & & & & & & & & & & & &  \\ 
	& & & & & & & & & & & & & & & &  & & & & & & & & & & & & & & & 
	\protect\gategroup{1}{2}{7}{8}{0.9em}{-} 
	\protect\gategroup{3}{3}{6}{6}{0.9em}{--} 
	\protect\gategroup{1}{9}{7}{14}{0.9em}{-} 
	\protect\gategroup{3}{10}{6}{13}{0.9em}{--} 
	\protect\gategroup{1}{15}{7}{22}{0.9em}{-} 
	\protect\gategroup{3}{17}{6}{20}{0.9em}{--} 
	\protect\gategroup{3}{24}{6}{27}{0.9em}{--} 
	\protect\gategroup{1}{23}{7}{29}{0.9em}{-} 
} 
	\end{equation*}
\vspace{-10mm}
	\caption{Encoded two-qubit Grover. The two-qubit single-query ($q=1$) Grover circuit encoded using the \([[4,2,2]]\) code for the marked state $\ket{01}$ is shown. The encoding step (left dashed box) prepares the encoded initial state $\ket{\overline{00}} = \frac{1}{\sqrt{2}} \left( | 0000 \rangle + | 1111 \rangle \right)$ (the 4-qubit GHZ state) from the physical initial state $\ket{0000}$. The encoded Grover circuit is implemented by converting each physical gate of the $n=2$ case of \cref{fig:grover-setup} into its logical counterpart, which is then converted into a physical $4$-qubit implementation (middle left and right dashed boxes); see \cref{app:circuit-constr} for details. $P=\text{diag}(1, i)$ is the phase gate. We post-select the measured results by decoding (right dashed box) and discarding any measurement outcome for which the code detects errors, i.e., does not result in one of the four decoded states $\{\ket{0000}, \ket{0010}, \ket{0111}, \ket{0101}\}$.}
		\label{fig:enCircuit}
\end{figure*}
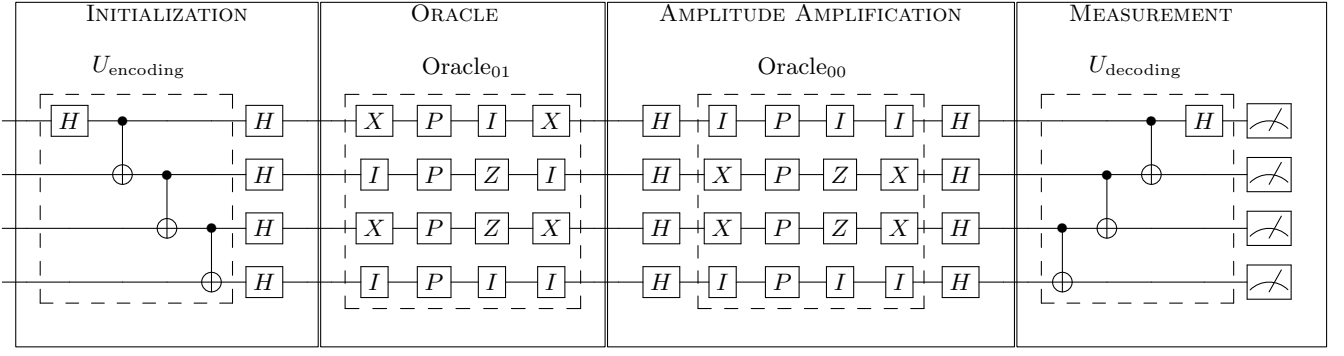

The $[[4,2,2]]$ code~\cite{Vaidman:1996vs} is the smallest possible qubit-based error detecting code~\cite{gottesman} and has been invoked for proof-of-principle demonstrations of quantum error detection~\cite{vuillotErrorDetectionHelpful}. Notably, it has been used to improve Clifford gate set fidelities~\cite{Harper:2019aa} and the performance of variational algorithms~\cite{urbanekErrorDetectionQuantum2020}. However, measurement error mitigation (MEM) played a dominant role in  Ref.~\cite{urbanekErrorDetectionQuantum2020}, and it is unclear if error detection alone would have improved performance in that work. Here, we compare the performance of the two-qubit Grover algorithm with and without the $[[4,2,2]]$ code and MEM. The unencoded version needs two qubits, while the encoded version requires four. To equalize resources, we simultaneously use two copies of the unencoded circuit and report the best fidelity of the two copies. We incorporate MEM using iterative Bayesian unfolding via the pyIBU package~\cite{srinivasanScalableMeasurementError2022} (see \cref{app:MEM} for details about MEM). Ultimately, we demonstrate a conclusive improvement in algorithmic performance due to quantum error detection.

\subsection{Encoding into the $[[4,2,2]]$  code}

The stabilizers of the $[[4,2,2]]$  code are $XXXX$ and $ZZZZ$. The logical operators of this code can be chosen as $\overline{X}_1=XIXI$, $\overline{X}_2=XXII$, $\overline{Z}_1=ZZII$, and $\overline{Z}_2=ZIZI$. Two-qubit Grover also requires the encoded Hadamard $\overline{H}$ and controlled-phase $\overline{\text{C}Z}$.
The deconstruction of the logical circuit into physical components is detailed in \cref{app:circuit-constr}, and the resultant encoded two-qubit Grover circuit is shown in \cref{fig:enCircuit}. 

The encoding and decoding circuits, $U_{\text{enc}}$ 
and $U_{\text{enc}}^{\dagger}$, are also depicted in \cref{fig:enCircuit}. The logical basis states are:
\bes
\label{eq:encoding}
\begin{align}
\ket{\overline{00}}= U_{\text{enc}} \ket{0000} = \frac{\ket{0000}+\ket{1111}}{\sqrt{2}} \\
\ket{\overline{01}}= U_{\text{enc}} \ket{0010} =\frac{\ket{0011}+\ket{1100}}{\sqrt{2}} \\
\ket{\overline{10}}= U_{\text{enc}} \ket{0111} =\frac{\ket{0101}+\ket{1010}}{\sqrt{2}} \\
\ket{\overline{11}}= U_{\text{enc}} \ket{0101} =\frac{\ket{1001}+\ket{0110}}{\sqrt{2}}
\end{align}
\ees
These are also the four possible marked states in the two-qubit Grover problem.
Consequently, after applying $U_{\text{enc}}^{\dagger}$ to decode the results, only  states from the set $\mathcal{I} = \{\ket{0000}, \ket{0010}, \ket{0111}, \ket{0101}\}$ could have arisen from valid logical states. Therefore, we postselect by removing any of the $12$ measurement outcomes that do not correspond to valid logical states, i.e., states in $\mathcal{I}^\perp = \{0,1\}^4\smallsetminus \mathcal{I}$. 

\subsection{Algorithmic error tomography }
\label{sec:es}

Even though we discard any states not in the set $\mathcal{I}$ during postselection, there is important information in such outcomes. They allow us to diagnose the frequency with which different Pauli errors appear at the end of the circuit for a given encoded marked state, i.e., to perform \emph{algorithmic error tomography} (AET), which we now describe.

Let us first consider the different ways errors can occur before the measurement outcome is obtained. In principle, errors can occur anywhere during the circuit, including between the decoding and measurement steps or even during the measurement. There can be multiple errors at multiple locations of arbitrary weight. Given an $[[n,k,d]]$ code $\mathcal{C}$, in AET, we treat all these errors as either the \emph{effective} errors that this code can detect or as logical errors, and assign a location to these errors as if they happened just before decoding. Namely, if $\ket{\overline{b}}\in\mathcal{C}$ is a code basis state, where $b\in\{0,1\}^k$, $E$ is an error, $U_{\text{dec}} = U_{\text{enc}}^{\dagger}$ is the decoding unitary for $\mathcal{C}$, and $\mathcal{M}$ denotes a projective measurement in the computational basis, then in AET we interpret each measured computational basis state $\ket{b'}$ as having arisen from 
$\mathcal{M} U_{\text{dec}}E \ket{\overline{b}}$, where $b'\in\{0,1\}^n$. If $E$ is an error the code can detect then $\ket{b'}=\mathcal{M} U_{\text{dec}}E \ket{\overline{b}}\in\mathcal{C}^\perp_E\subset\mathcal{C}^\perp$, while if $E$ is a logical error then $\ket{b'}=\mathcal{M} U_{\text{dec}}E \ket{\overline{b}}\in\mathcal{C}$. Here $\mathcal{C}\oplus \mathcal{C}^\perp = \mathcal{H}$, the full system Hilbert space, and $\mathcal{C}^\perp = \bigoplus_E \mathcal{C}^\perp_E$. We can now empirically compute the probability $p_{b'}$ of each measurement outcome $\ket{b'}$, and from here the probability $p_{E}$ of each error $E$ by constructing the \emph{error outcome table} $\{U_{\text{dec}}E \ket{\overline{b}}\}$ for all $b\in\{0,1\}^k$ and all the errors $E$ the code can detect. With this table in hand, we can check for each measurement outcome $\ket{b'}$ whether it is in the table (if so, it corresponds to an effective error the code can detect) or not (in which case it corresponds to a logical error, or no error). Suppose $\ket{b'}$ is in the error outcome table. In that case, we can further identify the error type it arose from (i.e., associate it to a particular \emph{error subspace} $\mathcal{C}^\perp_E$). By finding the empirical relative frequency of this error type, we can compute this error's probability $p_E$. For logical errors, the procedure is similar; we need to find the empirical relative frequency of $\ket{b'}$ arising from a given code basis state $\ket{b}$.

\begin{table}[t]
	\centering
	\begin{tabular}{||l|r|r|r|r||} 
		\hline
		$E$ & $U_{\text{dec}} E \ket{ \overline{00} }$ & $U_{\text{dec}} E\ket{ \overline{01}}$ & $U_{\text{dec}} E \ket{ \overline{10}}$ & $U_{\text{dec}} E \ket{ \overline{11}}$\\
		\hline
		$X_1$ & $\ket{ 0100}$ & $\ket{ 0110}$ & $\ket{ 0011}$ & $\ket{ 0001}$\\
		$X_2$ & $\ket{ 0110}$ & $\ket{ 0100}$ & $\ket{ 0001}$ & $\ket{ 0011}$\\
		$X_3$ & $\ket{ 0011}$ & $\ket{ 0001}$ & $\ket{ 0100}$ & $\ket{ 0110}$\\
		$X_4$ & $\ket{ 0001}$ & $\ket{ 0011}$ & $\ket{ 0110}$ & $\ket{ 0100}$\\
		$i Y_1$ & $\ket{ 1100}$ & $\ket{ 1110}$ & $\ket{ 1011}$ & $\ket{ 1001}$\\
		$i Y_2$ & $-\ket{ 1110}$ & $-\ket{ 1100}$ & $\ket{ 1001}$ & $\ket{ 1011}$\\
		$i Y_3$ & $-\ket{ 1011}$ & $\ket{ 1001}$ & $-\ket{ 1100}$ & $\ket{ 1110}$\\
		$i Y_4$ & $-\ket{ 1001}$ & $\ket{ 1011}$ & $\ket{ 1110}$ & $-\ket{ 1100}$\\
		$Z_1$ & $\ket{ 1000}$ & $\ket{ 1010}$ & $\ket{ 1111}$ & $\ket{ 1101}$\\
		$Z_2$ & $\ket{ 1000}$ & $\ket{ 1010}$ & $-\ket{ 1111}$ & $-\ket{ 1101}$\\
		$Z_3$ & $\ket{ 1000}$ & $-\ket{ 1010}$ & $\ket{ 1111}$ & $-\ket{ 1101}$\\
		$Z_4$ & $\ket{ 1000}$ & $-\ket{ 1010}$ & $-\ket{ 1111}$ & $\ket{ 1101}$\\
		\hline
	\end{tabular}
	\caption{Error outcome table for the $[[4,2,2]]$ code. The table describes the effect of single-qubit Pauli errors $E$ on each of the encoded computational basis elements $\ket{\overline{b_1b_2}}$. The $X$, $Y$, and $Z$-type errors map each $\ket{\overline{b_1b_2}}$ to a distinct subspace after decoding: $\mathcal{C}^\perp_X = \text{span}(\{ \ket{0100}, \ket{0110}, \ket{0011}, \ket{0001} \})$, $\mathcal{C}^\perp_Y = \text{span}(\{ \ket{1100}, \ket{1110}, \ket{1011}, \ket{1001} \})$, and $\mathcal{C}^\perp_Z = \text{span}(\{ \ket{1000}, \ket{1010}, \ket{1111}, \ket{1101} \})$, respectively, such that $\mathcal{C}^\perp = \mathcal{C}^\perp_X\oplus \mathcal{C}^\perp_Y\oplus \mathcal{C}^\perp_Z$. Thus, each subspace uniquely determines the error type, which we use to perform algorithmic error tomography.}
	\label{tab:es}
\end{table}

Let us now illustrate this formal description of AET using the $[[4,2,2]]$ code. In this case $\mc{C} = \text{span}(\{\ket{0000}, \ket{0010}, \ket{0111}, \ket{0101}\})$.
It is clear from \cref{eq:encoding} that the decoding $U_{\text{dec}} = U_{\text{enc}}^{\dagger}$ is a one-to-one map between encoded states  $\ket{\overline{b_{1}b_2}}$ and unencoded states: $U_{\text{dec}} \ket{\overline{b_1b_2}}\in\mathcal{C}$. 
\cref{tab:es} is the error outcome table for the $[[4,2,2]]$ code and shows $U_{\text{dec}} E \ket{\overline{b_1b_2}}$ for all single-qubit Pauli errors and logical computational basis elements. As the probability of each outcome (i.e., each row in \cref{tab:es}) depends on the error $E$, the bitstring observed after applying the decoding circuit tells us the frequency with which different single-qubit Pauli errors occurred. More specifically, since the $[[4,2,2]]$ code is a distance-$2$ code, we can only infer which weight-$1$ error type occurred ($X$, $Y$, or $Z$), but not which specific qubit was affected. In other words, we can associate each observed state $\ket{b'}$ ($b'\in\{0,1\}^4 \smallsetminus \mathcal{I}$) with an error subspace $\mathcal{C}^\perp_X$, $\mathcal{C}^\perp_Y$, or $\mathcal{C}^\perp_Z$. Each observed state $\ket{b'}$ will have an empirical probability $p_{b'} = N_{b'}/N_{\text{tot}}$, where $N_{b'}$ is the number of times $\ket{b'}$ is observed out of a total of $N_{\text{tot}}$ observations. The probability of single-qubit $X$-type errors, $p_X$, is the sum of $p_{b'}$ in the first four rows, etc.

\begin{figure*}[t]
	\centering
	\includegraphics[width=\linewidth]{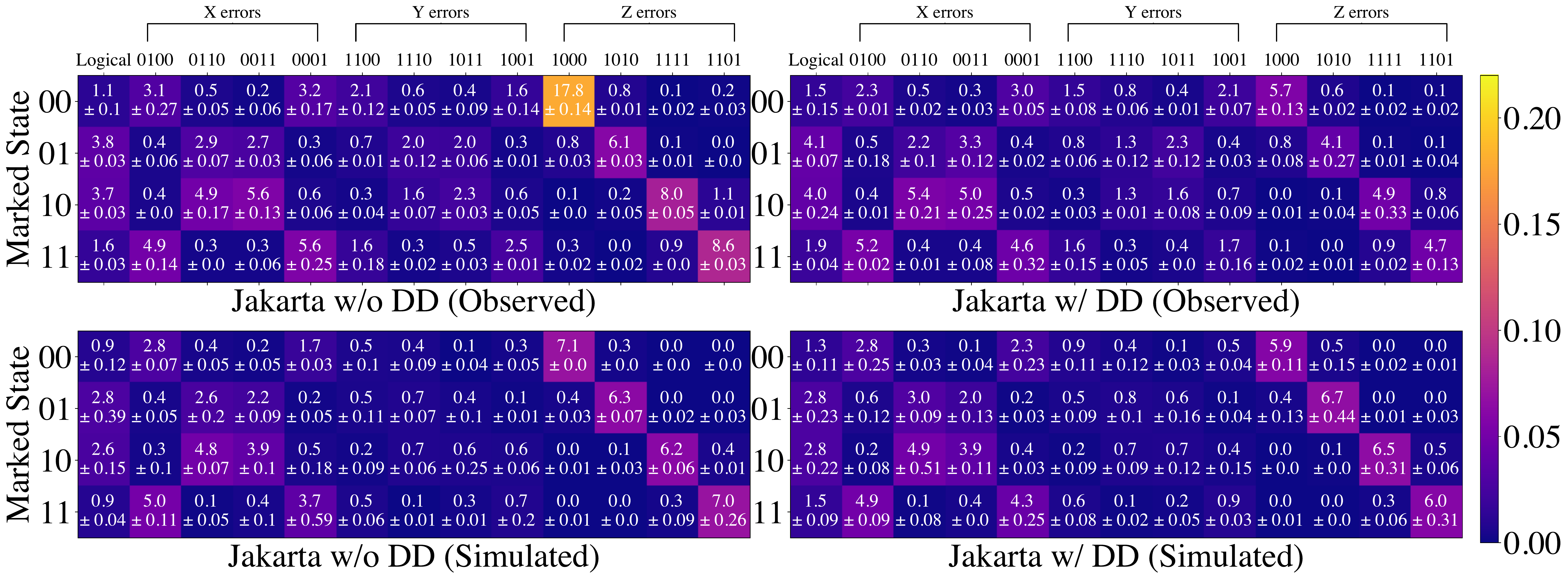}
	\caption{Algorithmic error tomography on Jakarta. The bitstring observed after $U_{\text{dec}}$ either corresponds to a marked entry or an error tabulated in \cref{tab:es}. Top: experimental results. The numbers in each box are the empirical percentage probabilities for detected $X$, $Y$, or $Z$-type errors, with $2\sigma$ standard deviation. Logical error percentage probabilities are shown in the first column of each table. Each row corresponds to a different marked state. The probabilities in each row do not sum to unity since we do not display the probability of obtaining the correct marked state. Left: without DD protection. Right: with DD protection. Bottom: the same for the simulated model.}
	\label{fig:et-dd}
\end{figure*}

\begin{figure}[h]
	\includegraphics[width=\columnwidth]{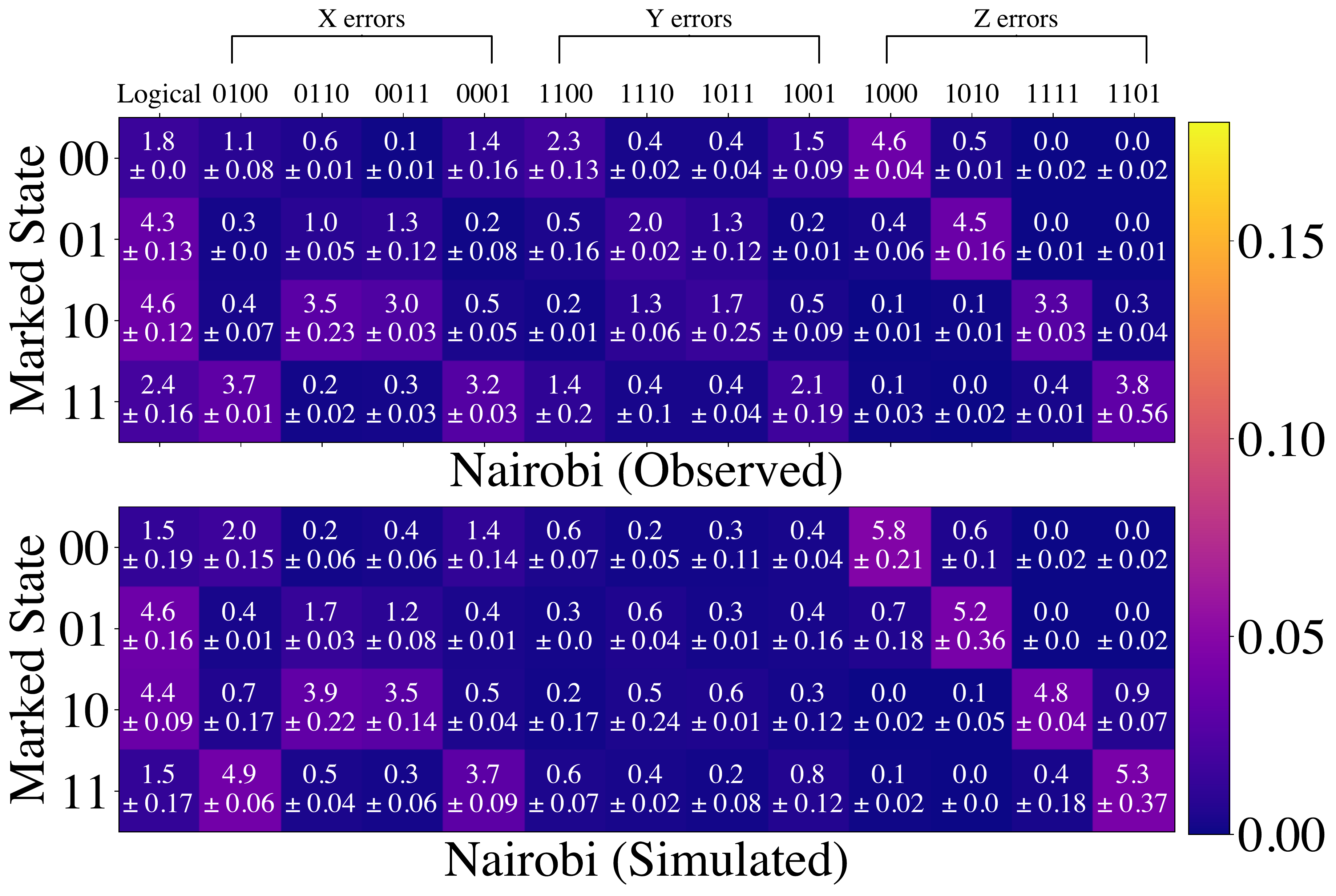}
	\caption{Algorithmic error tomography on Nairobi. Data entries are as in \cref{fig:et-dd}, except that only data without DD is shown. Good agreement is observed between the results of our error model and the experiment. In particular, compared to the error tomography table for Jakarta (\cref{fig:et-dd}), we do not observe an asymmetry in $Z$ errors across marked states.}
	\label{fig:et-nairobi}
	\end{figure}

\begin{figure}[t]
	\centering
	\includegraphics[width=\linewidth]{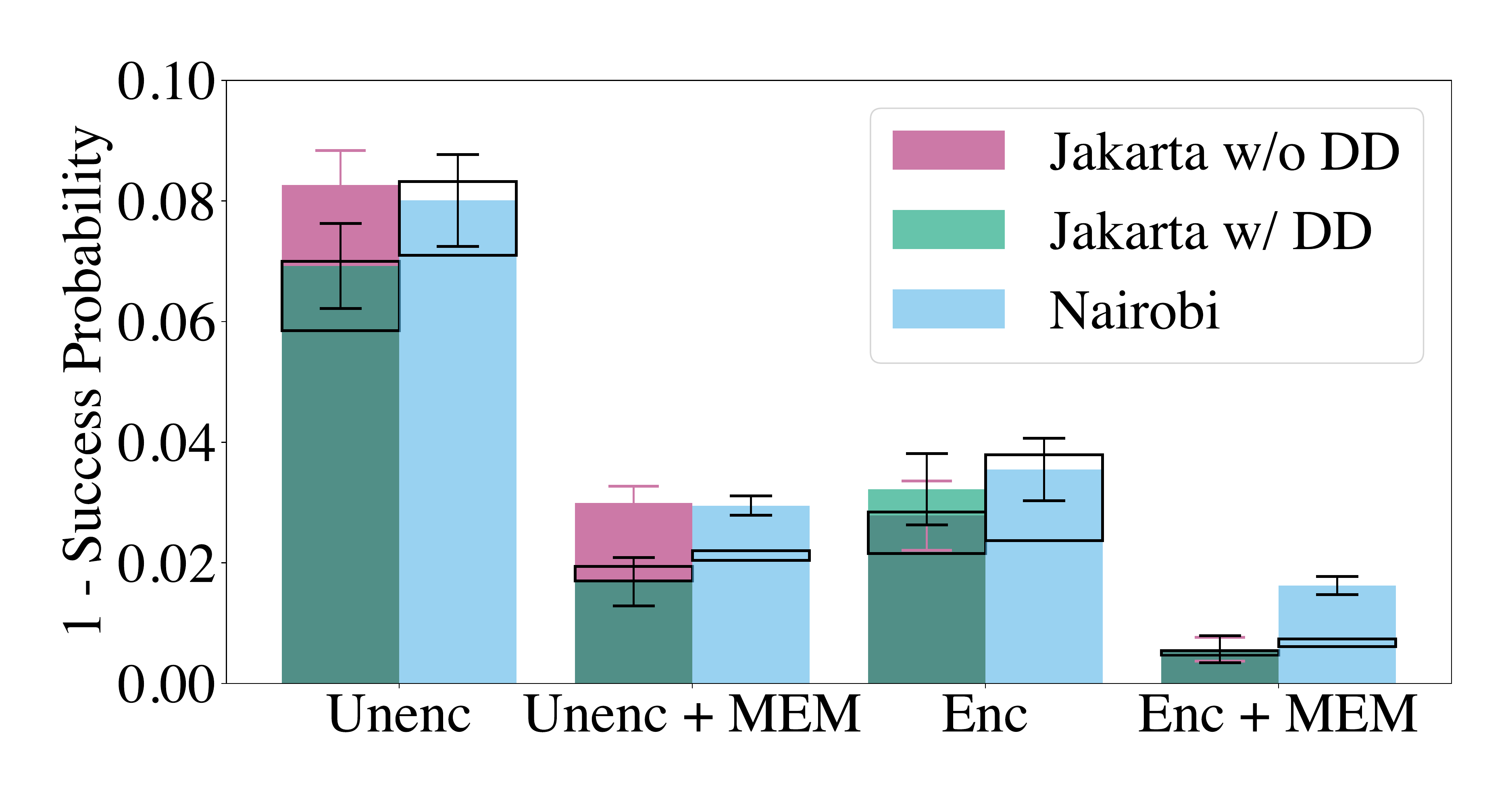}
	\caption{Two-qubit Grover results. Two-qubit single-query Grover failure probability results without (Unenc) and with (Enc) postselection using the $\left[ \left[ 4,2,2 \right]\right]$ code on Jakarta and Nairobi are shown. The transparent boxes represent the theoretically expected failure probabilities from the model described in \cref{sec:sim}, which does not include DD; their centers correspond to the average over marked states, and their boundaries correspond to 95\% confidence intervals after bootstrapping. The colored bars represent the experimental results (see the legend), and the experimental error bars (black for Jakarta with DD and Nairobi, or pink for Jakarta without DD) correspond to 95\% confidence intervals after bootstrapping. Dark green appears where the pink and light green colors (i.e., Jakarta with and without DD) overlap. In the Unenc case,  we run two identical copies of the two-qubit Grover problem to equalize resources with the Enc case and choose the copy with the highest success probability. Also shown are the results with MEM using iterative Bayesian unfolding (see~\cref{app:MEM} for details). Failure probabilities with and without DD protection are shown for Jakarta but not for Nairobi, where the simulated and observed error tomography and failure probabilities are in agreement (see~\cref{sec:2q-oqs-dd}). The presence of DD does not affect the success probability in the encoded implementation, and as a result, the pink bars are mostly hidden behind the green bars. However, the nature of detected errors, even in the encoded case, is affected by DD (see \cref{fig:et-dd}).  All data for different runs on the same QPU were collected on the same day; data from different QPUs were collected on different days.
	\label{fig:2q_fid}
	} 
\end{figure}

\begin{figure}[h]
\includegraphics[width=\columnwidth]{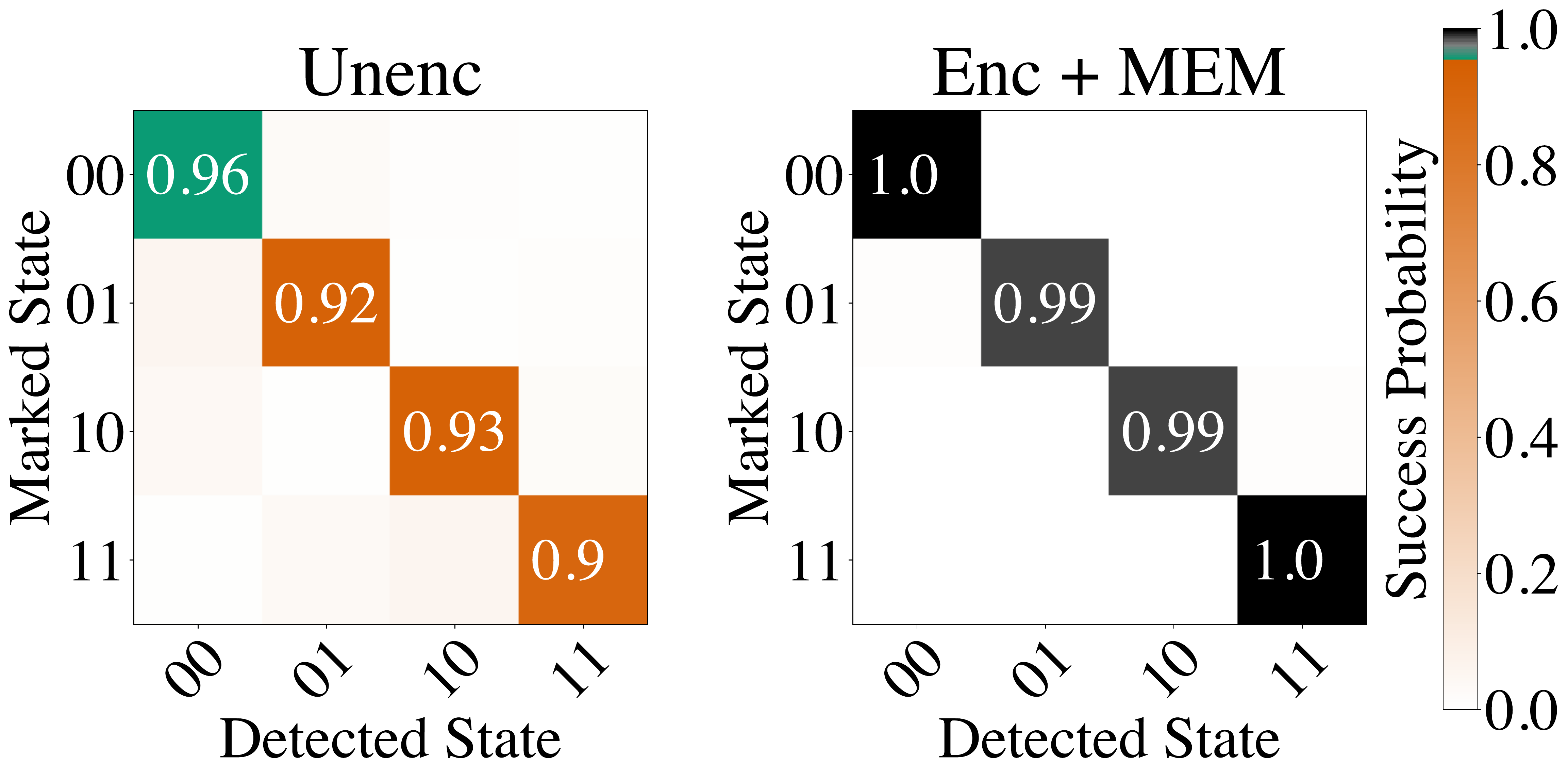}
\includegraphics[width=\columnwidth]{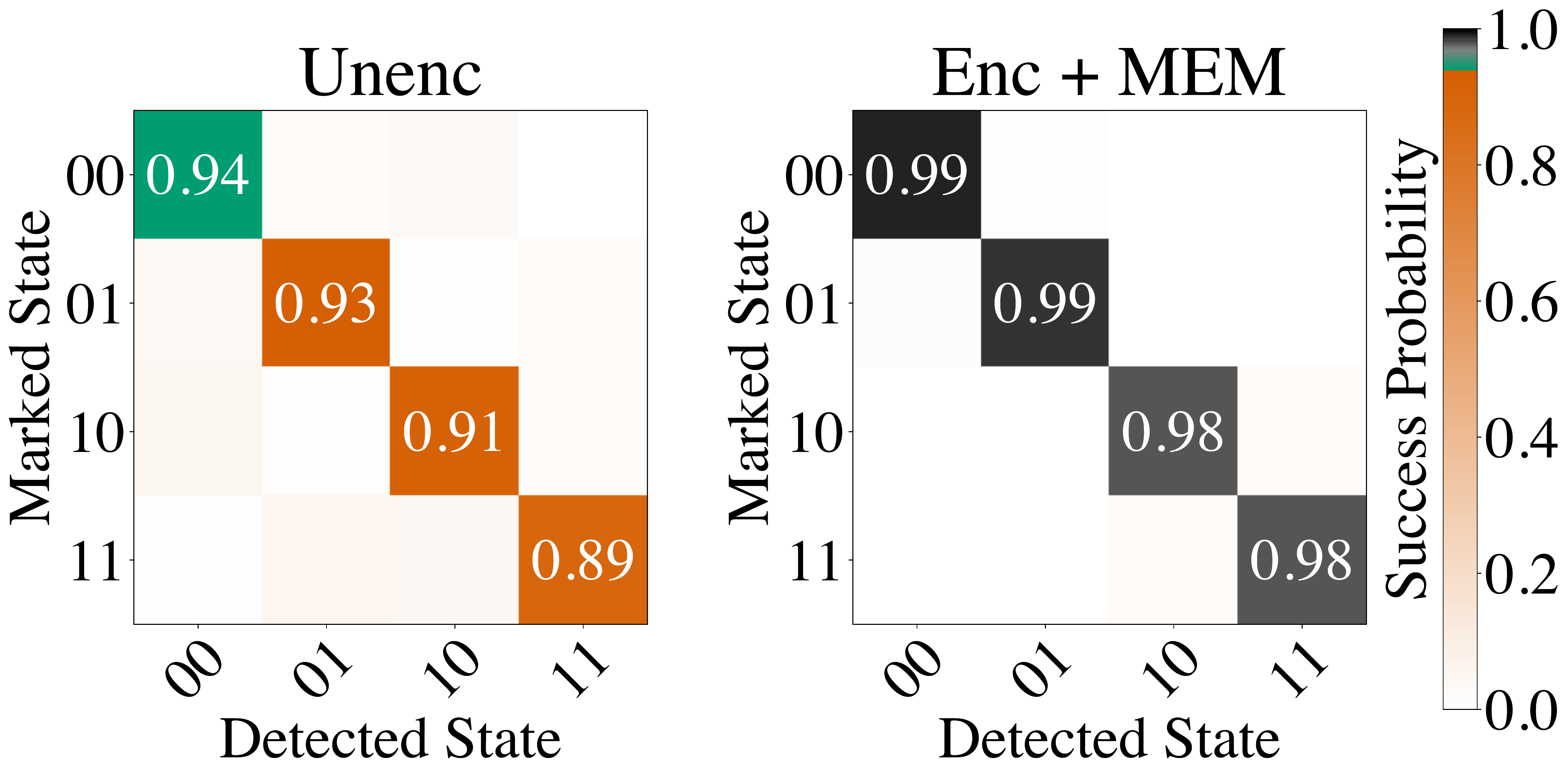}
\caption{Two-qubit Grover results for Jakarta (top) and Nairobi (bottom). The left and right panels show the output distribution for all possible oracles for two setups: unencoded and encoded with MEM. 
As in \cref{fig:2q_fid}, Unenc corresponds to two copies of unencoded two-qubit Grover, of which the best result is reported. Enc corresponds to the results encoded using the $[[ 4,2,2 ]]$ code. The Enc results are reported after postselection. 
Let $p_s(m,b,e)$ be the observed success probability for marked state $m$, detected state $b$, and experiment type $e\in\{\text{Enc,Unenc+MEM,Enc+MEM}\}$.
The success probability changes from orange to green when $p_s(m,b,e)>\max_{m,b}p_s(m,b,\text{Unenc})$. Error bars correspond to 95\% confidence intervals. 
\label{fig:2q-all-results}}
\end{figure}

\subsection{Results}

\subsubsection{Algorithmic error tomography}

\cref{fig:et-dd,fig:et-nairobi} show the results of AET after implementing or simulating the encoded two-qubit Grover algorithm on Jakarta and Nairobi, respectively. Each of the four panels corresponds to a different error outcome table, with the top row representing experiments and the bottom row representing simulations using the model of \cref{sec:sim}. The left and right columns of \cref{fig:et-dd} exclude or include DD, respectively. We did not use DD in the Nairobi case (\cref{fig:et-nairobi}).
Each row corresponds to a different encoded basis state within each error outcome table, i.e., encoded marked state. Other than the logical errors counted during postselection, all other columns represent outcomes ignored during the postselection step. We discuss these results in more detail in the following subsection, showing how AET allows us to identify and mitigate qubit crosstalk.

\subsubsection{DD protection and comparison with the open system model}
\label{sec:2q-oqs-dd}

Recall that we define the success probability $p_s$ as the probability of correctly identifying the marked element. We denote the empirical success probability obtained for a list of $N$ elements after $q$ oracle queries by $p_s^{\text{e}}(q,N)$. Our results are summarized in \cref{fig:2q_fid}, which shows the failure probability ($1-p_s^{\text{e}}(1,4)$) for the unencoded and the encoded implementations on two different QPUs. 

Before comparing the results with and without encoding,  we analyze whether the observed performance matches the model of \cref{sec:sim} in both cases. Let us focus first on Jakarta, where without DD, the empirical failure probabilities in the unencoded case are slightly higher than predicted; see the leftmost column of \cref{fig:2q_fid}. Fortunately, in the encoded case, Jakarta's failure probability overlaps with the prediction bands (\cref{fig:2q_fid}, third column from the left). However, a closer look at the detected errors via AET reveals a different discrepancy. The simulated results for Jakarta (bottom-left table in \cref{fig:et-dd}) do not match the empirical error profile (top-left table of~\cref{fig:et-dd}), which has significantly stronger $Z$ errors and also a state-dependent asymmetry in these errors. In other words, Jakarta does not match the simulations for unencoded or encoded circuits without DD. In contrast, for Nairobi, \cref{fig:et-nairobi} shows that
the AET simulation results agree with the empirically observed ones. This also holds for the simulated failure probabilities (\cref{fig:2q_fid}).

To investigate Jakarta's observed discrepancy, we first attempt to systematically amplify $p_D, T_1$, and $T_2$ by multiplying each quantity by a phenomenologically determined variable $\lambda_i$ (see \cref{app:optimization}). 
This leads to a better overlap between predicted and observed success probabilities but does not reproduce the AET asymmetry seen in~\cref{fig:et-dd}. This shows the limitations of the phenomenological model of \cref{sec:sim} and highlights the level of detail provided by AET.

However, the Jakarta discrepancy is effectively removed after the application of DD. Our two-qubit Grover implementation uses four qubits, leaving three inactive qubits in the 7-qubit QPUs used in our experiments. As there are no idle intervals in the unencoded two-qubit Grover circuit, we applied the XY4 sequence on the inactive qubits -- q$_2$,q$_4$, and q$_6$ (see \cref{app:device}). We applied the XY4 sequence to both the active and inactive qubits for the encoded case. Due to the relative sparsity of idle intervals in the two-qubit Grover circuits, we did not attempt to implement robust sequences, which require more pulses than XY4.

\cref{fig:2q_fid} shows how the failure probability and the rates of various detected errors on Jakarta are affected by the presence of DD. For the unencoded case (the first two columns from left of~\cref{fig:2q_fid}), DD improves the performance slightly, and the discrepancy between the predicted and observed failure probabilities is removed. The improvement by DD in the unencoded two-qubit Grover case is in concurrence with Refs.~\cite{tripathi2021suppression,Zeyuan:22}, which showed the efficacy of the XY4 sequence in suppressing static $ZZ$ crosstalk in superconducting qubits. In other words, these results confirm that $ZZ$ crosstalk -- which is well-documented for superconducting QCs~\cite{krantzQuantumEngineerGuide2019} -- likely contributes to the observed performance being slightly worse than expected from the model. 

Adding the XY4 sequence removes most of the empirical-theoretical discrepancies in both the magnitude and the asymmetry of the errors exhibited by the AET profiles, as seen by comparing the top and bottom right of~\cref{fig:et-dd}. With DD, the encoded circuits have a weaker state-wise asymmetry in $Z$-errors than seen in the left column of \cref{fig:et-dd}. Moreover, the DD-protected circuits more closely reproduce the distribution of detected errors predicted by the model of \cref{sec:sim} than the same circuits without DD. This observation -- that the agreement between our theoretical model and the experimental results improves under DD -- is further validated below.

The close agreement we found for Nairobi between our (crosstalk-free) model and the experimental results without DD or MEM (\cref{fig:et-nairobi} and the first and third from left columns of~\cref{fig:2q_fid}) suggests that crosstalk does not play a significant role in this QPU. 
\cref{fig:2q_fid} does exhibit a significant discrepancy between the model and the experimental Nairobi results when MEM is included (second and last columns of~\cref{fig:2q_fid}). As we show in \cref{app:MEM}, this discrepancy arises from the choice to mitigate readout errors using iterative Bayesian unfolding (IBU)~\cite{nachman2020unfolding}.

Finally, \cref{fig:2q-all-results} complements the first and last columns of \cref{fig:2q_fid}, as well as the AET results, and shows the output distributions for Jakarta and Nairobi for the two-qubit Grover case, with and without encoding and MEM.
The main observation is that for the unencoded case, the maximum success probability is obtained for the marked state $\ket{00}$, which is also the QPU's ground state; this is unsurprising given the dominance of amplitude damping errors. With encoding plus error mitigation, the overall performance increases and becomes independent of the marked state.

\subsubsection{Success probability: beyond break-even improvement}

We now focus on the effect of error detection on two-qubit Grover performance as seen in \cref{fig:2q_fid}. Due to the shallow circuit depth, even without any error detection, $p_s^{\text{e}}(1,4) \sim 93.0\%$ - already much higher than the classical success probability $p_s^{\text{C}}(1,4)=\frac{1}{2}$. Adding error detection improves the success probability to $\sim 96.0\%$. 
The effect of MEM is similar to that of error detection: the success probability increases to $\sim 97.0\%$. Combining error detection with MEM results in additional improvement: we obtain success probabilities of $\sim 98.5\%$ on Nairobi and $\sim 99.5\%$ on Jakarta. Due to error detection and MEM, Jakarta's success probabilities increase by an order of magnitude. 

This improvement over the unencoded case is non-trivial, considering that the $[[4,2,2]]$ code can only detect weight-1 errors, and the encoded circuit requires six two-qubit gates. In contrast, the unencoded version requires only two. The relatively high success probabilities we observe in the encoded case suggest that most errors, even those due to the two-qubit gates, manifest as weight-1 errors. This shows, albeit for a relatively small problem size, that error detection can more than offset the extra errors introduced due to increased circuit depth and complexity. 

We have demonstrated an \emph{algorithmic beyond break-even improvement} using error detection in the sense that the protected algorithm clearly outperforms its unprotected counterpart. Previous break-even improvements were at the individual gate level~\cite{Ofek:2016aa,Ryan-Anderson:22}. Here we have demonstrated such an improvement at the level of the execution of an entire algorithm, albeit of a fixed size. The holy grail is to demonstrate the implementation of an algorithm for a family of problem sizes at the logical level with higher fidelity than the same algorithm executed at the physical level. Achieving this in our setting would require increasing the problem and code sizes. The family of $[[2k+2,k,2]]$ subsystem quantum error detecting codes is an attractive option in this regard since all their logical operators can be chosen to be 2-local~\cite{Marvian-Lidar:16}, which simplifies the circuit design. An experimental implementation of such larger codes and problem sizes remains a coveted goal.

\begin{figure}[t]
	\includegraphics[width=\columnwidth]{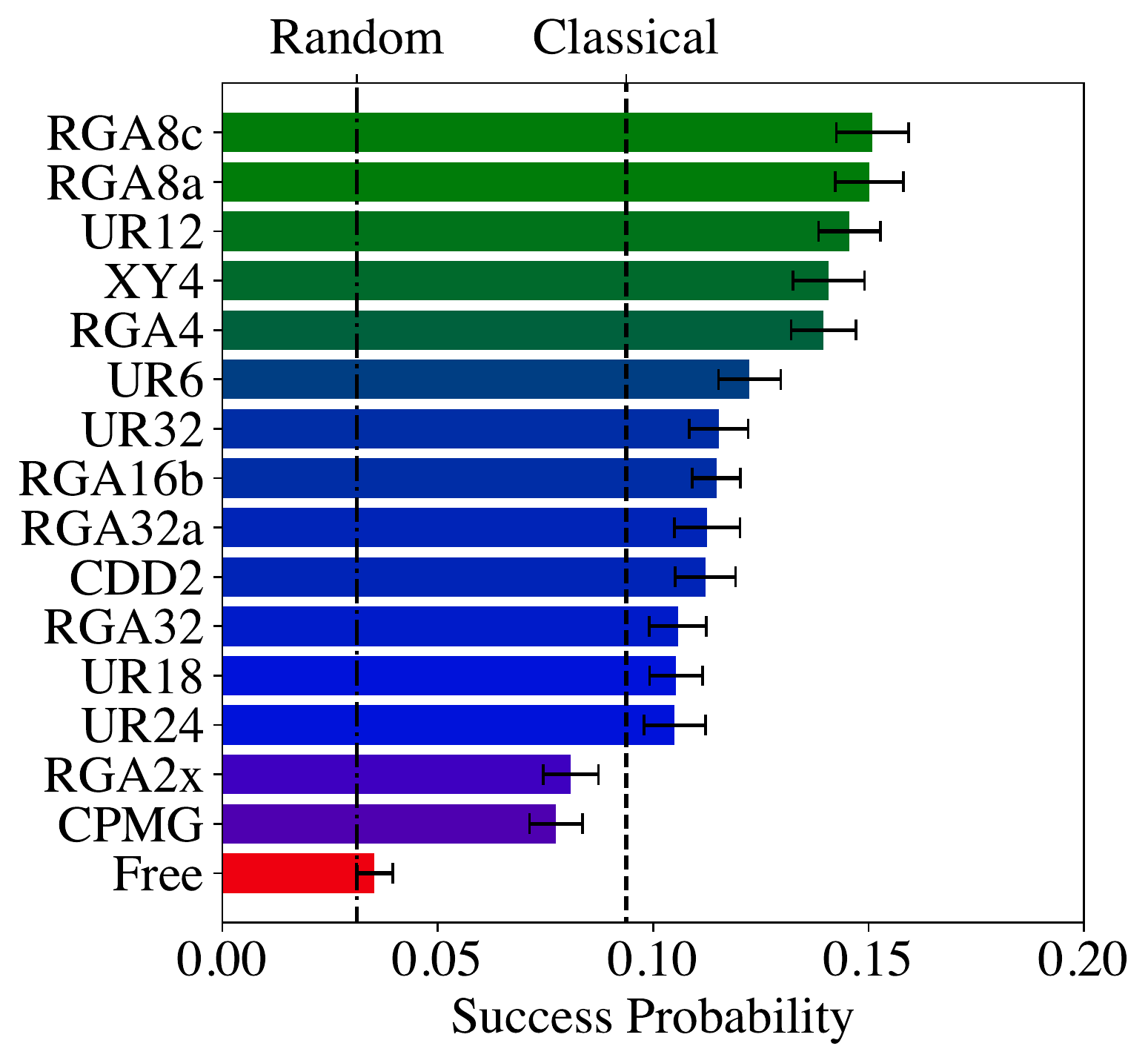}
	\caption{Performance of DD sequences. Average success probability for 5-qubit Grover with two oracle queries on Nairobi. The DD sequences are ranked in order of decreasing success probability. The two dotted lines represent success probabilities corresponding to a random and classical strategy, respectively. RGA8a and RGA8c are tied as the best-performing sequences. Free denotes the result of an unprotected implementation. Error bars correspond to 99\% confidence intervals. }
	\label{fig:dd-nairobi}
\end{figure}

\begin{figure*}[t]
	\includegraphics[width=\textwidth]{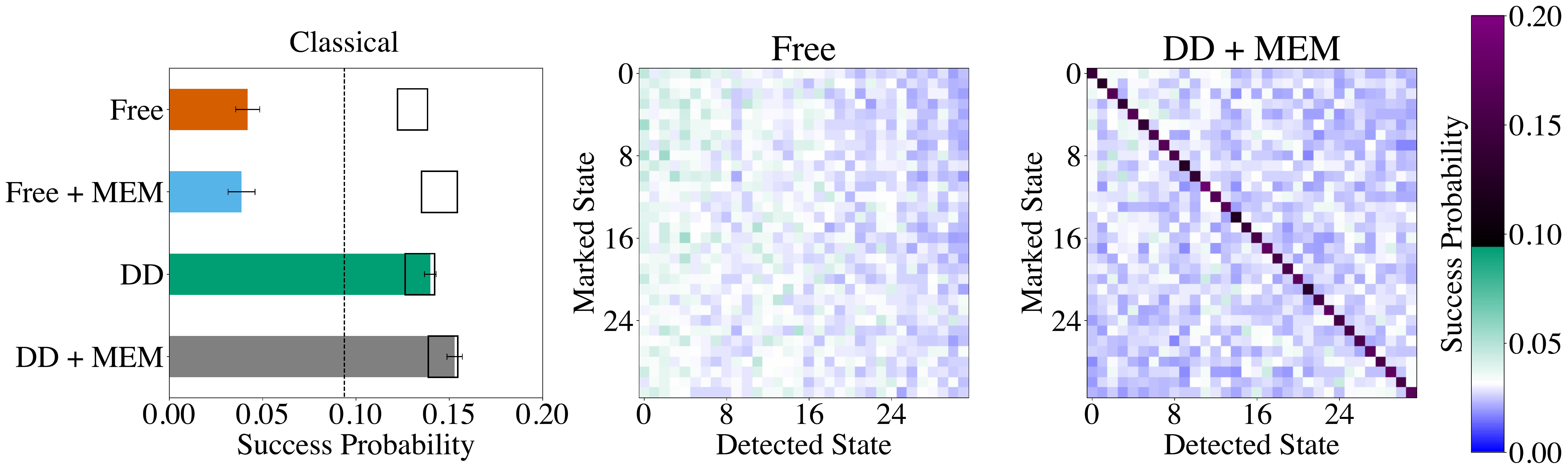}
	\caption{5-qubit Grover results on Nairobi. Left: average success probability with and without DD or MEM for 5-qubit Grover implemented on Nairobi. The boxes correspond to the theoretically expected success probabilities. The quantum oracle is queried twice; in the ideal case, the success probability is 0.602. The unprotected (Free) evolution is on par with a random guess, significantly worse than the optimal classical strategy (dashed vertical line), and just adding MEM does not change the result. In contrast, the DD-assisted implementation crosses the classical threshold, and the results improve even more with MEM, up to a success probability of $0.15$. Error bars correspond to 99\% confidence intervals.
    Middle and right: the complete input-output maps for all $2^5$ marked states, without and with DD + MEM, are shown. States are sorted by increasing Hamming weight; in the Free case, low Hamming weight states have a higher success probability (more green on the left). This is likely to be a consequence of amplitude damping (spontaneous emission), which favors the $\ket{0}$ state of each qubit. In the unprotected case (Free, middle), there is no discernible correlation between the input marked state and the output detected state. In the protected case (DD + MEM right), black-to-purple signifies better-than-classical success probability, and this threshold is crossed for all $32$ marked states. The DD sequence used here is RGA8a~\cite{Quiroz:2013fv}, which was the top-performing sequence in our DD survey (see \cref{fig:dd-nairobi}).}
	\label{fig:5q-compare}
	\end{figure*}

\begin{figure}[htbp]
		\centering
		\includegraphics[trim=20 20 50 50, clip,width=\linewidth]{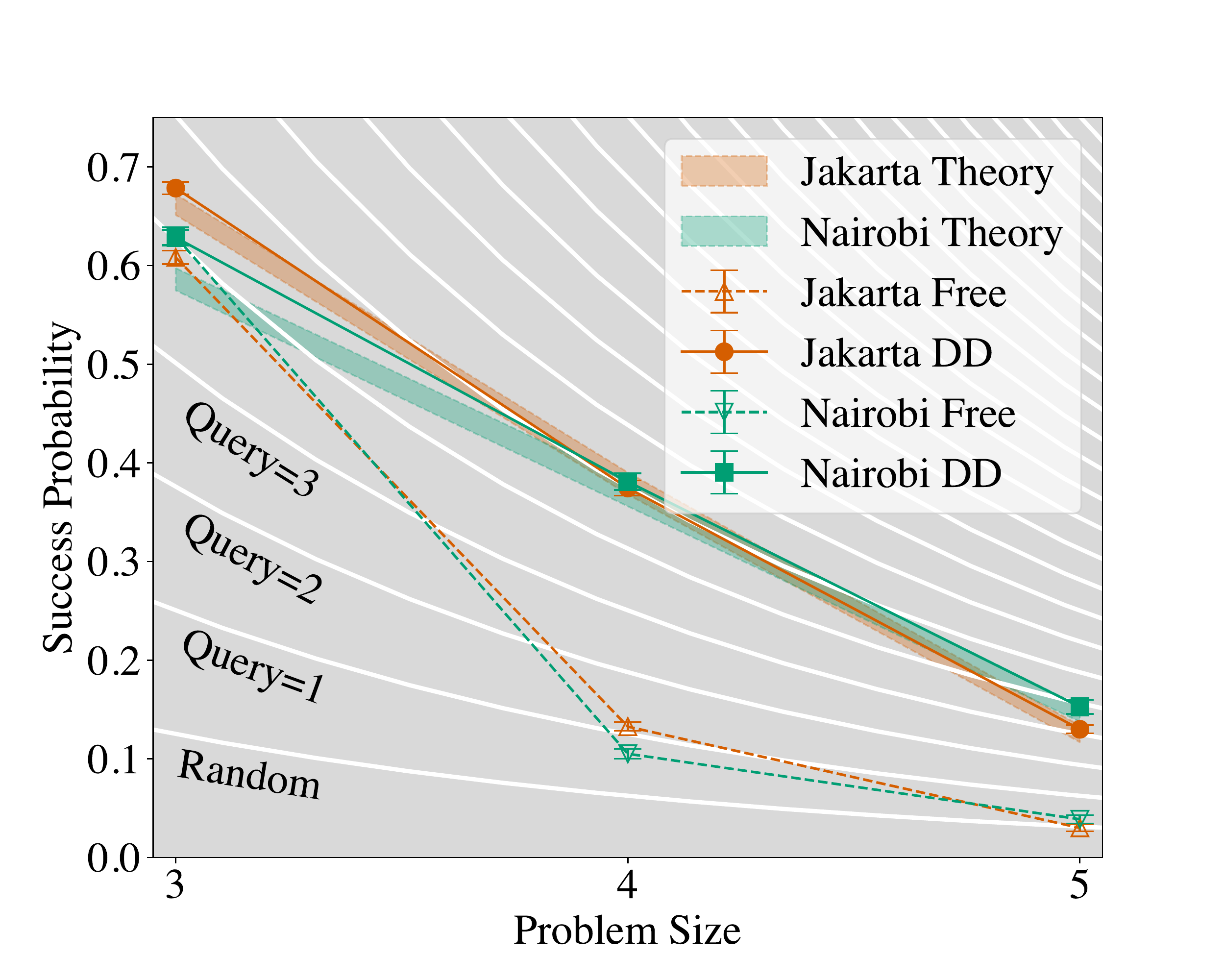}
		\caption{Success probabilities \textit{vs} problem size. Nairobi (green) and Jakarta (orange) success probabilities for $n\in\{3,4,5\}$ are shown for DD-protected and unprotected implementations. The translucent bands indicate the theoretically estimated success probabilities using the model described in \cref{sec:sim}. We performed $q=2$ queries to the quantum oracle in all cases. The ideal success probabilities are $0.945$, $0.908$, and $0.602$ for $n=3$, $4$, and $5$, respectively. The white lines correspond to the success probabilities for the classical strategy and random sampling from the unsorted list ($q=0$). Error bars correspond to 99\% confidence intervals.}
		  \label{fig:success_curves}
	\end{figure}

\section{3-qubit to 5-qubit Grover protected by dynamical decoupling}
\label{sec:3q_and_higher_results}

Crossing the classical threshold in Grover's search for an increasingly larger number of qubits is a meaningful goal, not only because the quadratic speedup offered by Grover's algorithm leads to a more dramatic improvement as the problem size increases but also because it becomes more challenging to realize the speedup experimentally as the controlled phase gate C$_{n-1}Z$ is an $n$-qubit entangling operation. In the implementation of Ref.~\cite{lubinskiApplicationOrientedPerformanceBenchmarks2021}, 5-qubit Grover required nearly a thousand two-qubit gates, and for 8-qubit Grover, nearly 15000 gates were used. Notably, this exponential increase in the number of two-qubit gates with problem size is because Ref.~\cite{lubinskiApplicationOrientedPerformanceBenchmarks2021} did not use ancilla qubits to make the circuits shallower. It is possible to implement C$_{n-1}Z$ with circuits where two-qubit gates scale linearly with $n$ (see \cref{app:circuit-constr}). Ref.~\cite{Zhang_2021} employed shallower circuits for C$_{n-1}Z$ and solved 5-qubit Grover with slightly better-than-random success probabilities but without better-than-classical performance. We use an efficient, ancilla-assisted implementation of generalized Toffoli-type gates~\cite{maslovAdvantagesUsingRelativephase2016, Zhang_2021} to implement C$_{n-1}Z$ (see \cref{app:circuit-constr}). Our implementation, which builds upon the circuits from Ref.~\cite{Zhang_2021}, uses $8$, $14$, and $22$ CNOTs for a single $C_{n-1}Z$ gate for $n=3$, $4$, and $5$, respectively. The deepest circuit we implement is two oracle queries for 5-qubit Grover, totaling $88$ CNOTs. Despite being far shallower than Ref.~\cite{lubinskiApplicationOrientedPerformanceBenchmarks2021}'s implementation,  this is still a relatively deep circuit; e.g., the quantum supremacy demonstration of Ref.~\cite{Arute:2019aa} and the algorithmic quantum speedup demonstration of Ref.~\cite{pokharel2022demonstration} involved circuits of depth up to $40$ and $44$, respectively. As we detail in this section, owing to error suppression via DD, we crossed the classical probability threshold for all problem sizes, including for 5-qubit Grover.

\subsection{Implementation with dynamical decoupling}

DD sequences are inserted into idle intervals of a quantum circuit using the ``decouple then compute strategy'' demonstrated in \cref{fig:dd-on-grover}, which shows the DD insertion scheme for a single query on a 4-qubit Grover circuit. In contrast to two-qubit Grover, where we restricted DD implementation to XY4, $n\geq3$ has ample idle intervals. Therefore we can implement robust sequences (RGA, CDD, UR) requiring more than four pulses. Each of these families has multiple members that are parameterized by the number of pulses in the sequence. We restrict our implementation to DD sequences with fewer than 32 pulses and, for each circuit, only consider sequences that we can fit in the idle intervals available in the quantum circuit.

At each problem size $n$, there are $2^n$ possible oracles, each corresponding to one marked state $\ket{b}$, where $b \in \{ 0,1 \}^n$.  We proceed as follows to avoid implementing this exponentially large set of oracles. Given $0 \leq k \leq n$, there are $\binom{n}{k}$ distinct bitstrings that are identical to $0^k 1^{n-k}$ up to qubit permutation. Recall that marked states differ only by whether $X$ or $I$ gates surround the C$_{n-1}Z$ gate. Thus, we only consider the $n+1$ oracles with marked states $\ket{0^k 1^{n-k}}$, $k\in\{0,\dots,n\}$. We then estimate the average success probability by computing 
\beq
\<p(n)\> = \frac{1}{2^n} \sum\limits_{k=0}^{n} \binom{n}{k} p(| 0^k 1^{n-k}\rangle) .
\eeq
We use $\<p(n)\>$ as the metric for selecting the optimal DD sequences among those we tested and to identify the experimentally optimal number of queries $q^e_{\text{opt}}$. Once the optimal DD sequence and $q^e_{\text{opt}}$ are identified for each $n$, we run the unprotected and the DD-protected Grover's algorithm again at $q^e_{\text{opt}}$, but this time for all $2^n$ oracles.

\subsection{Results}

\subsubsection{Optimal DD sequence and number of queries}

Our first goal is to identify the best DD sequence and $\qopt$. The determinations made in this step inform our choices for the next step. For conciseness, in this section, we focus on the results of the largest problem size we implemented, i.e., $n=5$.  \cref{app:DD-survey} shows the results for $3\le n\le 5$ on both Nairobi and Jakarta. The performance of various DD sequences for Nairobi for 5-qubit Grover are compared in \cref{fig:dd-nairobi} by computing $\<p(5)\>$. The unprotected evolution (Free) is marginally better than choosing an element randomly and does not cross the classical threshold. DD protection is necessary to cross this threshold, but the two-pulse sequences RGA2x and CPMG still result in worse-than-classical performance. The RGA and UR sequences perform well, particularly those with fewer than 12 pulses. RGA8a and RGA8c are tied as the best-performing sequences; we choose RGA8a for the next step, where we implement all of the $2^5$ oracles. The performance improvement seen due to robust sequences is consistent across problem sizes and devices, as detailed in the \cref{app:DD-survey}. 

We also use  $\<p(n)\>$ to identify the experimentally optimal number of oracle queries $q^e_{\text{opt}}$. The theoretically optimal number of repetitions for $n=3,4,5$ is $q_{\text{opt}}=2,3,4$, respectively. However, \cref{app:DD-survey} shows that in reality, the theoretically expected $q_{\text{opt}}$ often leads to worse performance than $q^e_{\text{opt}}$. For the DD-protected implementation, $q=2$ maximizes the success probability $p_s$ in all cases other than 5-qubit Grover on Jakarta, where the performance at $q=2$ is comparable to $q=1$. As DD protection is necessary to cross the classical threshold, for simplicity of analysis and to maximize $p_s$ we set $q=2$ from here on. 

\subsubsection{Better-than-classical performance}

\cref{fig:5q-compare} shows our results for the 5-qubit Grover problem on Nairobi with and without DD. Even in our relatively shallow-depth implementation, before error suppression via DD (whether Free or Free + MEM), the final results are indistinguishable from randomly guessing the marked state. The results change significantly when we implement DD. With DD, the classical threshold is crossed by all marked states. Adding MEM improves the results slightly, but only when accompanied by DD.

This dramatic improvement due to DD holds for other problem sizes as well. \cref{fig:success_curves} shows the success probabilities after two oracle queries on both devices for $3\le n\le 5$ (see \cref{app:PS} for the role of postselection in these results). At the two smaller problem sizes ($n=3,4$), the unprotected implementations are better than random sampling, but the success probability is relatively low. For $n=4$, the unprotected quantum Grover circuit does not exceed the classical single-query threshold. It is effectively on par with random sampling for $n=5$. In contrast, for all problem sizes, the DD-protected quantum strategy at  $q=2$ outperforms the classical strategy for $q \le 3$. DD-protected Grover performance at $n=3,4,5$ is equivalent to classical $q=4,5,3$ for Jakarta and $q=4,5,4$ for Nairobi, respectively. Thus, DD is essential in attaining a better-than-classical performance.

The translucent bands in \cref{fig:success_curves} and the boxes in \cref{fig:5q-compare} (left) show the theoretically expected results computed from the open system model with the IBMQE-supplied parameters. The success probability in this unprotected Grover case (dashed lines in \cref{fig:success_curves}) is considerably lower than the theoretical expectation. This discrepancy is likely due to crosstalk and non-Markovianity, which are well-documented for IBMQE's superconducting qubit-based QPUs. Once we use DD, the observed fidelities improve and are close to the theoretical predictions. This improvement is expected given DD's ability to reduce the effect of crosstalk~\cite{tripathi2021suppression,Zeyuan:22} and non-Markovian effects. 

With DD, the algorithmic performance approaches the expectations based on our error model. However, we emphasize that this model does not predict the QPU's performance under DD; it simply tells us what the performance would be if the reported calibration metrics corresponded to observed dynamics. The overlap between the theoretically predicted (translucent) and the DD-protected (solid) performance implies that DD successfully mitigates the errors that our simple model does not account for. However, the model does not provide an upper bound on the possible performance improvement due to error suppression. For instance, better-optimized sequences could suppress idle-time errors further, and dynamically corrected gates can suppress errors during operations~\cite{khodjasteh:080501,Khodjasteh:2010qd}. 

We note that the restriction to $n\le 5$ arose not because of circuit width but depth. In particular, we used two oracle queries, though theoretically $\qopt=4$ at $n=5$. The gap between the theoretically and experimentally optimal number of queries is expected to grow with problem size. As is true for any quantum algorithm, optimizing circuit compilation and increasing metrics such as $T_1$, $T_2$, and gate fidelities are all vital for scalability. 

\section{Discussion and Conclusions}
\label{sec:conclusion}

We implemented Grover's algorithm of various sizes on multiple superconducting qubit devices. To our knowledge, this is the largest successful demonstration of Grover's algorithm for which the quantum strategy outperforms its classical counterpart. For two-qubit Grover, we focused on error detection via the $[[4,2,2]]$ code and showed that it allowed us to achieve near-optimal performance. Along the way, we introduced the method of algorithmic error tomography. We showed that it provides a wealth of information complementary to previous protocols, such as gate set tomography or just measuring the success probability of an algorithm. We showed that error suppression via DD is essential in attaining better-than-classical performance for larger problem sizes.

Grover's algorithm is a demanding algorithm~\cite{lubinskiApplicationOrientedPerformanceBenchmarks2021} as it requires multiple implementations of C$_{n-1}Z$ --  a fully entangling operation. The superconducting trimon device~\cite{royProgrammableSuperconductingProcessor2020}, which prior to our results achieved the highest success probability for 3-qubit Grover,  is an example of algorithm-tailored hardware where C$_2Z$ is a native gate. Constructing hardware that can natively perform such entangling operations may be one path to realizing the full potential of Grover's algorithm. Still, it is desirable to achieve this goal with more general-purpose quantum hardware, as we have strived to do here.

Today's quantum experimentalists have various error mitigation tools at their disposal. Measurement error mitigation~\cite{maciejewskiMitigationReadoutNoise2020,nachman2020unfolding}, dynamical decoupling, zero noise extrapolation~\cite{kandalaErrorMitigationExtends2019}, and quantum error detection~\cite{Lidar-Brun:book} are parallel strategies that address different kinds of errors. Whether and which error mitigation method to employ must be decided based on the problem and available resources. In this work, we combined MEM with DD and quantum error detection. As expected, these strategies complement each other. However, we found that often, MEM only became useful after DD was employed. Dynamical decoupling, which arguably has the lowest resource overhead and requires no postprocessing, was the single most effective strategy in improving the performance of our implementation of Grover's algorithm. Our work adds to the growing literature~\cite{Pokharel2018,jurcevicDemonstrationQuantumVolume2021,pokharel2022demonstration,raviVAQEMVariationalApproach2021,tripathi2021suppression,DD-survey}  on the effectiveness of error suppression through DD.

While we demonstrated a crossing of the classical threshold at every problem size we tested, better-than-classical success probabilities are not enough to claim a provable quantum speedup~\cite{speedup}. Such a claim would require computing the scaling of the time-to-solution metric as a function of problem size and extending it to the largest possible problem size that can be embedded on the device. Here we could not go to the largest possible problem size as even at $n=5$, our circuit is quite deep -- two queries required $88$ two-qubit gates, and for a larger number of queries or qubits, we no longer observed a quantum advantage. Achieving quantum speedup for Grover search will require devices that can implement circuits much deeper than those used here without a catastrophic drop in fidelity. Recent results~\cite{campbell2019applyingquantum,s2020compilation} suggest that without significant improvements in the surface code implementation, the latter will not necessarily provide an advantage over the type of error suppression and mitigation methods we have explored here. Thus, our results are likely to be necessary (but not sufficient) stepping stones toward a quantum speedup for Grover's algorithm.

\appendix

\section{Classical success probability}
\label{app:classical-p_s}

Let $p_s^{\text{C}}(q,N)$ be the classical success probability after $q$ oracle queries for an unsorted list with $N$ elements. If the oracle is never consulted, then we are simply picking an element at random, and the probability of finding the marked element is $p_s^{\text{C}}(0,N) = 1/N$. At the other extreme, after $N-1$ queries with negative replies from the oracle, we are guaranteed to identify the marked element by selecting the last remaining element so that $p_s^{\text{C}}(N-1,N) = 1$. Clearly, the probability grows in proportion to $q$, so we may conclude that $p_s^{\text{C}}(q,N) = \frac{q+1}{N}$.

A complete argument is the following. 
Suppose there are $N$ elements. With zero queries, we pick the marked element with probability $p_s^{\text{C}}(0,N)=1/N$ and stop. With one query, either we already picked the correct element with probability $p_s^{\text{C}}(0,N)=1/N$ and are so informed by the oracle, or we are told this was the wrong element, so pick again from the remaining set of $N-1$ elements. The probability that the marked element was in the set of $N-1$ is $(N-1)/N$, and the probability of now picking the correct element is $1/(N-1)$: $p_s^{\text{C}}(1,N)=p_s^{\text{C}}(0,N)+(N-1)/N \times 1/(N-1) = 2/N$. With another query, we're now told whether our second pick was correct or wrong; if the latter, we pick again from the remaining set of $N-2$ elements. The probability that the marked element was in the set of $N-2$ is $(N-2)/N$, and the probability of now picking the correct element is $1/(N-2)$: $p_s^{\text{C}}(2,N)=p_s^{\text{C}}(1,N)+(N-2)/N\times 1/(N-2) = 3/N$. Each time we increase the probability of success by $1/N$. Thus, continuing, we have  $p_s^{\text{C}}(q,N)=p_s^{\text{C}}(q-1,N)+(N-q)/N\times 1/(N-q) = (q+1)/N$.

\section{Survey of dynamical decoupling sequences}
\label{app:DD-survey}

\begin{figure*}[htpb!]
	\subfigure[\ $n=3$, Nairobi]{\includegraphics[width=0.305 \textwidth ]{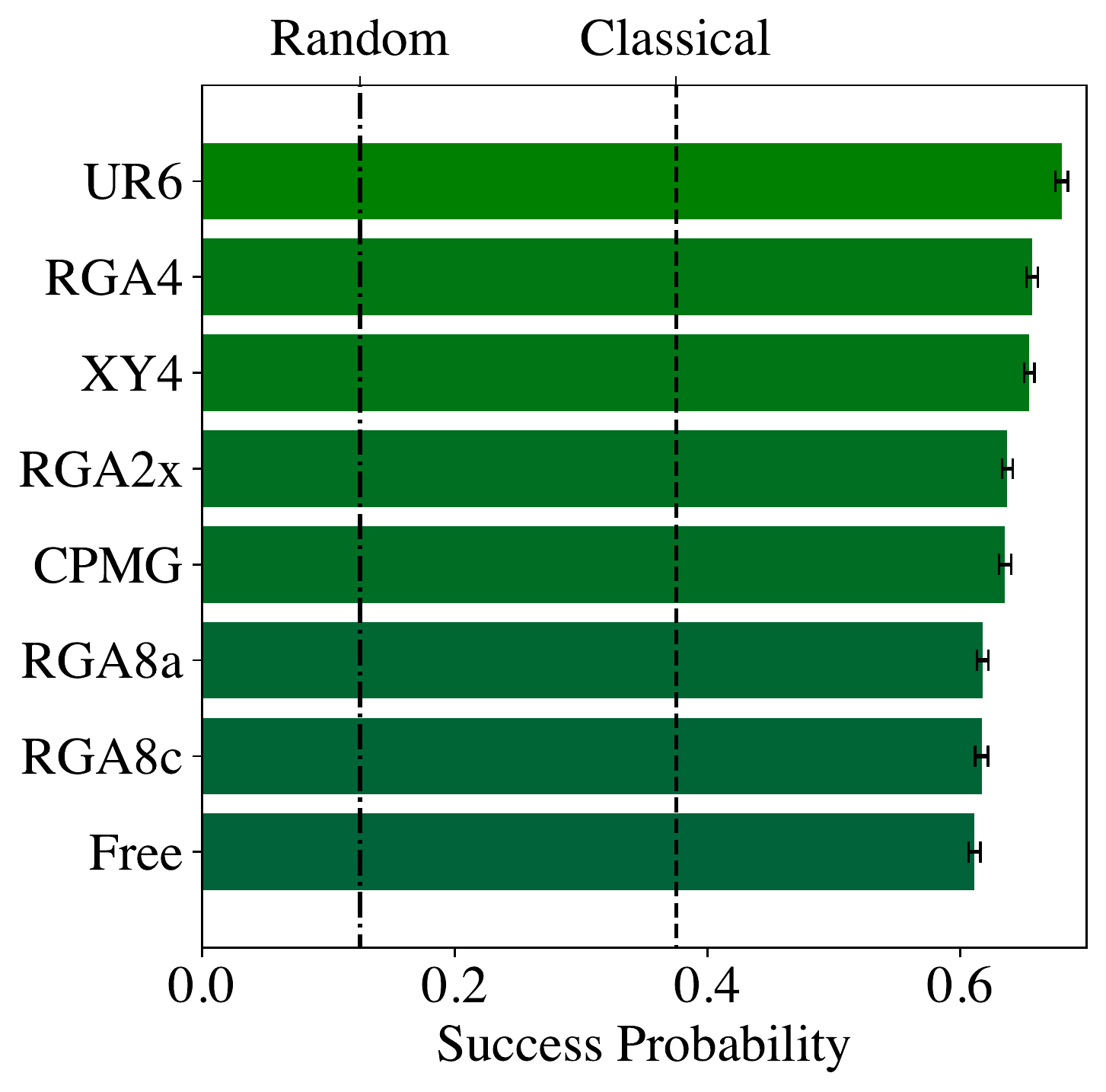}}
	\subfigure[\ $n=4$, Nairobi]{\includegraphics[width=0.32 \textwidth ]{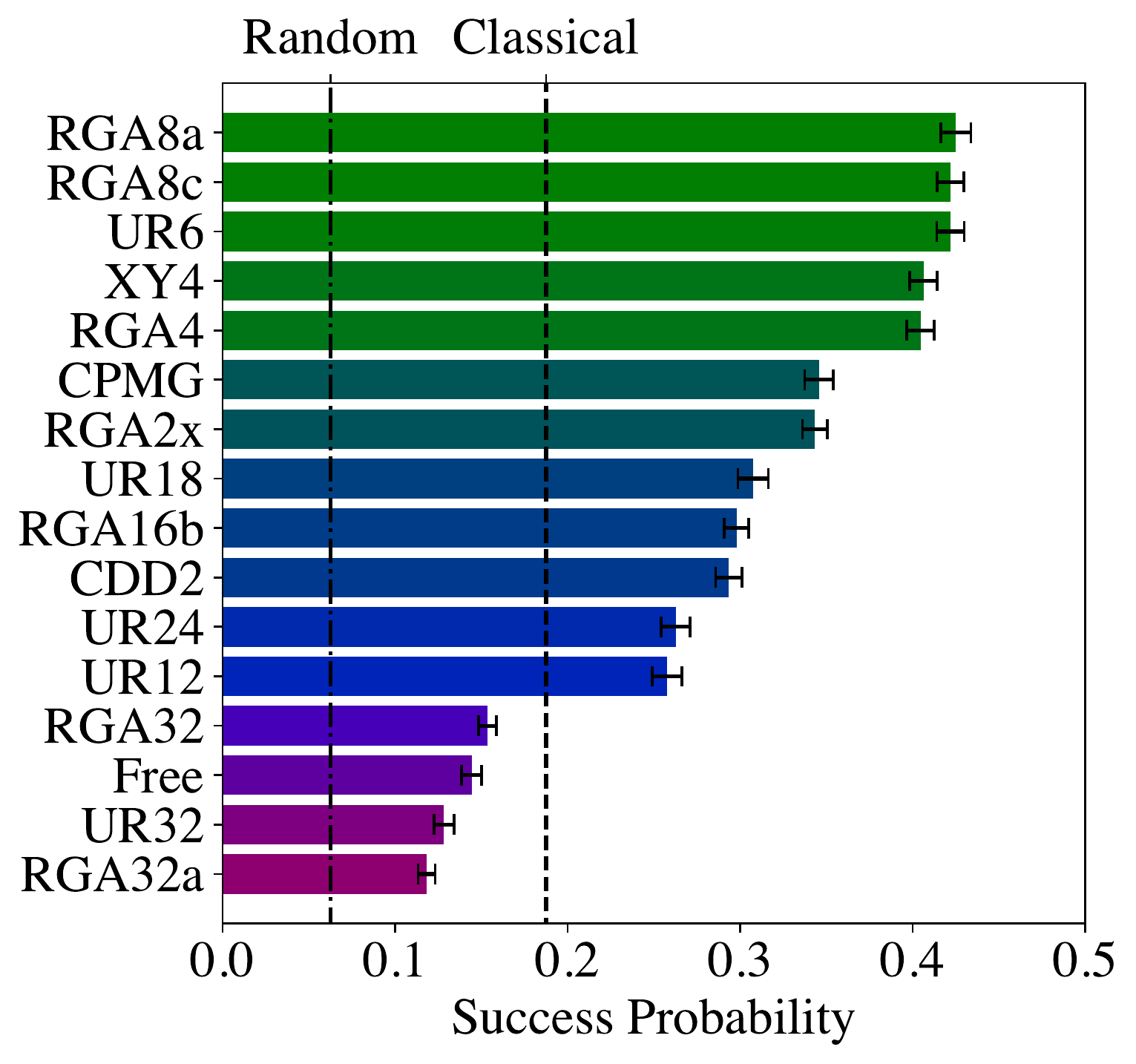}}
	\subfigure[\ $n=5$, Nairobi]{\includegraphics[width=0.32 \textwidth ]{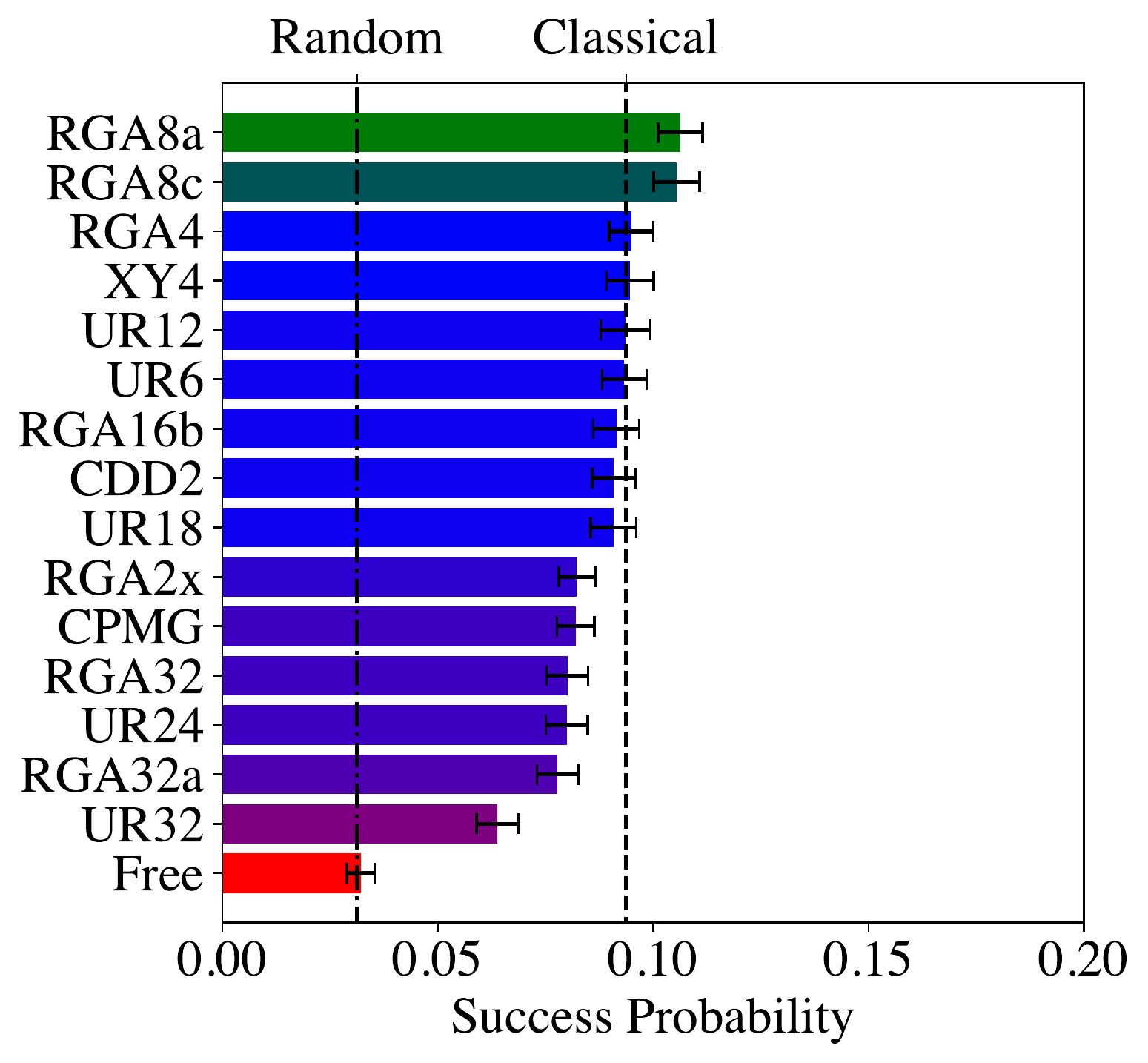}}
	\subfigure[\ $n=3$, Jakarta]{\includegraphics[width=0.305 \textwidth ]{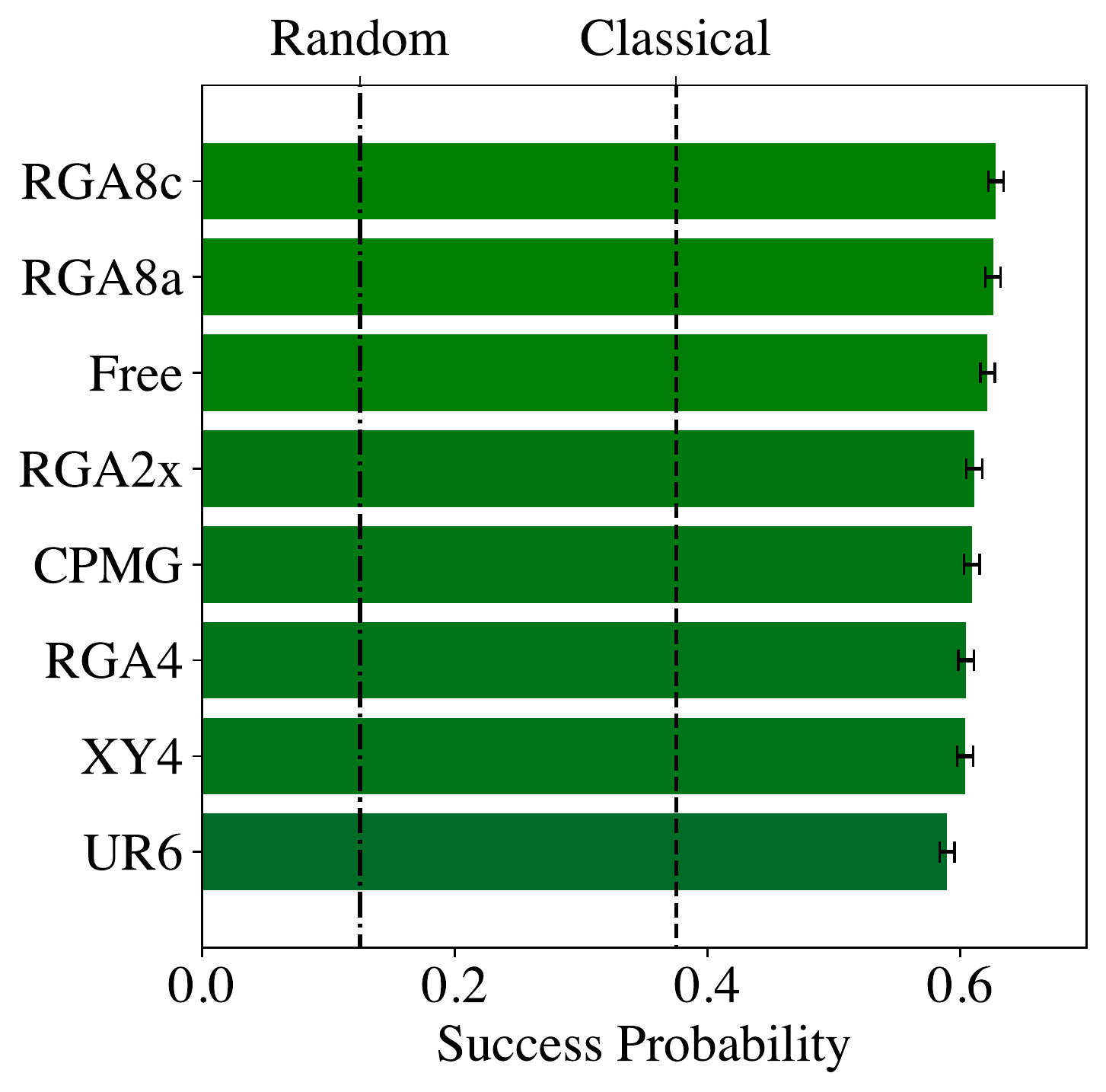}}
	\subfigure[\ $n=4$, Jakarta]{\includegraphics[width=0.32 \textwidth ]{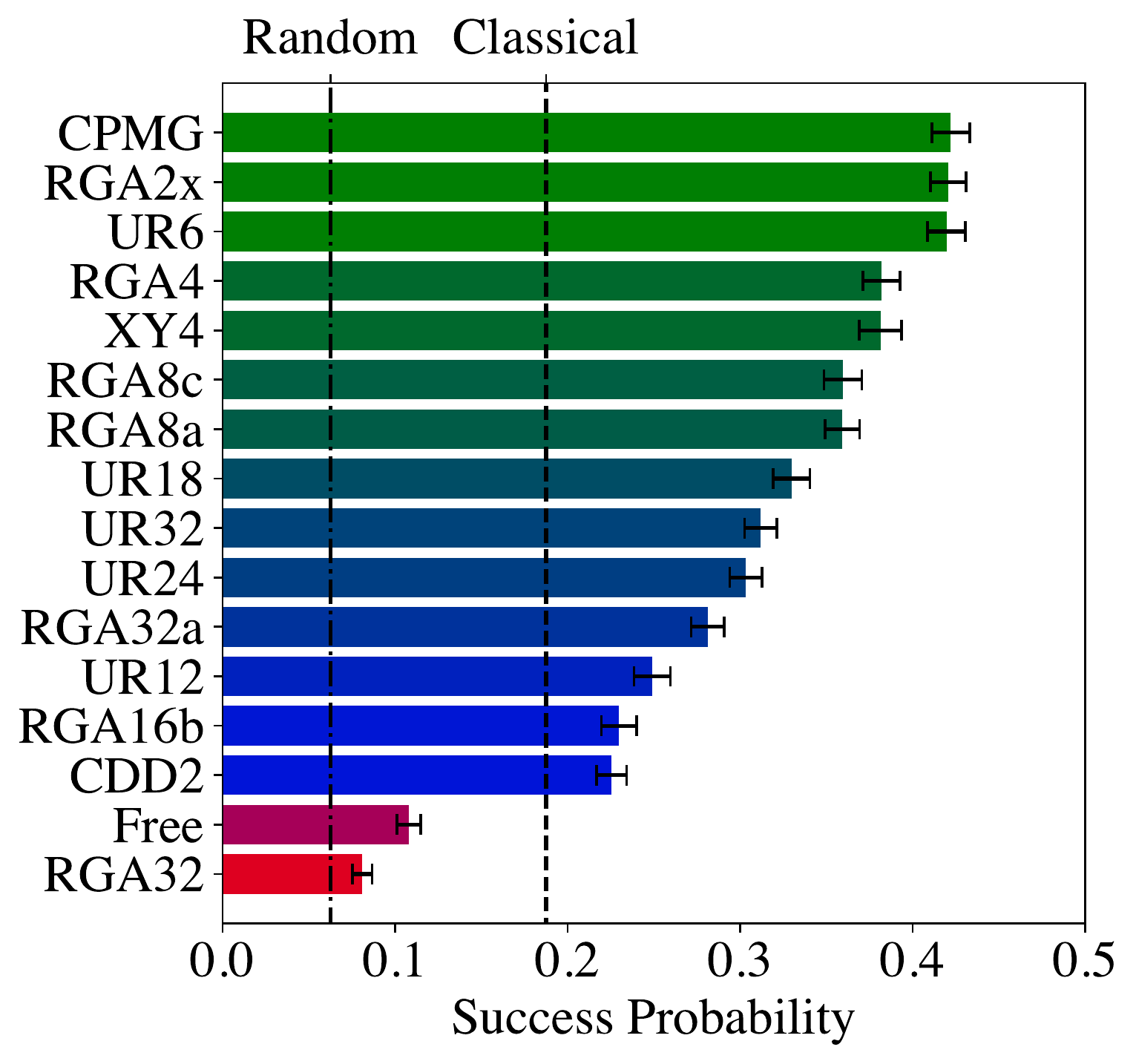}}
	\subfigure[\ $n=5$, Jakarta]{\includegraphics[width=0.32 \textwidth ]{ibm_nairobi_5q_rep-2_dd_comparison_ps.pdf}}
	\caption{Performance of DD sequences, expanding on the results shown in Fig.~\ref{fig:dd-nairobi}. Average success probability for  $n=3,4,5$  with two oracle queries on Jakarta (top) and Nairobi (bottom). The DD sequences are ranked in order of decreasing success probability. The two dotted lines represent success probabilities corresponding to a random and classical strategy, respectively. For $n>3$, the unprotected evolution (Free) is marginally better than choosing an element randomly and does not cross the classical threshold. DD protection is necessary to cross the classical threshold, and the RGA and UR sequences with fewer than 12 pulses are the best performers. Error bars correspond to 99\% confidence intervals.}
	\label{fig:dd-survey}
	\end{figure*}

	\begin{figure*}[htpb!]
		\includegraphics[width=\columnwidth]{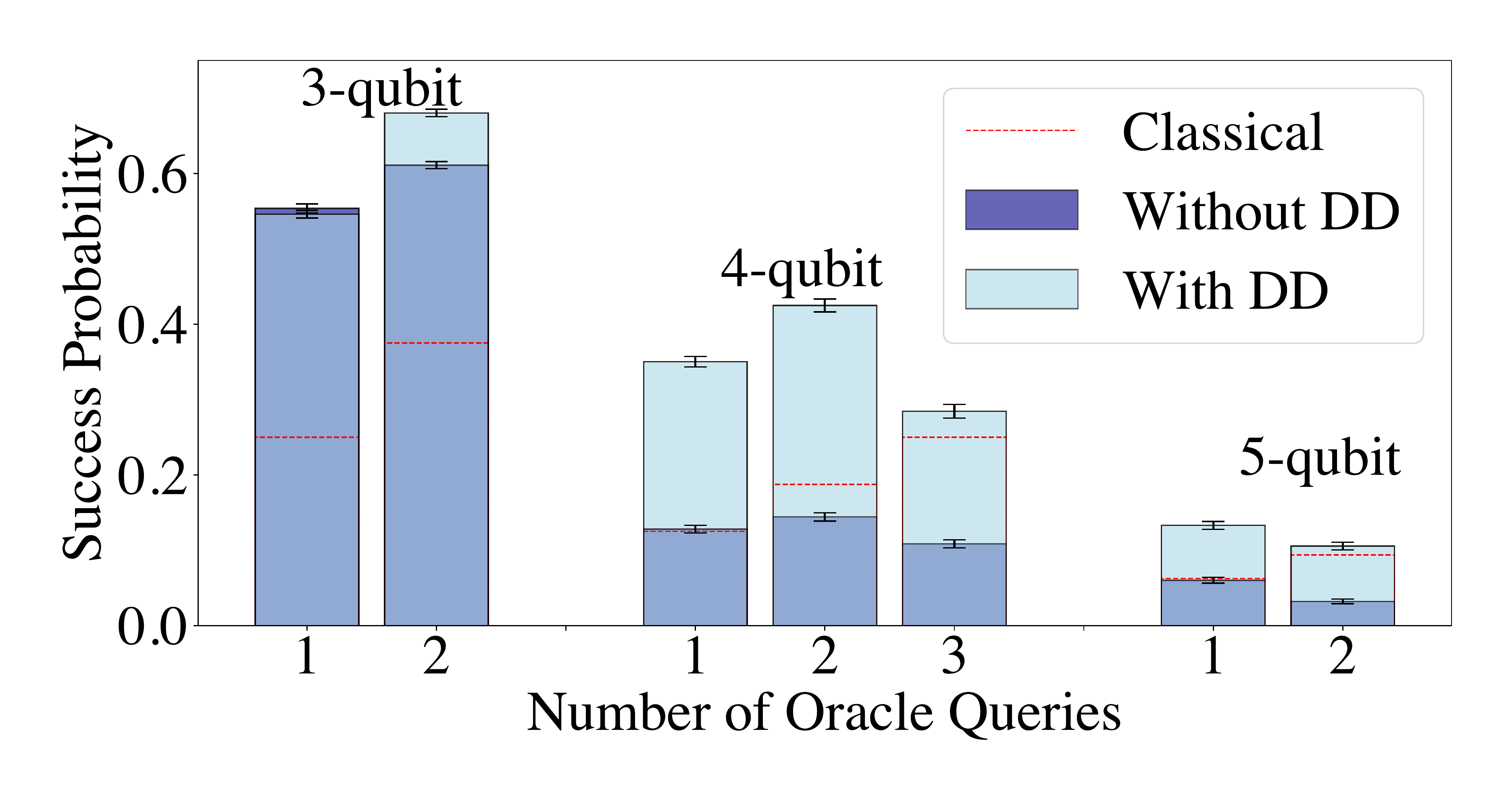}
		\includegraphics[width=\columnwidth]{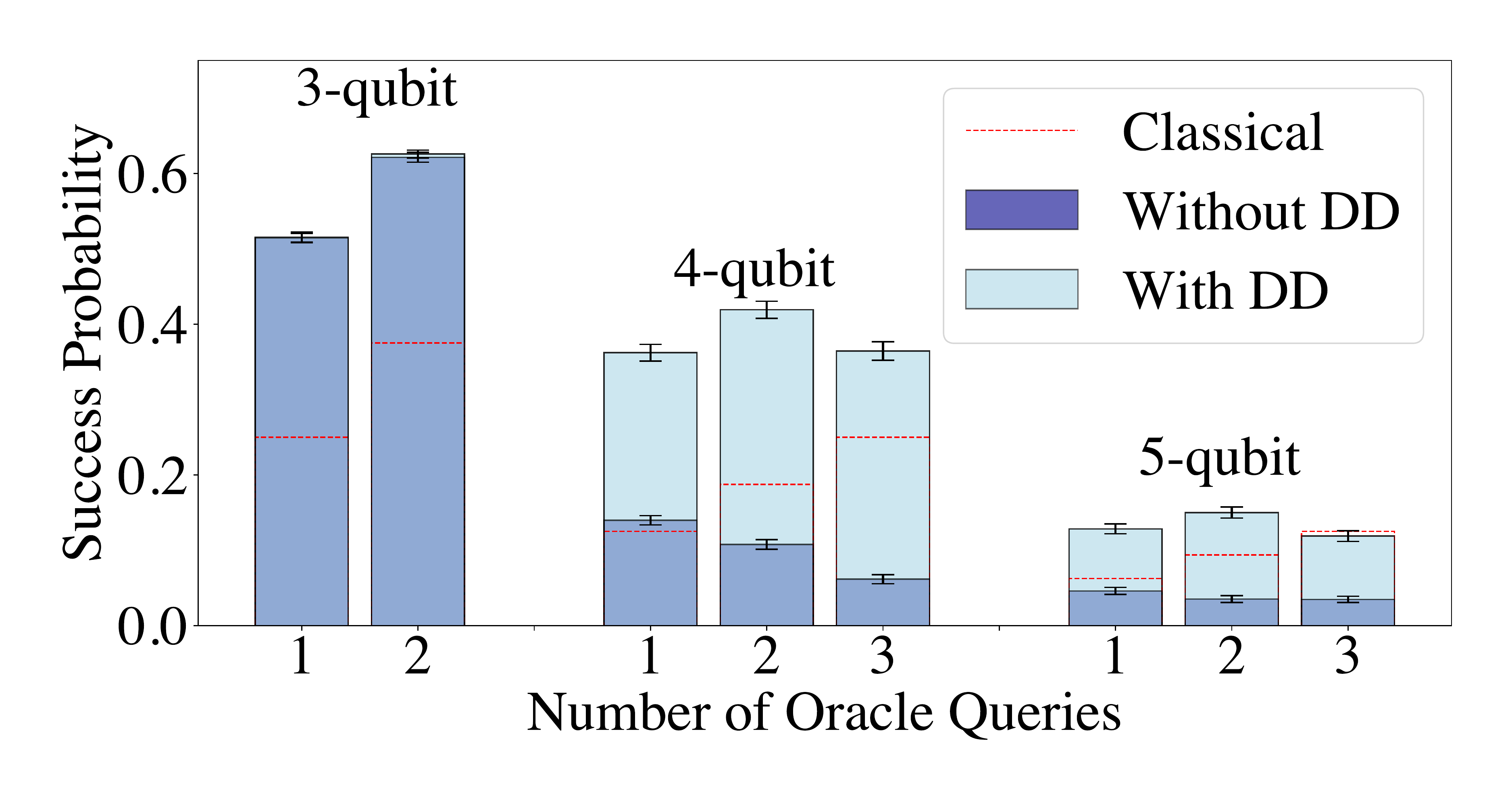}
		\caption{Performance under different oracle query numbers. Success probabilities are shown as a function of the number of oracle queries for Jakarta (left) and Nairobi (right). All results included MEM, and error bars represent 99\% confidence intervals. Dashed red lines correspond to the optimal classical success probability. Except for $n=3$, the classical threshold is crossed only with DD. In the main text, we set $q=2$, which is the optimal number of repetitions for all instances other than $n=5$ on Jakarta. Error bars correspond to 99\% confidence intervals. }
		\label{fig:reps_v_probs}
		\end{figure*}

In addition to the well-known XY4 and CPMG sequences, we consider three families of robust dynamical decoupling sequences. These sequences are expected to work well on a superconducting device with finite pulse width and flip-angle errors. The first sequence family is concatenated DD (CDD). CDD comprises recursively generated sequences by concatenating a base sequence such as the XY4 sequence. Formally, 
\bes
\begin{align} 
	\mathrm{XY}4/\mathrm{CDD}_{1} & \equiv Y-X-Y-X \\ 
	\mathrm{CDD}_{n} & \equiv \mathrm{XY} 4\left(\left[\mathrm{CDD}_{n-1}\right]\right) 
\end{align}
\ees
Here we could only proceed as far as $\mathrm{CDD}_{2}$, as the idle intervals in the circuit were too short to incorporate higher-order CDD sequences. 

The second family comprises the robust genetic algorithm (RGA) sequences~\cite{Quiroz:2013fv}. These were found by assuming a generic single-qubit error term and a numerical optimization using genetic algorithms. A subset of the sequences was enforced to be robust against flip-angle errors. Therefore, these sequences are called robust genetic algorithm sequences. Due to duration constraints, we only attempted sequences up to $32$ pulses, even though longer sequences were identified in Ref.~\cite{Quiroz:2013fv}.
\bes
\begin{align}
\mathrm{RGA}_{4} & \equiv \bar{Y}-X-\bar{Y}-X \\ 
\mathrm{RGA}_{4 p} & \equiv \bar{Y}-\bar{X}-\bar{Y}-\bar{X} \\ 
\mathrm{RGA}_{8 a} & \equiv X-\bar{Y}-X-\bar{Y}-Y-\bar{X}-Y-\bar{X} \\
\mathrm{RGA}_{8 c}/\mathrm{EDD} & \equiv X-Y-X-Y-Y-X-Y-X \\ 
\mathrm{RGA}_{16 b} & \equiv \mathrm{RGA}_{4 p}\left(\left[\mathrm{RGA}_{4 p}\right]\right) \\ 
\mathrm{RGA}_{32 a} &\equiv \mathrm{RGA}_{4}\left(\left[\mathrm{RGA}_{8 a}\right]\right) \\ 
\mathrm{RGA}_{32 c} & \equiv \mathrm{RGA}_{8c}\left(\left[\mathrm{RGA}_{4}\right]\right) . 
\end{align}
\ees
Here ${X}$ means a $\pi$-rotation about the $+x$ axis. In contrast, $\bar{X}$ means a $\pi$-rotation about the $-x$ axis (see Ref.~\cite{DD-survey} for a concise and detailed summary with more explicit definitions, including the effect of pulse width and the associated errors).

Finally, the third family is that with universally robust (UR) sequences~\cite{Genov:2017aa}. UR sequences are defined such that
\bes
\begin{align} 
\mathrm{UR}_{n} &=(\pi)_{\phi_{1}}-(\pi)_{\phi_{2}}-\ldots-(\pi)_{\phi_{n}} \\
\phi_{k} &=\frac{(k-1)(k-2)}{2} \Phi^{(n)}+(k-1) \phi_{2} \\ 
\Phi^{(4 m)} &=\frac{\pi}{m} \Phi^{(4 m+2)}=\frac{2 m \pi}{2 m+1} ,
\end{align}
\ees
where $(\pi)_{\phi}$ is rotation about the axis at an angle $\phi$ from the $+x$-axis. We choose $\phi_1 = 0$, and $\phi_2=\Phi(n)$ so that all UR$_n$ sequences are palindromic. Once again, we constrained our survey to sequences with up to $32$ pulses. 

Our results from testing these three robust families of DD sequences are shown in \cref{fig:dd-survey}. For $n>3$, almost all DD sequences improved the success probability, but even among the sequences tried, there was a considerable variation. Robust sequences with fewer than $12$ pulses per DD cycle were the best performers. The eventual decrease in the performance of sequences with an increasing number of pulses is to be expected as they are implemented using noisy gates, and there is a trade-off between the protection provided by DD and the accumulation of gate errors. The RGA8c and RGA8a sequences performed consistently well and are the only sequences to cross the classical threshold on Jakarta for $n=5$. RGA8c is also commonly known as the Eulerian DD (EDD) sequence, and RGA8a is a slightly modified version of EDD. These palindromic sequences are known to be robust against flip-angle and finite-width errors. The best sequence at each problem size is shown in \cref{tab:seq}.

\begin{table}[h]
    \centering
    \begin{tabular}{|l|l|l|}
    \hline
        Problem Size & Jakarta & Nairobi \\ \hline
        $n = 3$ & UR6 & RGA8a \\ \hline
        $n = 4$ & RGA8a  & UR6 \\ \hline
        $n = 5$ & RGA8c & RGA8a \\ \hline
    \end{tabular}
	\caption{The best-performing DD sequence at each problem size for both QPUs. These sequences were determined by implementing $n+1$ oracles of the form $0^k1^{n-k}$ for the $n$-qubit Grover problem.}
	\label{tab:seq}
\end{table}

	\begin{figure*}[!htpb]
		\subfigure[Jakarta, $n=3$]{\includegraphics[width=\columnwidth]{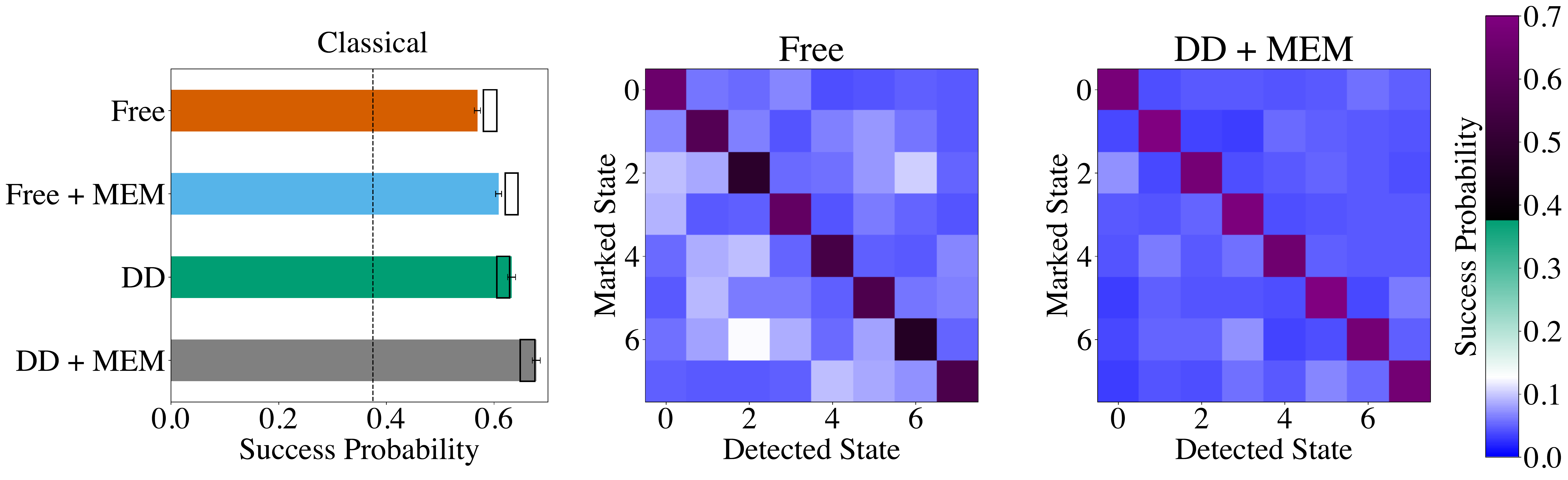}} 
		\subfigure[Nairobi, $n=3$]{\includegraphics[width=\columnwidth]{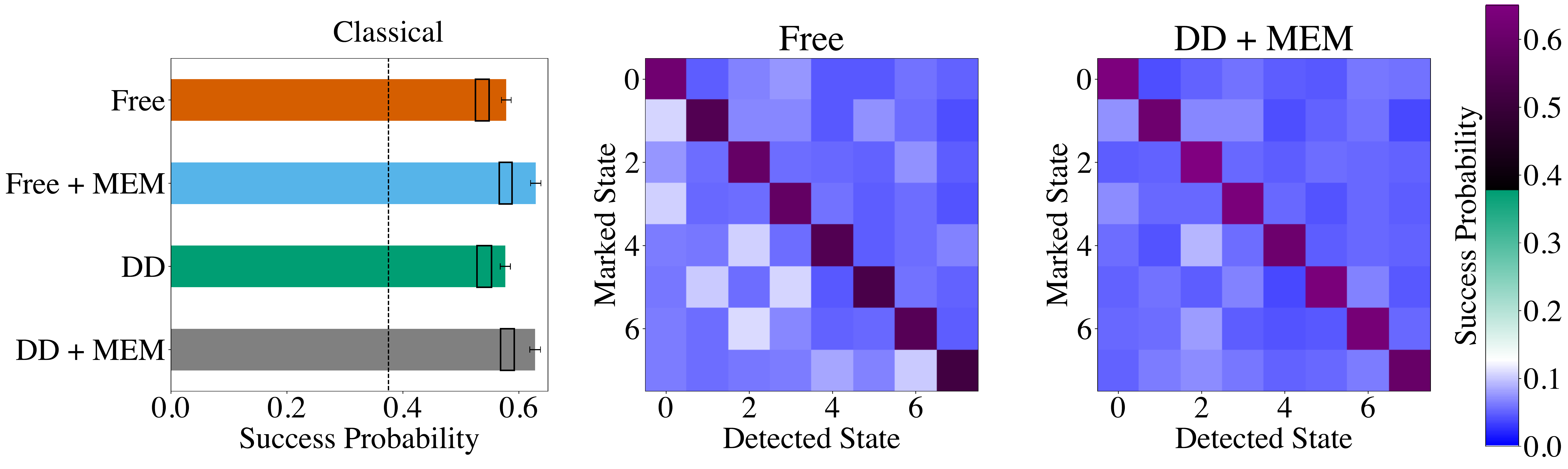}} 
		\subfigure[Jakarta, $n=4$]{\includegraphics[width=\columnwidth]{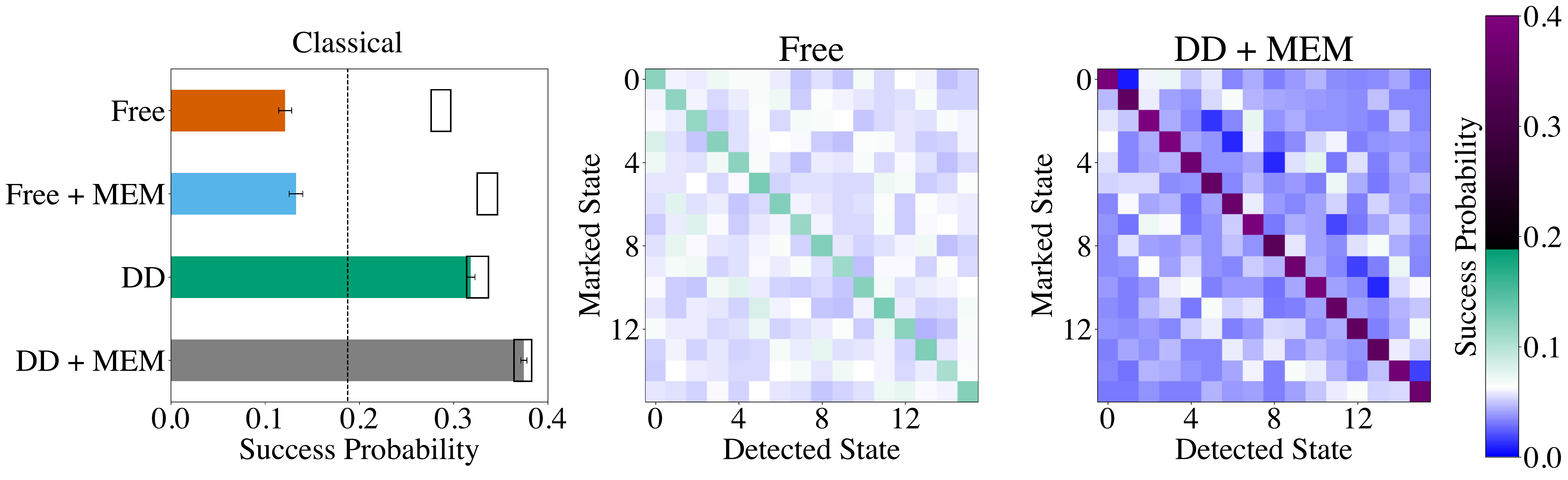}} 
		\subfigure[Nairobi, $n=4$]{\includegraphics[width=\columnwidth]{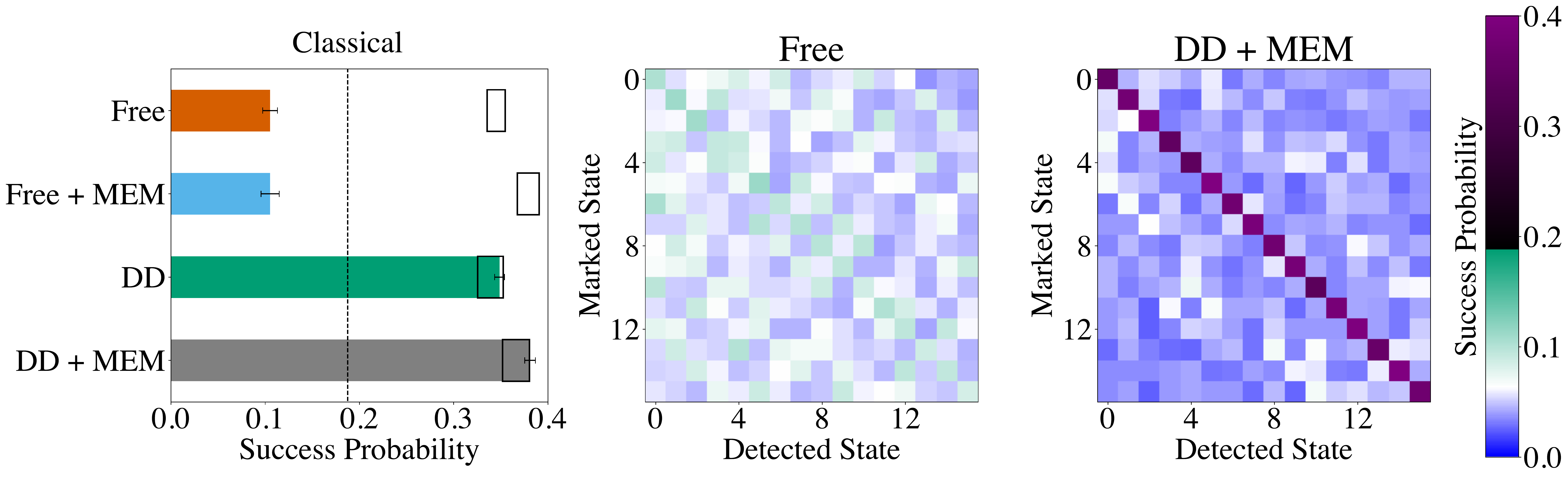}} 
		\subfigure[Jakarta, $n=5$]{\includegraphics[width=\columnwidth]{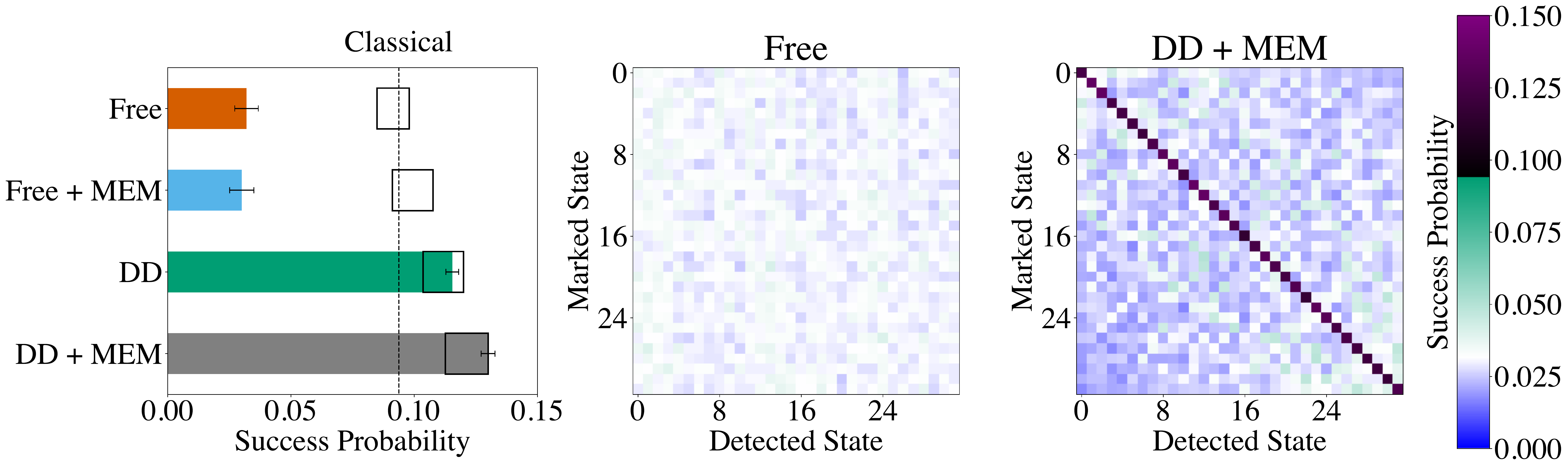}} 
		\subfigure[Nairobi, $n=5$]{\includegraphics[width=\columnwidth]{ibm_nairobi_5q_rep-2_dist_simdd.pdf}} 
\caption{3-qubit, 4-qubit, and 5-qubit Grover on Jakarta (left) and Nairobi (right) after two oracle queries, complementing \cref{fig:5q-compare}, which only shows Nairobi for $n=5$. Each row represents a problem size in ascending order. In a row, the horizontal bar plot on the left shows the success probability under no error suppression and mitigation (Free), with measurement error mitigation (Free + MEM), with DD protection (DD), and with DD protection and measurement error mitigation (DD + MEM). The dashed horizontal line and the boxes represent the classical and the theoretically expected success probability, respectively. The second and third columns show the input-output map for Free and DD + MEM, highlighting the improvement offered by these strategies. The states are sorted by increasing Hamming weight. The transition from green to black occurs at the classical success probability threshold. With DD protection, the classical threshold is crossed in all cases.  }		
\label{fig:jakarta-all+nairobi-all}
		\end{figure*}

\cref{fig:reps_v_probs} shows the experimental success probabilities for the unprotected and DD-protected Grover circuits for all queries $q$. Here, only the best DD sequence from the survey above (listed in \cref{tab:seq}) is used in each case. Theoretically, for $n=3,4,5$, $\qopt = 2,3,4$ respectively. Unfortunately, for 5-qubit Grover, software restrictions prevented us from going beyond $q=2$ and $3$ on Jakarta and Nairobi, respectively. However, it is already clear that the experimentally optimal value was reached in both cases. Recall that we restricted our results to $q=2$ oracle queries in the main text. For $n=3$, this is both experimentally and theoretically optimal. For Nairobi, two queries have the highest experimental success probability for all problem sizes. For Jakarta and $n=5$, a single query has a slightly higher success probability, but the difference between $q=1$ and $q=2$ is not substantial. Overall, for simplicity of analysis, in the main text, we focused only on results for $q=2$.

Finally, \cref{fig:jakarta-all+nairobi-all} shows the results for all oracles at two queries using the DD sequence found from the survey above. The results are qualitatively identical on both devices. We have already clarified that DD is necessary to cross the classical threshold. One might suspect that majority voting may suffice to declare a detected state as the marked state if it is the mode of its corresponding probability distribution. However, even under this criterion, for 5-qubit Grover, there is no way to detect the marked state without DD.

\section{Device Specifications}
\label{app:device}

\begin{figure}[h]
\includegraphics[trim = 80 70 70 70, clip,width=0.7\columnwidth]{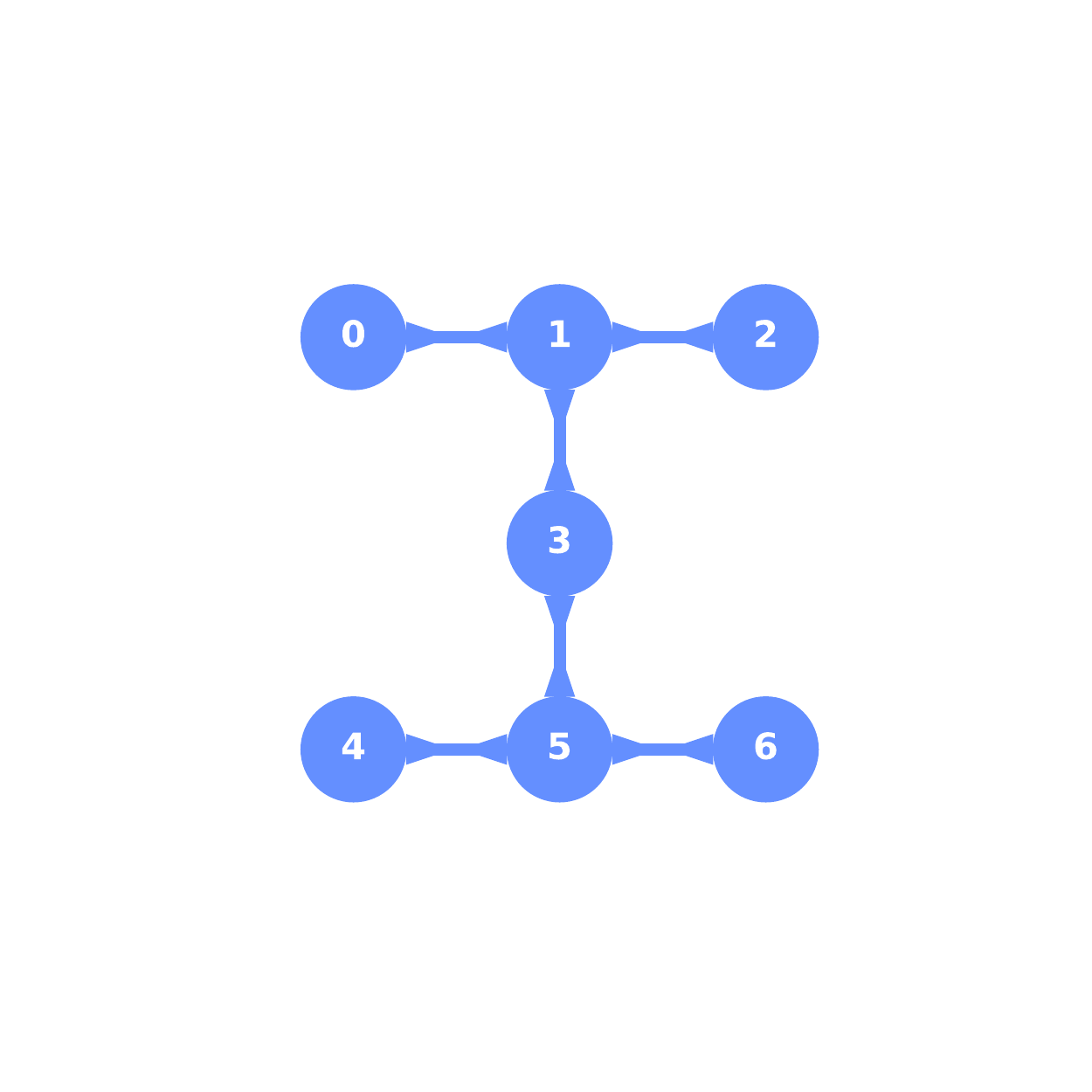}
\caption{Device connectivity. Jakarta and Nairobi devices are built using the IBM Quantum Falcon r5.11H processors and have seven qubits. 
\label{fig:gate-map}}
\end{figure}

Jakarta and Nairobi are 7-qubit QPUs comprising the IBMQE Falcon r5.11 processors~\cite{ibm}. On these transmon qubit-based devices, single qubit gates are performed by driving a DRAG pulse, and two-qubit gates are implemented using echoed cross-resonance gates. They have quantum volumes of 16 and 32, respectively~\cite{IBMQuantum2022}. The qubit connectivity for these QPUs is shown in \cref{fig:gate-map}. \cref{tab:config} shows the gate errors, readout errors and the $T_1$ and $T_2$ times for both devices.  The qubits used for each $n$ are listed below in increasing order: q$_0$, q$_1$, q$_3$, and q$_5$ for our encoded two-qubit Grover implementation; q$_0$, q$_1$, and q$_2$ for 3-qubit Grover; q$_0$, q$_2$, q$_3$, q$_5$ as the main qubits and q$_1$ as the ancilla for 4-qubit Grover;
q$_0$, q$_2$, q$_3$, q$_5$, q$_6$ as the main qubits and q$_1$ as the ancilla for 5-qubit Grover.

\begin{table}[h]
    \centering
    \begin{tabular}{|l|c|c|c|c|c|c|}
    \hline
        ~ & \multicolumn{2}{c}{Jakarta}  & & \multicolumn{2}{c}{Nairobi} &  \\ \hline
        ~ & Min & Mean & Max & Min & Mean & Max \\ \hline
        $T_1$ ($\mu$s)  & 74.54 & 140.36 & 179.61 & 98.73  & 135.52  & 179.93 \\ \hline
        $T_2$ ($\mu$s) & 23.02  & 48.74 & 106.67 & 40.52 &  74.58 & 136.91 \\ \hline
        1QG Error (\%) & 0.02 & 0.03 & 0.03 & 0.03 & 0.04 & 0.07 \\ \hline
        2QG Error (\%) & 0.6  & 0.79 & 1.08 & 0.52 & 1.0 &  1.61 \\ \hline
        1QG Duration ($\mu$s) & 0.04 & 0.04 & 0.04 & 0.04 & 0.04 & 0.04 \\ \hline
        2QG Duration ($\mu$s) & 0.23 & 0.34 & 0.54 & 0.24 &  0.31 &  0.43 \\ \hline
        RO Error (\%) & 1.73 & 3.44 & 5.12 & 2.27 &  3.43 &  4.68 \\ \hline
        RO Duration ($\mu$s) & 5.35 & 5.35 & 5.35 & 5.35 & 5.35 & 5.35 \\ \hline
    \end{tabular}
\caption{Device specifications for Jakarta and Nairobi on July 14, 2022. 1QG, 2QG, and RO denote 1-qubit gate, two-qubit gate, and readout, respectively~\cite{IBMQuantum2022}.}
\label{tab:config}
\end{table}

\section{Measurement Error Mitigation}
\label{app:MEM}

\begin{figure*}[htpb!]
	\includegraphics[width=\textwidth]{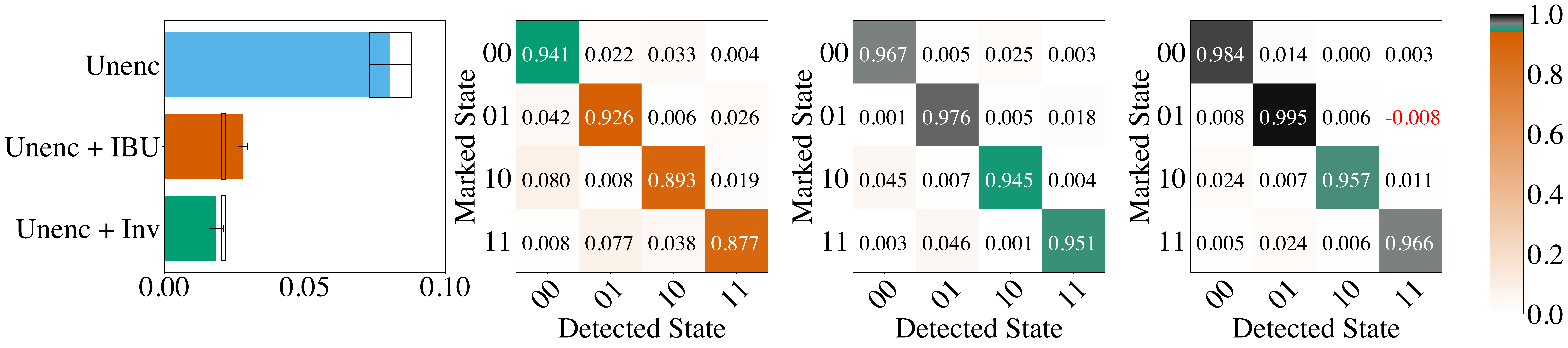}
	\caption{ Nairobi and MEM for unencoded two-qubit Grover. The horizontal bar plot on the left shows the failure probability for the unencoded two-qubit Grover algorithm for the unmitigated data and under two MEM techniques: iterative Bayesian unfolding (IBU) and response matrix inversion (Inv). The three heat maps show the output distribution under no mitigation, IBU, and Inv, going from left to right. The input marked states are on the vertical axis, and the printed numbers on the diagonal represent $p_{s}$ for the respective marked states. The off-diagonal elements in the output distribution are written explicitly to emphasize the presence of negative probabilities under Inv in the right-most figure.}
	\label{fig:nairobi-mem-unenc}
\end{figure*}

\begin{figure*}[htpb!]
	\includegraphics[width=\textwidth]{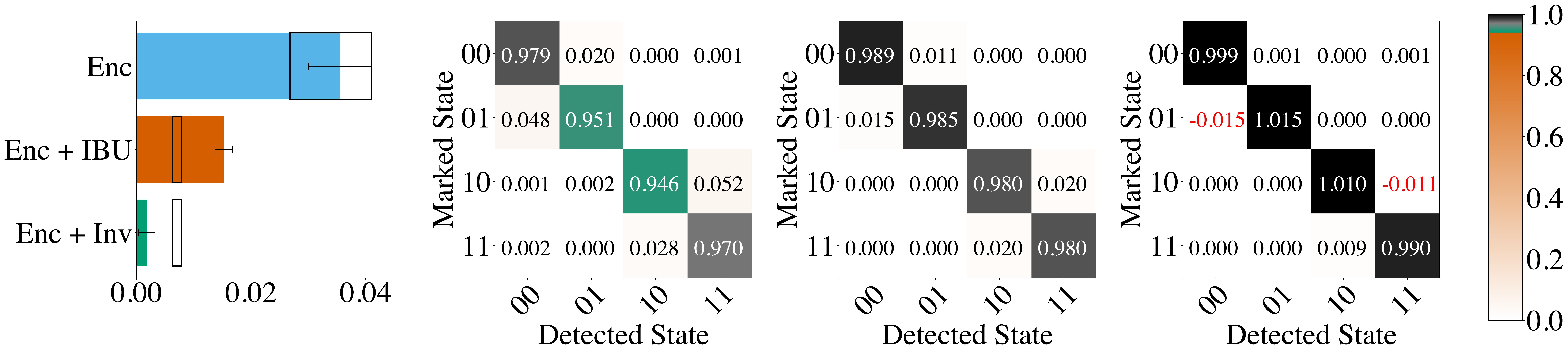}
	\caption{ Same as \cref{fig:nairobi-mem-unenc} but for encoded two-qubit Grover. }
	\label{fig:nairobi-mem-enc}
\end{figure*}

Measurement error mitigation relies on data collected from quantum detector tomography~\cite{maciejewskiMitigationReadoutNoise2020} and classical post-processing to address systematic measurement errors. The calibration experiments involve the preparation of computational basis states $| j \rangle$  and acquiring the response matrix $M$ with elements \( m_{kj} = \text{probability} (\text{prepare } | j \rangle | \text{measure bitstring } k ) \). $M$ is used  to extract the ``true'' mitigated probability vector \(\vec{t} = f(\vec{p}, M)\)  from the observed probability vector $\vec{p}$. Different error mitigation methods differ in how they define \(f\). The most commonly used MEM method, response matrix inversion (Inv), sets $\vec{t} = M^{-1} \vec{p}$.  This is a frequently used MEM strategy~\cite{jurcevicDemonstrationQuantumVolume2021,mooneyGenerationVerification27qubit2021, mooneyWholeDeviceEntanglement2021,barronMeasurementErrorMitigation2020,bravyiMitigatingMeasurementErrors2021}. However, Inv is inherently flawed in the sense that $M^{-1}$ need not be stochastic, and as a result $\vec{t}$ can have negative elements. Ref.~\cite{nachman2020unfolding} notes that iterative Bayesian unfolding (IBU) -- a well-established method to correct detector defects in high-energy physics experiments -- can address readout noise without compromising on stochasticity. IBU is a simplified form of the expectation-maximization method from machine learning that maximizes the likelihood function. In particular, starting with a prior ``truth spectrum'' $\vec{t}^0$, the error mitigated distribution $\vec{t}^n$ is obtained by repeatedly applying Bayes' rule to get
\begin{equation}
    t_i^{n} = \sum\limits_{j} \frac{M_{ji} t^{n-1}_{i}}{\sum\limits_{k} M_{jk} t_{k}^{n-1}} p_j.
    \label{eq:ibu}
\end{equation}
We rely on IBU, in particular, the pyIBU package~\cite{srinivasanScalableMeasurementError2022} to perform MEM to avoid dealing with negative probabilities. 
In Ref.     ~\cite{srinivasanScalableMeasurementError2022}, the optimal $n$ at which the iteration stops is determined by placing a lower limit on the $\ell_1$-distance between $\vec{t}^n$ and $\vec{t}^{n-1}$.

In \cref{fig:2q_fid}, there is a slight discrepancy in the observed and predicted fidelities in the Unenc + MEM and Enc + MEM columns for Nairobi: the success probability is lower than expected. The bar plots in \cref{fig:nairobi-mem-unenc,fig:nairobi-mem-enc} show that this discrepancy is reduced if the mitigated distribution is computed using Inv instead of IBU.
However, Inv leads to unphysical results, as seen in the rightmost output distribution map in \cref{fig:nairobi-mem-enc} where $p_{s} > 1$ for the marked states $\ket{01}$ and $\ket{10}$. In particular, when using Inv, the terms in $\vec{t}$ sum to 1 but the elements $t_i$ are not guaranteed to be in $[0,1]$. The off-diagonal terms under Enc + Inv in \cref{fig:nairobi-mem-enc} are indeed negative. So while Enc + Inv has the best reported average success probability, which is also higher than our model predicts, the underlying output distribution is unphysical. IBU, on the other hand, always returns a valid probability distribution but does not fully invert the response matrix $M$. The discrepancy in simulated and observed values of Nairobi's measurement error mitigated failure probabilities reflects this. Despite its limitations, we err on the side of caution and use IBU as the default MEM method. A more systematic and recent critique of MEM is provided in Ref.~\cite{Quek:2022}.

\section{Extracting the depolarizing parameter from gate errors}
\label{app:depolarizing}

IBMQE devices are calibrated daily, and for each gate, the gate error $e_g$ and gate time $\tau_g$ are reported. The associated $T_1$ and $T_2$ times are also reported for each qubit. As $e_g$ is extracted in presence of thermal relaxation errors $\mathcal{R} = \Phi \circ \mathcal{A}$, $p_D = f(e_g, \tau_g, T_1, T_2)$. In order to extract $p_D$ from $e_g$, we assume that
\begin{equation}
    e_g = 1 -  F(\mathcal{D} \circ \mathcal{R}).
\end{equation}
Using  
\begin{equation}
    F(\mathcal{D} \circ \mathcal{R}) = (1 - p_{D}) F(\mathcal{R}) + p_{D} F(\mathbb{ D } \circ R),
\end{equation}
where $\mathbb{D}$ is the completely depolarizing channel, we get
\begin{subequations}
    \begin{align}
        1 - e_g & = F(\mathcal{D} \circ \mathcal{R}) \\
            & = (1 - p_{D}) F(\mathcal{R}) + p_{D} F(\mathbb{ D } \circ R)\\
            & = (1 - p_{D}) F(\mathcal{R}) + p_{D} F(\mathbb{ D })\\ 
            & = (1 - p_{D}) F(\mathcal{R}) + \frac{p_{D}}{d}.
    \end{align}
\end{subequations}
Here, $d=4^n$ for $n$ qubits. Consequently,
\begin{equation}
    p_{D} = d \frac{F(\mathcal{R}) - 1 + e_g}{d F(\mathcal{R})-1}   . 
\end{equation}
If we assume that there are no relaxation errors and only depolarizing noise affects the gate error $e_g$, then $p_D = (d/d-1) e_g$.

\section{Circuit Construction}
\label{app:circuit-constr}

Recall that implementing the $n$-qubit Grover's algorithm requires the C$_{n-1}Z$ gate, the only multi-qubit gate necessary for both the oracle and the amplitude amplification step. We provide circuit diagrams for how each $n$-qubit controlled phase gate, C$_{n-1}Z$, was transpiled. We rely on previously known circuit designs for our circuit construction, particularly the circuits used by Ref.~\cite{Zhang_2021}. 

\subsection{Two-qubit Grover circuits}
\label{app:circuit-2q}
For two-qubit Grover, $\mathrm{C}Z$ does not require transpilation as $\mathrm{C}Z = H \cdot \mathrm{C}X \cdot H$. However, a few nuances must be considered when constructing the encoded two-qubit Grover circuits. 
There are three components to a Grover circuit with marked element $m = b_{1} b_{2}$: initialization into the state $| \psi \rangle$, oracle query $O_{m}$, and amplitude amplification $A$. More precisely,
\bes
\begin{align}
| \psi \rangle &= \frac{1}{4} \sum\limits_{ b_{i}, b_{j} = \{ 0,1 \}  } | b_{j} \rangle \otimes | b_{k} \rangle = H \otimes H | 00 \rangle \\
O_{m} &= (X^{1-b_{1}} \otimes X^{1-b_{2}}) \cdot \text{C}Z \cdot (X^{1-b_{1}} \otimes X^{1-b_{2}})\\
A &=  H \otimes H \cdot O_{00} \cdot H \otimes H .
\end{align}
\ees
To convert these circuits into their logical counterparts, we note that $\overline{X}_1=XIXI$, $\overline{X}_2=XXII$, $\overline{Z}_1=ZZII$,  $\overline{Z}_2=ZIZI$, $H^{\otimes 4}= \overline{\text{SWAP}}_{12} (\overline{H}_1 \overline{H}_2)$ and $P^{\otimes 4} = (\overline{Z}_1 \overline{Z}_2) \overline{\text{C}Z}$. It is also helpful to notice that $[\text{SWAP}_{12},  U \otimes U] = 0$ for any unitary $U$ and $[\text{SWAP}_{12}, \text{C}Z]=0$. Moreover, $\text{SWAP}_{12} \ket{b_1 b_1} = \ket{b_1 b_1}$ and $[\text{SWAP}_{12}, O_{00}] =0$ as $O_{00}$ only has operators of the form $U \otimes U$ and C$Z$. Consequently,
\begin{subequations}
	\begin{align}
		\ket{ \overline{\psi}} &=  \overline{H} \otimes \overline{H} \ket{\overline{00}} \\
		&= \overline{H} \otimes \overline{H} \cdot \overline{\text{SWAP}_{12}}\ket{\overline{00}} \\
 		&= \overline{\text{SWAP}_{12}} \cdot\overline{H} \otimes \overline{H} 
 		\ket{\overline{00}} \\
 		&= H^{\otimes 4} U_{\text{enc}} \ket{00} .
	\end{align}
\end{subequations}
Implementing the encoded Grover oracle is straightforward and does not involve any two-qubit operations:
\beq
\overline{O_{m}} = \overline{X^{1-b_{1}}} \otimes \overline{X^{1-b_{2}}} \cdot IZZI \cdot P^{\otimes 4}  \cdot \overline{X^{1-b_{1}}} \otimes \overline{X^{1-b_{2}}}.
\eeq

Lastly,
\begin{subequations}
	\begin{align}
		A & = \overline{H} \otimes \overline{H} \cdot \overline{O_{00}} \cdot \overline{H} \otimes \overline{H} \\
		&  = \overline{H} \otimes \overline{H} \cdot \overline{O_{00}} \cdot \overline{H} \otimes \overline{H} \cdot \overline{\text{SWAP}_{12}} \cdot \overline{\text{SWAP}_{12}}  \\
		& = \overline{\text{SWAP}_{12}} \cdot \overline{H} \otimes \overline{H} \cdot \overline{O_{00}} \cdot \overline{\text{SWAP}_{12}} \cdot \overline{H} \otimes \overline{H} \\
		& = H^{\otimes 4} \cdot \overline{O_{00}} \cdot H^{\otimes 4}\\
		& =  H^{\otimes 4} \cdot IXXI \cdot IZZI \cdot P^{\otimes 4}  \cdot IXXI \cdot H^{\otimes 4} . 
	\end{align}
\end{subequations} 
The corresponding circuits for the marked state $| 01 \rangle $ are shown in \cref{fig:enCircuit}.

\subsection{3-qubit to 5-qubit Grover circuits}
\label{app:circuit-nq}
The problem of transpilation increases in complexity with problem size. C$_{n-1}Z$ can be achieved by finding a circuit decomposition for the $n$-qubit Toffoli gate C$_{n-1}X$. It is known that the three-qubit Toffoli gate, C$_2X$, can be implemented using six CNOTs~\cite{nielsen2010quantum}. However, this requires a fully connected architecture. As no fully connected group of three qubits can be found in the QPUs we used, we rely on the $8$-CNOT decomposition~\cite{Zhang_2021} of C$_2Z$ shown in \cref{eq:CCZ}. 

\begin{equation}
\small
\Qcircuit @C=0.7em @R=0.7em @!R {
& \ctrl{1} & \qw & && \gate{T^\dag} & \ctrl{1} & \qw & \ctrl{1} & \qw & \ctrl{1} & \qw & \qw & \ctrl{1} & \qw & \qw  \\
& \ctrl{1} & \qw && = \hspace{0.7cm} & \gate{T^\dag} & \targ &  \ctrl{1} & \targ & \ctrl{1} & \targ & \gate{T} & \ctrl{1} & \targ & \ctrl{1} & \qw \\
&  \control \qw& \qw &&& \gate{T^\dag} & \qw & \targ & \gate{T^\dag} & \targ & \gate{T} & \qw & \targ & \gate{T} & \targ & \qw \\
}
\label{eq:CCZ}
\end{equation}

Here $T = Z^{1/4}$. For C$_3Z$ and C$_4Z$ we use relative-phase Toffoli gates~\cite{maslovAdvantagesUsingRelativephase2016}. Breaking down C$_{k}Z$ using C$_aZ$ and C$_bY$ such that $a + b = k + c$ allows for C$_{k}Z$ to be implemented with fewer CNOTs as long as we use $c$ ancillas. In our construction, we only use one ancilla for C$_3Z$ and C$_4Z$. C$_2Y$ is shown in \cref{eq:C2Y},

\begin{equation}
\small
	\Qcircuit @C=0.7em @R=1em {
	& \ctrl{1} & \qw &&&& \qw  & \qw  & \qw & \ctrl{2} & \qw & \qw & \qw & \ctrl{2} & \qw \\
	& \ctrl{1} & \qw &&& = \hspace{0.7cm} & \qw & \ctrl{1} & \qw & \qw & \qw & \ctrl{1} & \qw & \qw & \qw \\
	& \gate{Y} \qw& \qw &&&&  \gate{G} & \targ & \gate{G} & \targ & \gate{G^\dag} & \targ & \gate{G^\dag} & \control \qw & \qw \\
	}
	\label{eq:C2Y}
\end{equation}
where $G = R_y(\pi/4)$, and C$_3Y$ is shown in \cref{eq:C3Y}:\\
\begin{widetext}
	\begin{equation}
		\small
		\Qcircuit @C=0.7em @R=1.3em {
		& \ctrl{1} & \qw &&&& \qw & \qw & \qw & \qw & \qw & \ctrl{3} & \qw & \qw & \qw & \ctrl{3} & \qw & \qw & \qw & \qw & \qw & \qw & \qw & \qw & \ctrl{1} & \qw \\
		& \ctrl{1} & \qw &&&& \qw & \qw & \qw & \qw & \qw & \qw & \qw & \ctrl{2} & \qw & \qw & \qw & \ctrl{2} & \qw & \qw & \qw & \qw & \qw & \qw & \ctrl{1} & \qw \\
		& \ctrl{1} & \qw &&& \raisebox{0.4cm}{=} \hspace{0.9cm} & \qw & \qw & \ctrl{1} & \qw & \qw & \qw & \qw & \qw & \qw & \qw & \qw & \qw & \qw & \qw & \qw & \ctrl{1} & \qw & \qw & \ctrlo{1} & \qw \\
		& \gate{Y} \qw& \qw &&&& \gate{H} & \gate{T} & \targ & \gate{T^\dag} & \gate{H} & \targ & \gate{T} & \targ & \gate{T^\dag} & \targ & \gate{T} & \targ & \gate{T^\dag} & \gate{H} & \gate{T} & \targ & \gate{T^\dag} & \gate{H} & \gate{-iZ} & \qw \\
		}
			\label{eq:C3Y}
		\end{equation}
\end{widetext}

Finally, using the relative-phase Toffoli gates, C$_3Z$ can be written as in \cref{eq:c3z}:
\begin{equation}
	\small
	\Qcircuit @C=0.7em @R=1.3em {
	& \ctrl{1} & \qw &&&&&&& \ctrl{1} & \qw & \ctrl{1} & \qw \\
	& \ctrl{2} & \qw &&&&&&& \ctrl{1} & \qw & \ctrl{1} & \qw \\
	& \qw \qw& \qw &&& =\hspace{0.5cm} & \hspace{0.4cm} |0\rangle &&& \gate{Y} & \ctrl{1} & \gate{Y^\dag} & \qw \\
	& \ctrl{1} & \qw &&&&&&& \qw & \ctrl{1} & \qw & \qw \\
	& \control \qw & \qw &&&&&&&  \qw & \control \qw & \qw & \qw   
	}
	\label{eq:c3z}
\end{equation}
and likewise, C$_4Z$can be constructed as in \cref{eq:c4z}:
\begin{equation}
	\small
	\Qcircuit @C=0.7em @R=1.3em {
	& \ctrl{1} & \qw &&&&&&& \ctrl{1} & \qw & \ctrl{1} & \qw \\
	& \ctrl{1} & \qw &&&&&&& \ctrl{1} & \qw & \ctrl{1} & \qw \\
	& \ctrl{2} & \qw &&&&&&& \ctrl{1} & \qw & \ctrl{1} & \qw \\
	& \qw \qw& \qw &&& \raisebox{0.5cm}{=} \hspace{0.5cm} & \hspace{0.4cm} |0\rangle &&& \gate{Y} & \ctrl{1} & \gate{Y^\dag} & \qw \\
	& \ctrl{1} & \qw &&&&&&& \qw & \ctrl{1} & \qw & \qw \\
	& \control \qw & \qw &&&&&&&  \qw & \control \qw & \qw & \qw   
	}
	\label{eq:c4z}
	\end{equation}

	This scheme -- where relative phase Toffoli gates~\cite{maslovAdvantagesUsingRelativephase2016} are sewn together to generate a circuit for $\mathrm{C}_{n-1}Z$ ($n > k + 2$) -- can be generalized. 	In particular, $\mathrm{C}_{n-1}Z$ can be implemented using C$_2Y$, C$_2Y^{\dagger}$ and  $\mathrm{C}_{n-2}Z$, which in turn uses $\mathrm{C}_{n-3}Z$. As a result of this recursion, the number of CNOTS for a $\mathrm{C}_{n-1}Z$ circuit is
	\begin{equation}
		\#(\mathrm{C}_{n-1}Z) = 2 (n-3) \  \#(\mathrm{C}_2Y) + \#(\mathrm{C}_2Z).
	\end{equation}
	Thus, the number of CNOTs required to implement a single query of $n$-qubit Grover scales as $O(n)$. At the same time, the number of necessary ancillas is $n-2$, i.e., it also scales linearly with $n$. As we did by using \cref{eq:c4z}, this linear scaling of ancillas could be avoided by considering $\mathrm{C}_kY$ with $k>2$ while increasing the number of CNOTs. Whether entangling fewer qubits by allowing for deeper circuits is worthwhile will depend on the QPU architecture under consideration. 
	Note that the theoretical optimal number of queries $\qopt = O(2^{n/2})$ so at $\qopt$, the number of CNOTs scales as $O(2^{n/2}n)$ where the exponential component will dominate. However, as we noted before, the experimentally allowed number of queries before decoherence takes over might be less than $\qopt$.

\section{Open system model optimization}
\label{app:optimization}

\begin{figure}[h]
	\centering
	\includegraphics[width=\columnwidth]{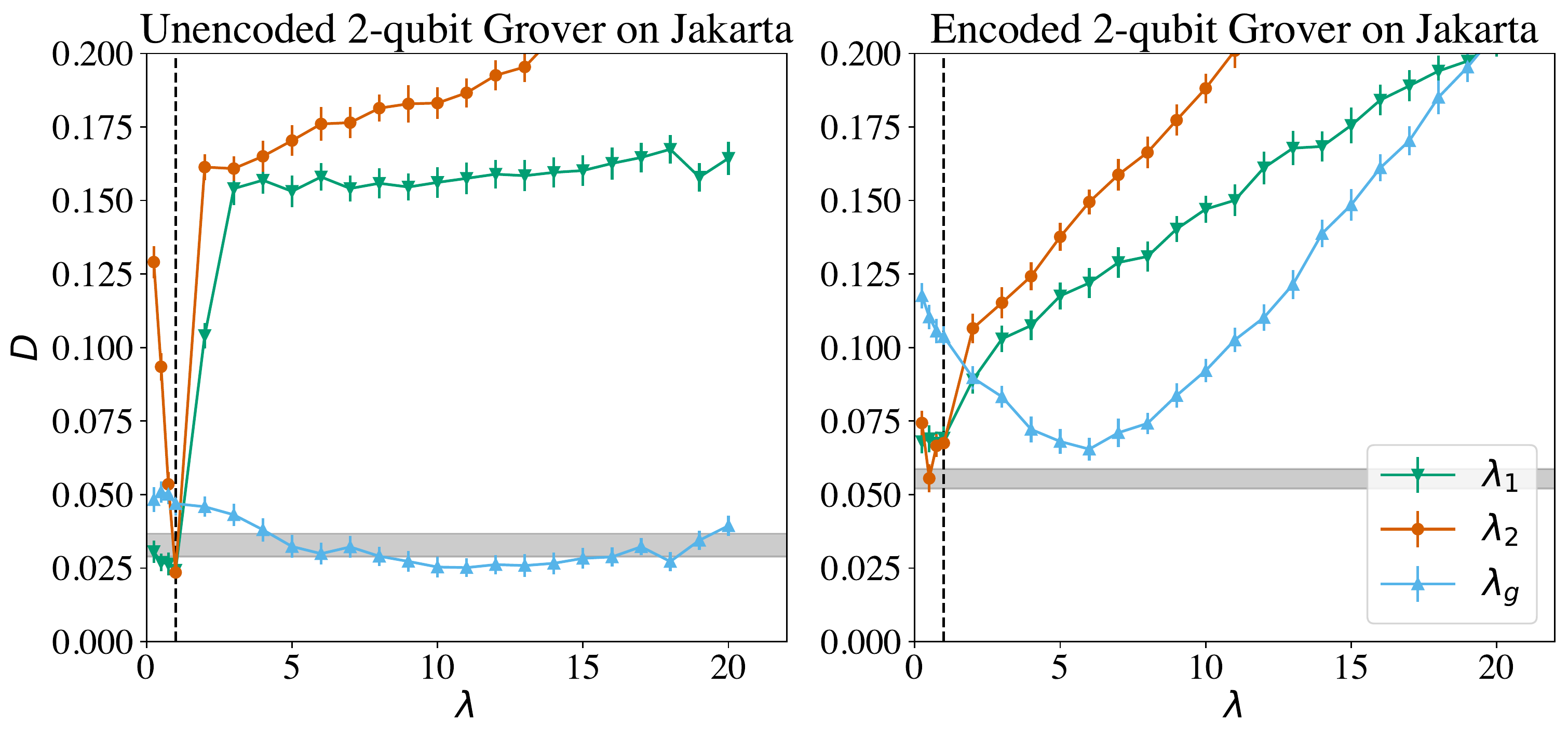}
	\includegraphics[width=\columnwidth]{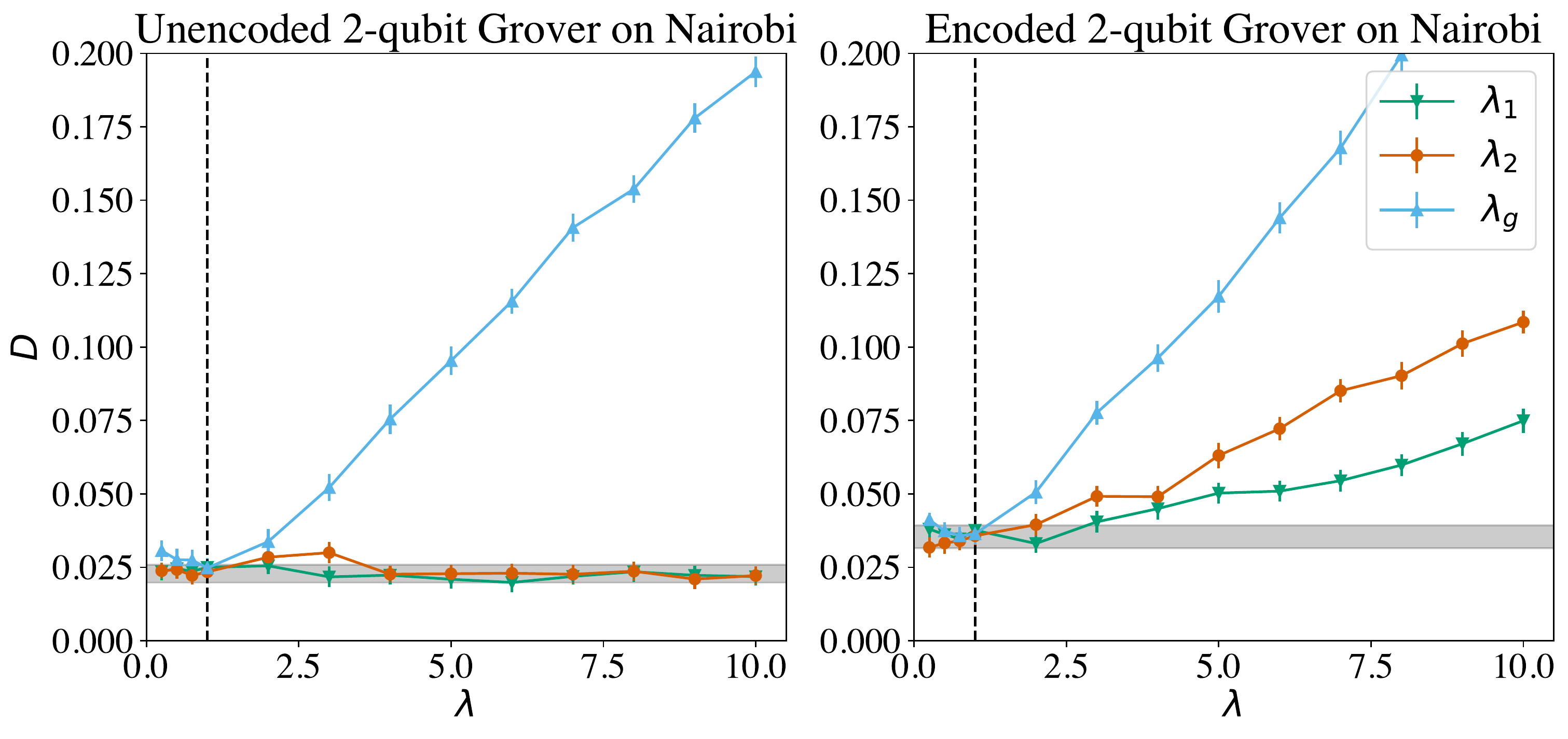}
	\caption{
		Scaling calibration metrics.
		The $\ell_1$-norm based distance $D$ between the observed and simulated outputs for two-qubit Grover is shown as a function of the scaling parameters $\lambda_i$. The model parameters are scaled by setting $T_1 \rightarrow \lambda_1^{-1} T_1$, $T_2 \rightarrow \lambda_2^{-1} T_2$ and $p_D \rightarrow \lambda_g p_D$. 
		The left and right columns represent unencoded and encoded implementations, respectively. The dotted black line represents the default setting $\lambda_i=1$. 
		The grey band in the Jakarta case (top row) is $D$ for the DD-protected circuit versions. For Nairobi (bottom row), even without error suppression, the fit between theory and experiment is already quite good, and our optimization aims to improve the fit further. Thus, in the bottom row (Nairobi), the grey band corresponds to the distance between simulated and observed at $\lambda_i = 1$. 
	}
	  \label{fig:optimization-scan}
\end{figure}

\begin{figure}[t]
	\includegraphics[width=\columnwidth]{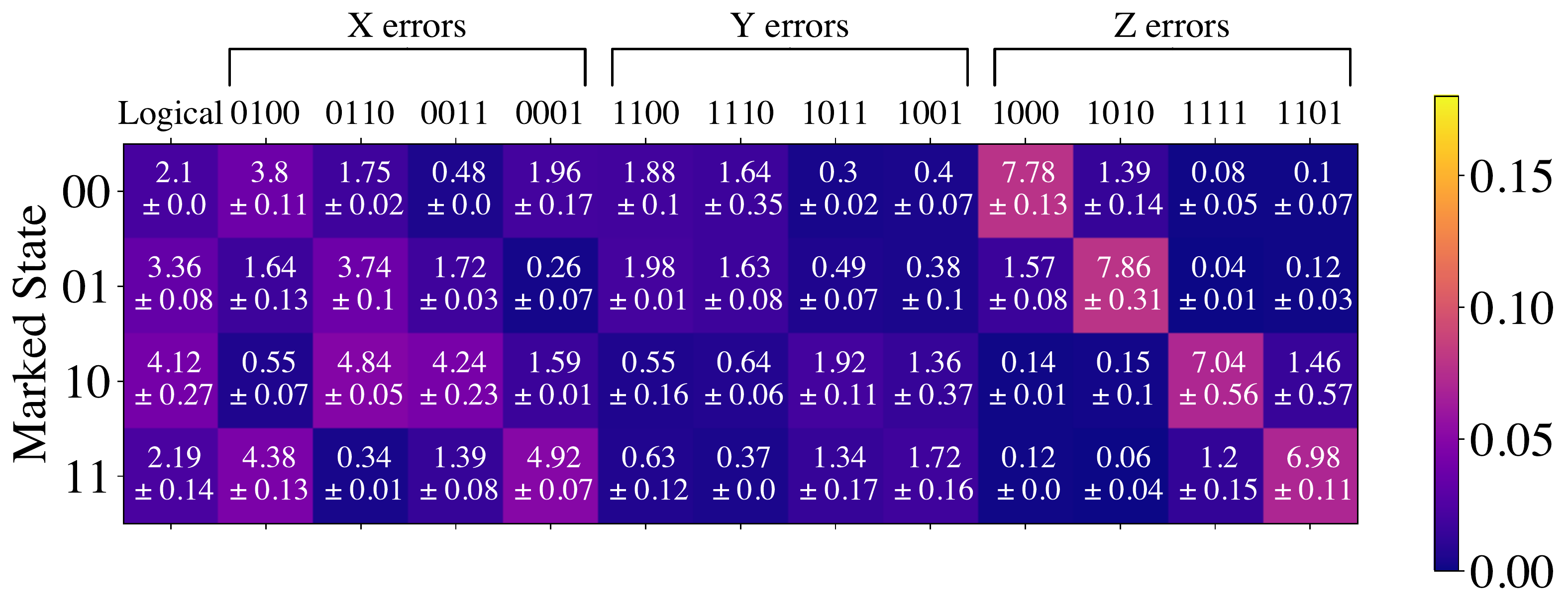}
	\caption{Algorithmic error tomography on Jakarta for optimized parameters. The plot shows the results of AET on the encoded two-qubit Grover algorithm after setting $(\lambda_1, \lambda_2, \lambda_g) = (1,1,6)$. Each row of the error tomography table corresponds to a marked state, and each column represents logical errors and $X, Y, Z$ type errors. Even after rescaling the model parameters, compared to the error tomography table for Jakarta (\cref{fig:et-dd}), we do not see an asymmetry in $Z$ errors across marked states. 
	\label{fig:et-opt-jakarta}}
	\end{figure}

The model presented in \cref{sec:sim} only accounts for amplitude damping, dephasing, and depolarization, which are a subset of the errors in a superconducting device. Notably, this model does not include crosstalk, leakage to higher energy levels, and non-Markovian system-environment interaction. As we saw in the main text, introducing error suppression improves the agreement between our model and the observations. However, in the absence of DD, the model results often provide an upper bound on the performance, and the device can perform worse than the model's predictions.

It is plausible to ask if rescaling the calibration metrics provided by IBMQE results in an improved phenomenological model and better predictability of the device performance.
To do this, we quantify model predictability by computing the $\ell_1$-norm based distance between the experimentally observed probability distribution $\mathbf{p}^m$ and the simulated distribution $\mathbf{q}^m$ for marked states $\ket{m}$. We define
\begin{equation}
    D = \frac{1}{2^{n}} \sum_{m=0}^{2^n-1} D^m \ , \quad D^m = \frac{1}{2}\sum_{i=0}^{2^n-1} |p^m_i - q^m_i|,
\end{equation}
such that $D$ is the distance between the probability distributions averaged over all the marked states $\ket{m}$. $D=0$ implies a perfect match between the simulated and the observed distributions, and $D=1$ means completely distinguishable distributions. We then consider $D$ as a function of rescaling parameters $\lambda_i$ such that $T_1 \rightarrow \lambda_1^{-1} T_1$, $T_2 \rightarrow \lambda_2^{-1} T_2$ and $p_D \rightarrow \lambda_g p_D$. 

To evaluate the feasibility of this approach, we focus on the smallest problem size, $n=2$. 
The most significant error contribution comes from the depolarizing channel, as $D$ is most responsive to $\lambda_g$. With this in mind, we start from the default setting $\lambda_i=1$, then optimize $\lambda_g, \lambda_1, \lambda_2$ in that order, using the optimal value from the previous scan as the input for the next parameter. 

\cref{fig:optimization-scan} shows the distance $D$ as a function of $\lambda_i$. Recall that in the absence of DD, we reported discrepancies in both the AET of the encoded circuit (\cref{fig:et-dd}) and the failure probabilities of the unencoded circuit (\cref{fig:2q_fid}) for Jakarta. To directly compare the result of optimizing $\lambda$ to the effectiveness of error suppression, \cref{fig:optimization-scan} (top row) also shows the distance $D$ between theory and experiment after DD. 

For the unencoded experiment, increasing $p_D$ by a factor of $\lambda_g=10$ suffices for the model to match the experiment. $D$ is minimized at $\lambda_1, \lambda_2=1$, i.e., there is no change from the default setting for $T_1$ and $T_2$. However, even with the optimized values (i.e., setting $\lambda_g=10$, $\lambda_1, \lambda_2=1$), the new model barely improves upon activating DD, corresponding to the grey band. Similarly, for the encoded two-qubit Grover on Jakarta, only optimizing $p_D$ leads to a noticeable decrease in $D$, with the best fit found at $\lambda_g =6$. The optimized model does match the success probability acquired by activating DD (grey band, top right). However, AET at $\lambda_g=6$, shown in \cref{fig:et-opt-jakarta}, does not exhibit the state-wise asymmetry seen in \cref{fig:et-dd}. A similar optimization on Nairobi confirms our earlier observation: the model that uses the provided calibration metrics fits with the observed results, and Nairobi does not benefit from parameter rescaling.

To summarize, rescaling the calibration metrics to improve predictability is mildly effective: the agreement with the experimental $p_s$ results improves by increasing the depolarization. However, the optimized model does not match the AET results for Jakarta. A more sophisticated model that includes qubit crosstalk and leakage to/from higher energy levels appears necessary for a better agreement. However, as highlighted in the main text, introducing error suppression via DD provides a reliable way to simultaneously improve performance and agreement with a model that accounts only for Markovian amplitude damping, dephasing, and depolarization.

\section{Postselection on 4-qubit and 5-qubit Grover}
\label{app:PS}

\begin{figure}[b]
	\centering
	\includegraphics[width=\linewidth]{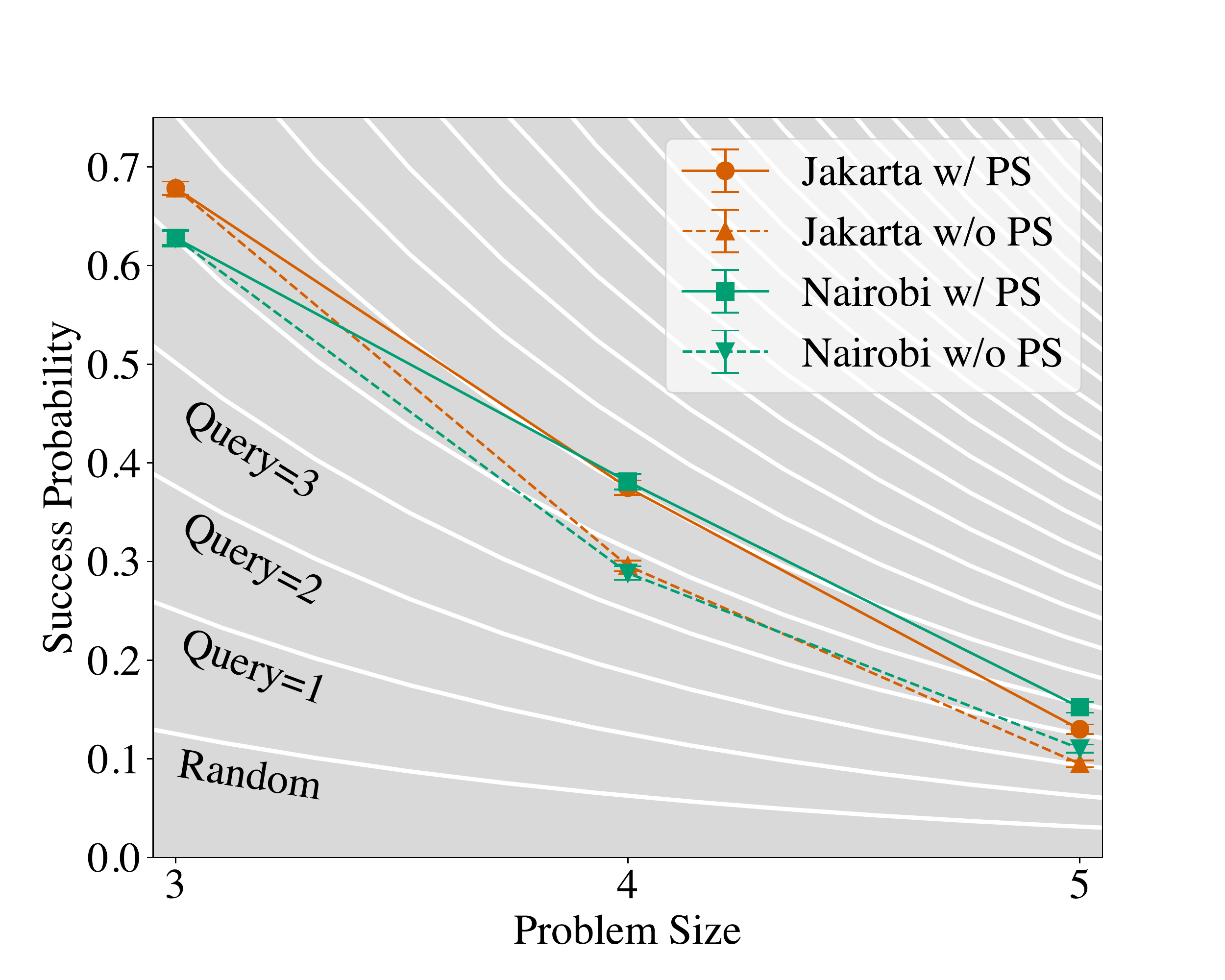}
	\caption{Postselection for 4-qubit and 5-qubit Grover: Success probabilities are shown with and without postselection for Nairobi (green) and Jakarta (orange). Here we only consider the DD-protected circuits. 3-qubit Grover does not undergo postselection. For 4-qubit and 5-qubit Grover, we postselect to count only the experiments for which the ancilla qubit (q1 in \cref{fig:gate-map}) is in $\ket{0}$. The white lines correspond to the success probabilities for the classical strategy and random sampling from the unsorted list ($q=0$). With postselection, for all problem sizes, the DD-protected quantum $q=2$ strategy outperforms the classical strategy for $q \le 3$. Without postselection, the better-than-classical requirement is met by all implementations other than 5-qubit Grover on Jakarta, where we achieve a breakeven. Error bars correspond to 99\% confidence intervals.}
	  \label{fig:postselection}
\end{figure}

The circuits for $\mathrm{C}_3Z$ and $\mathrm{C}_4Z$ require 5 and 6 qubits, respectively [see \cref{eq:c3z,eq:c4z}], with one ancilla qubit used to link the Toffoli and relative-phase Toffoli gates. The ancilla qubit is initialized in $\ket{0}$ and should be in that state at the end of the algorithm. Consequently, when measured in the $Z$-basis, observing the ancilla qubit in $\ket{1}$ implies that an error occurred in the implementation of the $\mathrm{C}_3Z$ or $\mathrm{C}_4Z$ gate. We postselect the 4-qubit and 5-qubit Grover experiments by only considering experiments for which the ancilla qubit (q1 in \cref{fig:gate-map}) is measured to be in state $\ket{0}$. \cref{fig:postselection} shows the effect of postselection on success probabilities. The better-than-classical performance for Nairobi holds even if postselection is avoided. However, without postselection, we see a tie with the classical result for 5-qubit Grover on Jakarta at $q=2$. Overall, there is a non-trivial increase in success probabilities due to postselection. Using ancilla qubits for postselection is convenient, as these ancilla qubits also allow for much shallower circuits for implementing $\mathrm{C}_{n-1}Z$ (recall \cref{app:circuit-constr}). All the results we report in other sections are after postselection, including the survey of DD sequences.

\bibliography{refs.bib,grover-es.bib}

\begin{thebibliography}{65}%
\makeatletter
\providecommand \@ifxundefined [1]{%
 \@ifx{#1\undefined}
}%
\providecommand \@ifnum [1]{%
 \ifnum #1\expandafter \@firstoftwo
 \else \expandafter \@secondoftwo
 \fi
}%
\providecommand \@ifx [1]{%
 \ifx #1\expandafter \@firstoftwo
 \else \expandafter \@secondoftwo
 \fi
}%
\providecommand \natexlab [1]{#1}%
\providecommand \enquote  [1]{``#1''}%
\providecommand \bibnamefont  [1]{#1}%
\providecommand \bibfnamefont [1]{#1}%
\providecommand \citenamefont [1]{#1}%
\providecommand \href@noop [0]{\@secondoftwo}%
\providecommand \href [0]{\begingroup \@sanitize@url \@href}%
\providecommand \@href[1]{\@@startlink{#1}\@@href}%
\providecommand \@@href[1]{\endgroup#1\@@endlink}%
\providecommand \@sanitize@url [0]{\catcode `\\12\catcode `\$12\catcode
  `\&12\catcode `\#12\catcode `\^12\catcode `\_12\catcode `\%12\relax}%
\providecommand \@@startlink[1]{}%
\providecommand \@@endlink[0]{}%
\providecommand \url  [0]{\begingroup\@sanitize@url \@url }%
\providecommand \@url [1]{\endgroup\@href {#1}{\urlprefix }}%
\providecommand \urlprefix  [0]{URL }%
\providecommand \Eprint [0]{\href }%
\providecommand \doibase [0]{http://dx.doi.org/}%
\providecommand \selectlanguage [0]{\@gobble}%
\providecommand \bibinfo  [0]{\@secondoftwo}%
\providecommand \bibfield  [0]{\@secondoftwo}%
\providecommand \translation [1]{[#1]}%
\providecommand \BibitemOpen [0]{}%
\providecommand \bibitemStop [0]{}%
\providecommand \bibitemNoStop [0]{.\EOS\space}%
\providecommand \EOS [0]{\spacefactor3000\relax}%
\providecommand \BibitemShut  [1]{\csname bibitem#1\endcsname}%
\let\auto@bib@innerbib\@empty
\bibitem [{\citenamefont {Grover}(1997)}]{Grover:97a}%
  \BibitemOpen
  \bibfield  {author} {\bibinfo {author} {\bibfnamefont {L.~K.}\ \bibnamefont
  {Grover}},\ }\href {http://link.aps.org/doi/10.1103/PhysRevLett.79.325}
  {\bibfield  {journal} {\bibinfo  {journal} {Phys. Rev. Lett.}\ }\textbf
  {\bibinfo {volume} {79}},\ \bibinfo {pages} {325} (\bibinfo {year}
  {1997})}\BibitemShut {NoStop}%
\bibitem [{\citenamefont {Bennett}\ \emph {et~al.}(1997)\citenamefont
  {Bennett}, \citenamefont {Bernstein}, \citenamefont {Brassard},\ and\
  \citenamefont {Vazirani}}]{Bennett:1997lh}%
  \BibitemOpen
  \bibfield  {author} {\bibinfo {author} {\bibfnamefont {C.}~\bibnamefont
  {Bennett}}, \bibinfo {author} {\bibfnamefont {E.}~\bibnamefont {Bernstein}},
  \bibinfo {author} {\bibfnamefont {G.}~\bibnamefont {Brassard}}, \ and\
  \bibinfo {author} {\bibfnamefont {U.}~\bibnamefont {Vazirani}},\ }\bibfield
  {booktitle} {\emph {\bibinfo {booktitle} {SIAM Journal on Computing}},\
  }\href {\doibase 10.1137/S0097539796300933} {\bibfield  {journal} {\bibinfo
  {journal} {SIAM Journal on Computing}\ }\textbf {\bibinfo {volume} {26}},\
  \bibinfo {pages} {1510} (\bibinfo {year} {1997})}\BibitemShut {NoStop}%
\bibitem [{\citenamefont {D\"{u}rr}\ \emph {et~al.}(2006)\citenamefont
  {D\"{u}rr}, \citenamefont {Heiligman}, \citenamefont {Hoyer},\ and\
  \citenamefont {Mhalla}}]{durrQuantumQueryComplexity2006}%
  \BibitemOpen
  \bibfield  {author} {\bibinfo {author} {\bibfnamefont {C.}~\bibnamefont
  {D\"{u}rr}}, \bibinfo {author} {\bibfnamefont {M.}~\bibnamefont {Heiligman}},
  \bibinfo {author} {\bibfnamefont {P.}~\bibnamefont {Hoyer}}, \ and\ \bibinfo
  {author} {\bibfnamefont {M.}~\bibnamefont {Mhalla}},\ }\href
  {https://doi.org/10.1137/050644719} {\bibfield  {journal} {\bibinfo
  {journal} {SIAM Journal on Computing}\ }\textbf {\bibinfo {volume} {35}},\
  \bibinfo {pages} {1310} (\bibinfo {year} {2006})}\BibitemShut {NoStop}%
\bibitem [{\citenamefont {Magniez}\ \emph {et~al.}(2007)\citenamefont
  {Magniez}, \citenamefont {Santha},\ and\ \citenamefont
  {Szegedy}}]{magniezQuantumAlgorithmsTriangle2007}%
  \BibitemOpen
  \bibfield  {author} {\bibinfo {author} {\bibfnamefont {F.}~\bibnamefont
  {Magniez}}, \bibinfo {author} {\bibfnamefont {M.}~\bibnamefont {Santha}}, \
  and\ \bibinfo {author} {\bibfnamefont {M.}~\bibnamefont {Szegedy}},\ }\href
  {\doibase 10.1137/050643684} {\bibfield  {journal} {\bibinfo  {journal} {SIAM
  Journal on Computing}\ } (\bibinfo {year} {2007}),\
  10.1137/050643684}\BibitemShut {NoStop}%
\bibitem [{\citenamefont {Lubinski}\ \emph {et~al.}(2021)\citenamefont
  {Lubinski}, \citenamefont {Johri}, \citenamefont {Varosy}, \citenamefont
  {Coleman}, \citenamefont {Zhao}, \citenamefont {Necaise}, \citenamefont
  {Baldwin}, \citenamefont {Mayer},\ and\ \citenamefont
  {Proctor}}]{lubinskiApplicationOrientedPerformanceBenchmarks2021}%
  \BibitemOpen
  \bibfield  {author} {\bibinfo {author} {\bibfnamefont {T.}~\bibnamefont
  {Lubinski}}, \bibinfo {author} {\bibfnamefont {S.}~\bibnamefont {Johri}},
  \bibinfo {author} {\bibfnamefont {P.}~\bibnamefont {Varosy}}, \bibinfo
  {author} {\bibfnamefont {J.}~\bibnamefont {Coleman}}, \bibinfo {author}
  {\bibfnamefont {L.}~\bibnamefont {Zhao}}, \bibinfo {author} {\bibfnamefont
  {J.}~\bibnamefont {Necaise}}, \bibinfo {author} {\bibfnamefont {C.~H.}\
  \bibnamefont {Baldwin}}, \bibinfo {author} {\bibfnamefont {K.}~\bibnamefont
  {Mayer}}, \ and\ \bibinfo {author} {\bibfnamefont {T.}~\bibnamefont
  {Proctor}},\ }\href {http://arxiv.org/abs/2110.03137} {\enquote {\bibinfo
  {title} {{Application-Oriented Performance Benchmarks for Quantum
  Computing}},}\ } (\bibinfo {year} {2021}),\ \Eprint
  {http://arxiv.org/abs/2110.03137} {arXiv:2110.03137 [quant-ph]} \BibitemShut
  {NoStop}%
\bibitem [{\citenamefont {Roy}\ \emph {et~al.}(2020)\citenamefont {Roy},
  \citenamefont {Hazra}, \citenamefont {Kundu}, \citenamefont {Chand},
  \citenamefont {Patankar},\ and\ \citenamefont
  {Vijay}}]{royProgrammableSuperconductingProcessor2020}%
  \BibitemOpen
  \bibfield  {author} {\bibinfo {author} {\bibfnamefont {T.}~\bibnamefont
  {Roy}}, \bibinfo {author} {\bibfnamefont {S.}~\bibnamefont {Hazra}}, \bibinfo
  {author} {\bibfnamefont {S.}~\bibnamefont {Kundu}}, \bibinfo {author}
  {\bibfnamefont {M.}~\bibnamefont {Chand}}, \bibinfo {author} {\bibfnamefont
  {M.~P.}\ \bibnamefont {Patankar}}, \ and\ \bibinfo {author} {\bibfnamefont
  {R.}~\bibnamefont {Vijay}},\ }\href
  {https://journals.aps.org/prapplied/abstract/10.1103/PhysRevApplied.14.014072}
  {\bibfield  {journal} {\bibinfo  {journal} {Phys. Rev. Applied}\ }\textbf
  {\bibinfo {volume} {14}},\ \bibinfo {pages} {014072} (\bibinfo {year}
  {2020})}\BibitemShut {NoStop}%
\bibitem [{\citenamefont {Figgatt}\ \emph {et~al.}(2017)\citenamefont
  {Figgatt}, \citenamefont {Maslov}, \citenamefont {Landsman}, \citenamefont
  {Linke}, \citenamefont {Debnath},\ and\ \citenamefont
  {Monroe}}]{figgattComplete3QubitGrover2017}%
  \BibitemOpen
  \bibfield  {author} {\bibinfo {author} {\bibfnamefont {C.}~\bibnamefont
  {Figgatt}}, \bibinfo {author} {\bibfnamefont {D.}~\bibnamefont {Maslov}},
  \bibinfo {author} {\bibfnamefont {K.~A.}\ \bibnamefont {Landsman}}, \bibinfo
  {author} {\bibfnamefont {N.~M.}\ \bibnamefont {Linke}}, \bibinfo {author}
  {\bibfnamefont {S.}~\bibnamefont {Debnath}}, \ and\ \bibinfo {author}
  {\bibfnamefont {C.}~\bibnamefont {Monroe}},\ }\href
  {https://www.nature.com/articles/s41467-017-01904-7} {\bibfield  {journal}
  {\bibinfo  {journal} {Nat. Commun.}\ }\textbf {\bibinfo {volume} {8}},\
  \bibinfo {pages} {1} (\bibinfo {year} {2017})}\BibitemShut {NoStop}%
\bibitem [{\citenamefont {Zhang}\ \emph {et~al.}(2021)\citenamefont {Zhang},
  \citenamefont {Rao}, \citenamefont {Yu}, \citenamefont {Lim},\ and\
  \citenamefont {Korepin}}]{Zhang_2021}%
  \BibitemOpen
  \bibfield  {author} {\bibinfo {author} {\bibfnamefont {K.}~\bibnamefont
  {Zhang}}, \bibinfo {author} {\bibfnamefont {P.}~\bibnamefont {Rao}}, \bibinfo
  {author} {\bibfnamefont {K.}~\bibnamefont {Yu}}, \bibinfo {author}
  {\bibfnamefont {H.}~\bibnamefont {Lim}}, \ and\ \bibinfo {author}
  {\bibfnamefont {V.}~\bibnamefont {Korepin}},\ }\href
  {https://doi.org/10.1007%2Fs11128-021-03165-2} {\bibfield  {journal}
  {\bibinfo  {journal} {{Quant. Inf. Proc.}}\ }\textbf {\bibinfo {volume}
  {20}},\ \bibinfo {pages} {233} (\bibinfo {year} {2021})}\BibitemShut
  {NoStop}%
\bibitem [{\citenamefont {Vaidman}\ \emph {et~al.}(1996)\citenamefont
  {Vaidman}, \citenamefont {Goldenberg},\ and\ \citenamefont
  {Wiesner}}]{Vaidman:1996vs}%
  \BibitemOpen
  \bibfield  {author} {\bibinfo {author} {\bibfnamefont {L.}~\bibnamefont
  {Vaidman}}, \bibinfo {author} {\bibfnamefont {L.}~\bibnamefont {Goldenberg}},
  \ and\ \bibinfo {author} {\bibfnamefont {S.}~\bibnamefont {Wiesner}},\ }\href
  {\doibase 10.1103/PhysRevA.54.R1745} {\bibfield  {journal} {\bibinfo
  {journal} {Physical Review A}\ }\textbf {\bibinfo {volume} {54}},\ \bibinfo
  {pages} {R1745} (\bibinfo {year} {1996})}\BibitemShut {NoStop}%
\bibitem [{\citenamefont {{Gottesman}}(1997)}]{gottesman}%
  \BibitemOpen
  \bibfield  {author} {\bibinfo {author} {\bibfnamefont {D.}~\bibnamefont
  {{Gottesman}}},\ }\emph {\bibinfo {title} {{Stabilizer codes and quantum
  error correction}}},\ \href@noop {} {Ph.D. thesis},\ \bibinfo  {school}
  {California Institute of Technology} (\bibinfo {year} {1997}),\ \Eprint
  {http://arxiv.org/abs/arXiv:quant-ph/9705052} {arXiv:quant-ph/9705052}
  \BibitemShut {NoStop}%
\bibitem [{\citenamefont {Kandala}\ \emph {et~al.}(2019)\citenamefont
  {Kandala}, \citenamefont {Temme}, \citenamefont {C{\'o}rcoles}, \citenamefont
  {Mezzacapo}, \citenamefont {Chow},\ and\ \citenamefont
  {Gambetta}}]{kandalaErrorMitigationExtends2019}%
  \BibitemOpen
  \bibfield  {author} {\bibinfo {author} {\bibfnamefont {A.}~\bibnamefont
  {Kandala}}, \bibinfo {author} {\bibfnamefont {K.}~\bibnamefont {Temme}},
  \bibinfo {author} {\bibfnamefont {A.~D.}\ \bibnamefont {C{\'o}rcoles}},
  \bibinfo {author} {\bibfnamefont {A.}~\bibnamefont {Mezzacapo}}, \bibinfo
  {author} {\bibfnamefont {J.~M.}\ \bibnamefont {Chow}}, \ and\ \bibinfo
  {author} {\bibfnamefont {J.~M.}\ \bibnamefont {Gambetta}},\ }\href {\doibase
  10.1038/s41586-019-1040-7} {\bibfield  {journal} {\bibinfo  {journal}
  {Nature}\ }\textbf {\bibinfo {volume} {567}},\ \bibinfo {pages} {491}
  (\bibinfo {year} {2019})}\BibitemShut {NoStop}%
\bibitem [{\citenamefont {Nachman}\ \emph {et~al.}(2020)\citenamefont
  {Nachman}, \citenamefont {Urbanek}, \citenamefont {de~Jong},\ and\
  \citenamefont {Bauer}}]{nachman2020unfolding}%
  \BibitemOpen
  \bibfield  {author} {\bibinfo {author} {\bibfnamefont {B.}~\bibnamefont
  {Nachman}}, \bibinfo {author} {\bibfnamefont {M.}~\bibnamefont {Urbanek}},
  \bibinfo {author} {\bibfnamefont {W.~A.}\ \bibnamefont {de~Jong}}, \ and\
  \bibinfo {author} {\bibfnamefont {C.~W.}\ \bibnamefont {Bauer}},\ }\href
  {https://www.nature.com/articles/s41534-020-00309-7} {\bibfield  {journal}
  {\bibinfo  {journal} {npj Quantum Information}\ }\textbf {\bibinfo {volume}
  {6}},\ \bibinfo {pages} {84} (\bibinfo {year} {2020})}\BibitemShut {NoStop}%
\bibitem [{\citenamefont {Blume-Kohout}\ \emph {et~al.}(2013)\citenamefont
  {Blume-Kohout}, \citenamefont {Gamble}, \citenamefont {Nielsen},
  \citenamefont {Mizrahi}, \citenamefont {Sterk},\ and\ \citenamefont
  {Maunz}}]{blume2013robust}%
  \BibitemOpen
  \bibfield  {author} {\bibinfo {author} {\bibfnamefont {R.}~\bibnamefont
  {Blume-Kohout}}, \bibinfo {author} {\bibfnamefont {J.~K.}\ \bibnamefont
  {Gamble}}, \bibinfo {author} {\bibfnamefont {E.}~\bibnamefont {Nielsen}},
  \bibinfo {author} {\bibfnamefont {J.}~\bibnamefont {Mizrahi}}, \bibinfo
  {author} {\bibfnamefont {J.~D.}\ \bibnamefont {Sterk}}, \ and\ \bibinfo
  {author} {\bibfnamefont {P.}~\bibnamefont {Maunz}},\ }\href
  {http://arXiv.org/abs/1310.4492} {\bibfield  {journal} {\bibinfo  {journal}
  {arXiv:1310.4492}\ } (\bibinfo {year} {2013})}\BibitemShut {NoStop}%
\bibitem [{\citenamefont {Merkel}\ \emph {et~al.}(2013)\citenamefont {Merkel},
  \citenamefont {Gambetta}, \citenamefont {Smolin}, \citenamefont {Poletto},
  \citenamefont {C{\'o}rcoles}, \citenamefont {Johnson}, \citenamefont {Ryan},\
  and\ \citenamefont {Steffen}}]{Merkel:2013aa}%
  \BibitemOpen
  \bibfield  {author} {\bibinfo {author} {\bibfnamefont {S.~T.}\ \bibnamefont
  {Merkel}}, \bibinfo {author} {\bibfnamefont {J.~M.}\ \bibnamefont
  {Gambetta}}, \bibinfo {author} {\bibfnamefont {J.~A.}\ \bibnamefont
  {Smolin}}, \bibinfo {author} {\bibfnamefont {S.}~\bibnamefont {Poletto}},
  \bibinfo {author} {\bibfnamefont {A.~D.}\ \bibnamefont {C{\'o}rcoles}},
  \bibinfo {author} {\bibfnamefont {B.~R.}\ \bibnamefont {Johnson}}, \bibinfo
  {author} {\bibfnamefont {C.~A.}\ \bibnamefont {Ryan}}, \ and\ \bibinfo
  {author} {\bibfnamefont {M.}~\bibnamefont {Steffen}},\ }\href {\doibase
  10.1103/PhysRevA.87.062119} {\bibfield  {journal} {\bibinfo  {journal}
  {Physical Review A}\ }\textbf {\bibinfo {volume} {87}},\ \bibinfo {pages}
  {062119} (\bibinfo {year} {2013})}\BibitemShut {NoStop}%
\bibitem [{\citenamefont {Viola}\ and\ \citenamefont {Lloyd}(1998)}]{Viola:98}%
  \BibitemOpen
  \bibfield  {author} {\bibinfo {author} {\bibfnamefont {L.}~\bibnamefont
  {Viola}}\ and\ \bibinfo {author} {\bibfnamefont {S.}~\bibnamefont {Lloyd}},\
  }\href {https://link.aps.org/doi/10.1103/PhysRevA.58.2733} {\bibfield
  {journal} {\bibinfo  {journal} {Phys. Rev. A}\ }\textbf {\bibinfo {volume}
  {58}},\ \bibinfo {pages} {2733} (\bibinfo {year} {1998})}\BibitemShut
  {NoStop}%
\bibitem [{\citenamefont {Viola}\ \emph {et~al.}(1999)\citenamefont {Viola},
  \citenamefont {Knill},\ and\ \citenamefont {Lloyd}}]{Viola:99}%
  \BibitemOpen
  \bibfield  {author} {\bibinfo {author} {\bibfnamefont {L.}~\bibnamefont
  {Viola}}, \bibinfo {author} {\bibfnamefont {E.}~\bibnamefont {Knill}}, \ and\
  \bibinfo {author} {\bibfnamefont {S.}~\bibnamefont {Lloyd}},\ }\href
  {http://link.aps.org/doi/10.1103/PhysRevLett.82.2417} {\bibfield  {journal}
  {\bibinfo  {journal} {Physical Review Letters}\ }\textbf {\bibinfo {volume}
  {82}},\ \bibinfo {pages} {2417} (\bibinfo {year} {1999})}\BibitemShut
  {NoStop}%
\bibitem [{\citenamefont {Zanardi}(1999)}]{Zanardi:1999fk}%
  \BibitemOpen
  \bibfield  {author} {\bibinfo {author} {\bibfnamefont {P.}~\bibnamefont
  {Zanardi}},\ }\href
  {http://www.sciencedirect.com/science/article/pii/S0375960199003655}
  {\bibfield  {journal} {\bibinfo  {journal} {Physics Letters A}\ }\textbf
  {\bibinfo {volume} {258}},\ \bibinfo {pages} {77} (\bibinfo {year}
  {1999})}\BibitemShut {NoStop}%
\bibitem [{\citenamefont {Vitali}\ and\ \citenamefont
  {Tombesi}(1999)}]{Vitali:99}%
  \BibitemOpen
  \bibfield  {author} {\bibinfo {author} {\bibfnamefont {D.}~\bibnamefont
  {Vitali}}\ and\ \bibinfo {author} {\bibfnamefont {P.}~\bibnamefont
  {Tombesi}},\ }\href {https://link.aps.org/doi/10.1103/PhysRevA.59.4178}
  {\bibfield  {journal} {\bibinfo  {journal} {Physical Review A}\ }\textbf
  {\bibinfo {volume} {59}},\ \bibinfo {pages} {4178} (\bibinfo {year}
  {1999})}\BibitemShut {NoStop}%
\bibitem [{\citenamefont {Genov}\ \emph {et~al.}(2017)\citenamefont {Genov},
  \citenamefont {Schraft}, \citenamefont {Vitanov},\ and\ \citenamefont
  {Halfmann}}]{Genov:2017aa}%
  \BibitemOpen
  \bibfield  {author} {\bibinfo {author} {\bibfnamefont {G.~T.}\ \bibnamefont
  {Genov}}, \bibinfo {author} {\bibfnamefont {D.}~\bibnamefont {Schraft}},
  \bibinfo {author} {\bibfnamefont {N.~V.}\ \bibnamefont {Vitanov}}, \ and\
  \bibinfo {author} {\bibfnamefont {T.}~\bibnamefont {Halfmann}},\ }\href
  {\doibase 10.1103/PhysRevLett.118.133202} {\bibfield  {journal} {\bibinfo
  {journal} {Physical Review Letters}\ }\textbf {\bibinfo {volume} {118}},\
  \bibinfo {pages} {133202} (\bibinfo {year} {2017})}\BibitemShut {NoStop}%
\bibitem [{\citenamefont {Khodjasteh}\ and\ \citenamefont
  {Lidar}(2005)}]{Khodjasteh:2005xu}%
  \BibitemOpen
  \bibfield  {author} {\bibinfo {author} {\bibfnamefont {K.}~\bibnamefont
  {Khodjasteh}}\ and\ \bibinfo {author} {\bibfnamefont {D.~A.}\ \bibnamefont
  {Lidar}},\ }\href {http://link.aps.org/doi/10.1103/PhysRevLett.95.180501}
  {\bibfield  {journal} {\bibinfo  {journal} {Physical Review Letters}\
  }\textbf {\bibinfo {volume} {95}},\ \bibinfo {pages} {180501} (\bibinfo
  {year} {2005})}\BibitemShut {NoStop}%
\bibitem [{\citenamefont {Quiroz}\ and\ \citenamefont
  {Lidar}(2013)}]{Quiroz:2013fv}%
  \BibitemOpen
  \bibfield  {author} {\bibinfo {author} {\bibfnamefont {G.}~\bibnamefont
  {Quiroz}}\ and\ \bibinfo {author} {\bibfnamefont {D.~A.}\ \bibnamefont
  {Lidar}},\ }\href {http://link.aps.org/doi/10.1103/PhysRevA.88.052306}
  {\bibfield  {journal} {\bibinfo  {journal} {Phys. Rev. A}\ }\textbf {\bibinfo
  {volume} {88}},\ \bibinfo {pages} {052306} (\bibinfo {year}
  {2013})}\BibitemShut {NoStop}%
\bibitem [{IBM(2022)}]{IBMQuantum2022}%
  \BibitemOpen
  \href {https://quantum-computing.ibm.com/} {\enquote {\bibinfo {title} {{{IBM
  Quantum}}},}\ } (\bibinfo {year} {2022})\BibitemShut {NoStop}%
\bibitem [{\citenamefont {Tripathi}\ \emph {et~al.}(2022)\citenamefont
  {Tripathi}, \citenamefont {Chen}, \citenamefont {Khezri}, \citenamefont
  {Yip}, \citenamefont {Levenson-Falk},\ and\ \citenamefont
  {Lidar}}]{tripathi2021suppression}%
  \BibitemOpen
  \bibfield  {author} {\bibinfo {author} {\bibfnamefont {V.}~\bibnamefont
  {Tripathi}}, \bibinfo {author} {\bibfnamefont {H.}~\bibnamefont {Chen}},
  \bibinfo {author} {\bibfnamefont {M.}~\bibnamefont {Khezri}}, \bibinfo
  {author} {\bibfnamefont {K.-W.}\ \bibnamefont {Yip}}, \bibinfo {author}
  {\bibfnamefont {E.~M.}\ \bibnamefont {Levenson-Falk}}, \ and\ \bibinfo
  {author} {\bibfnamefont {D.~A.}\ \bibnamefont {Lidar}},\ }\href {\doibase
  10.1103/PhysRevApplied.18.024068} {\bibfield  {journal} {\bibinfo  {journal}
  {Physical Review Applied}\ }\textbf {\bibinfo {volume} {18}},\ \bibinfo
  {pages} {024068} (\bibinfo {year} {2022})}\BibitemShut {NoStop}%
\bibitem [{\citenamefont {Zhou}\ \emph {et~al.}(2022)\citenamefont {Zhou},
  \citenamefont {Sitler}, \citenamefont {Oda}, \citenamefont {Schultz},\ and\
  \citenamefont {Quiroz}}]{Zeyuan:22}%
  \BibitemOpen
  \bibfield  {author} {\bibinfo {author} {\bibfnamefont {Z.}~\bibnamefont
  {Zhou}}, \bibinfo {author} {\bibfnamefont {R.}~\bibnamefont {Sitler}},
  \bibinfo {author} {\bibfnamefont {Y.}~\bibnamefont {Oda}}, \bibinfo {author}
  {\bibfnamefont {K.}~\bibnamefont {Schultz}}, \ and\ \bibinfo {author}
  {\bibfnamefont {G.}~\bibnamefont {Quiroz}},\ }\href
  {https://arxiv.org/abs/2208.05978} {\enquote {\bibinfo {title} {Quantum
  crosstalk robust quantum control},}\ } (\bibinfo {year} {2022}),\ \Eprint
  {http://arxiv.org/abs/2208.05978} {arXiv:2208.05978 [quant-ph]} \BibitemShut
  {NoStop}%
\bibitem [{\citenamefont {Boyer}\ \emph {et~al.}(1998)\citenamefont {Boyer},
  \citenamefont {Brassard}, \citenamefont {Hoyer},\ and\ \citenamefont
  {Tapp}}]{Boyer:96}%
  \BibitemOpen
  \bibfield  {author} {\bibinfo {author} {\bibfnamefont {M.}~\bibnamefont
  {Boyer}}, \bibinfo {author} {\bibfnamefont {G.}~\bibnamefont {Brassard}},
  \bibinfo {author} {\bibfnamefont {P.}~\bibnamefont {Hoyer}}, \ and\ \bibinfo
  {author} {\bibfnamefont {A.}~\bibnamefont {Tapp}},\ }\href
  {http://dx.doi.org/10.1002/(SICI)1521-3978(199806)46:4/5<493::AID-PROP493>3.0.CO;2-P}
  {\bibfield  {journal} {\bibinfo  {journal} {Fortschritte der Physik}\
  }\textbf {\bibinfo {volume} {46}},\ \bibinfo {pages} {493} (\bibinfo {year}
  {1998})}\BibitemShut {NoStop}%
\bibitem [{\citenamefont {Biham}\ \emph {et~al.}(1999)\citenamefont {Biham},
  \citenamefont {Biham}, \citenamefont {Biron}, \citenamefont {Grassl},\ and\
  \citenamefont {Lidar}}]{Biham:1999ye}%
  \BibitemOpen
  \bibfield  {author} {\bibinfo {author} {\bibfnamefont {E.}~\bibnamefont
  {Biham}}, \bibinfo {author} {\bibfnamefont {O.}~\bibnamefont {Biham}},
  \bibinfo {author} {\bibfnamefont {D.}~\bibnamefont {Biron}}, \bibinfo
  {author} {\bibfnamefont {M.}~\bibnamefont {Grassl}}, \ and\ \bibinfo {author}
  {\bibfnamefont {D.~A.}\ \bibnamefont {Lidar}},\ }\href
  {http://link.aps.org/doi/10.1103/PhysRevA.60.2742} {\bibfield  {journal}
  {\bibinfo  {journal} {Physical Review A}\ }\textbf {\bibinfo {volume} {60}},\
  \bibinfo {pages} {2742} (\bibinfo {year} {1999})}\BibitemShut {NoStop}%
\bibitem [{\citenamefont {Blank}\ \emph {et~al.}(2020)\citenamefont {Blank},
  \citenamefont {Park}, \citenamefont {Rhee},\ and\ \citenamefont
  {Petruccione}}]{blankQuantumClassifierTailored2020}%
  \BibitemOpen
  \bibfield  {author} {\bibinfo {author} {\bibfnamefont {C.}~\bibnamefont
  {Blank}}, \bibinfo {author} {\bibfnamefont {D.~K.}\ \bibnamefont {Park}},
  \bibinfo {author} {\bibfnamefont {J.-K.~K.}\ \bibnamefont {Rhee}}, \ and\
  \bibinfo {author} {\bibfnamefont {F.}~\bibnamefont {Petruccione}},\ }\href
  {\doibase 10.1038/s41534-020-0272-6} {\bibfield  {journal} {\bibinfo
  {journal} {npj Quantum Information}\ }\textbf {\bibinfo {volume} {6}},\
  \bibinfo {pages} {1} (\bibinfo {year} {2020})}\BibitemShut {NoStop}%
\bibitem [{\citenamefont {Rivas}\ and\ \citenamefont
  {Huelga}(2012)}]{rivas_open_2012}%
  \BibitemOpen
  \bibfield  {author} {\bibinfo {author} {\bibfnamefont {A.}~\bibnamefont
  {Rivas}}\ and\ \bibinfo {author} {\bibfnamefont {S.~F.}\ \bibnamefont
  {Huelga}},\ }\href {https://doi.org/10.1007/978-3-642-23354-8} {\emph
  {\bibinfo {title} {Open {Quantum} {Systems}: {An} {Introduction}}}},\
  {SpringerBriefs} in {Physics}\ (\bibinfo  {publisher} {Springer-Verlag},\
  \bibinfo {address} {Berlin Heidelberg},\ \bibinfo {year} {2012})\BibitemShut
  {NoStop}%
\bibitem [{\citenamefont {Nielsen}(2002)}]{nielsenSimpleFormulaAverage2002}%
  \BibitemOpen
  \bibfield  {author} {\bibinfo {author} {\bibfnamefont {M.~A.}\ \bibnamefont
  {Nielsen}},\ }\href {\doibase 10.1016/S0375-9601(02)01272-0} {\bibfield
  {journal} {\bibinfo  {journal} {Physics Letters A}\ }\textbf {\bibinfo
  {volume} {303}},\ \bibinfo {pages} {249} (\bibinfo {year}
  {2002})}\BibitemShut {NoStop}%
\bibitem [{\citenamefont {Paz-Silva}\ and\ \citenamefont
  {Lidar}(2013)}]{Paz-Silva:2013tt}%
  \BibitemOpen
  \bibfield  {author} {\bibinfo {author} {\bibfnamefont {G.~A.}\ \bibnamefont
  {Paz-Silva}}\ and\ \bibinfo {author} {\bibfnamefont {D.~A.}\ \bibnamefont
  {Lidar}},\ }\href {https://www.nature.com/articles/srep01530} {\bibfield
  {journal} {\bibinfo  {journal} {Sci. Rep.}\ }\textbf {\bibinfo {volume}
  {3}},\ \bibinfo {pages} {1530} (\bibinfo {year} {2013})}\BibitemShut
  {NoStop}%
\bibitem [{\citenamefont {Suter}\ and\ \citenamefont
  {{\'A}lvarez}(2016)}]{Suter:2016aa}%
  \BibitemOpen
  \bibfield  {author} {\bibinfo {author} {\bibfnamefont {D.}~\bibnamefont
  {Suter}}\ and\ \bibinfo {author} {\bibfnamefont {G.~A.}\ \bibnamefont
  {{\'A}lvarez}},\ }\href
  {https://link.aps.org/doi/10.1103/RevModPhys.88.041001} {\bibfield  {journal}
  {\bibinfo  {journal} {Rev. Mod. Phys.}\ }\textbf {\bibinfo {volume} {88}},\
  \bibinfo {pages} {041001} (\bibinfo {year} {2016})}\BibitemShut {NoStop}%
\bibitem [{\citenamefont {West}\ \emph
  {et~al.}(2010{\natexlab{a}})\citenamefont {West}, \citenamefont {Lidar},
  \citenamefont {Fong},\ and\ \citenamefont {Gyure}}]{West:10}%
  \BibitemOpen
  \bibfield  {author} {\bibinfo {author} {\bibfnamefont {J.~R.}\ \bibnamefont
  {West}}, \bibinfo {author} {\bibfnamefont {D.~A.}\ \bibnamefont {Lidar}},
  \bibinfo {author} {\bibfnamefont {B.~H.}\ \bibnamefont {Fong}}, \ and\
  \bibinfo {author} {\bibfnamefont {M.~F.}\ \bibnamefont {Gyure}},\ }\href
  {\doibase 10.1103/PhysRevLett.105.230503} {\bibfield  {journal} {\bibinfo
  {journal} {Phys. Rev. Lett.}\ }\textbf {\bibinfo {volume} {105}},\ \bibinfo
  {pages} {230503} (\bibinfo {year} {2010}{\natexlab{a}})}\BibitemShut
  {NoStop}%
\bibitem [{\citenamefont {Ng}\ \emph {et~al.}(2011)\citenamefont {Ng},
  \citenamefont {Lidar},\ and\ \citenamefont {Preskill}}]{Ng:2011dn}%
  \BibitemOpen
  \bibfield  {author} {\bibinfo {author} {\bibfnamefont {H.~K.}\ \bibnamefont
  {Ng}}, \bibinfo {author} {\bibfnamefont {D.~A.}\ \bibnamefont {Lidar}}, \
  and\ \bibinfo {author} {\bibfnamefont {J.}~\bibnamefont {Preskill}},\ }\href
  {http://link.aps.org/doi/10.1103/PhysRevA.84.012305} {\bibfield  {journal}
  {\bibinfo  {journal} {Phys. Rev. A}\ }\textbf {\bibinfo {volume} {84}},\
  \bibinfo {pages} {012305} (\bibinfo {year} {2011})}\BibitemShut {NoStop}%
\bibitem [{\citenamefont {Jurcevic}\ \emph {et~al.}(2021)\citenamefont
  {Jurcevic}, \citenamefont {{Javadi-Abhari}}, \citenamefont {Bishop},
  \citenamefont {Lauer}, \citenamefont {Bogorin}, \citenamefont {Brink},
  \citenamefont {Capelluto}, \citenamefont {G{\"u}nl{\"u}k}, \citenamefont
  {Itoko}, \citenamefont {Kanazawa}, \citenamefont {Kandala}, \citenamefont
  {Keefe}, \citenamefont {Krsulich}, \citenamefont {Landers}, \citenamefont
  {Lewandowski}, \citenamefont {McClure}, \citenamefont {Nannicini},
  \citenamefont {Narasgond}, \citenamefont {Nayfeh}, \citenamefont {Pritchett},
  \citenamefont {Rothwell}, \citenamefont {Srinivasan}, \citenamefont
  {Sundaresan}, \citenamefont {Wang}, \citenamefont {Wei}, \citenamefont
  {Wood}, \citenamefont {Yau}, \citenamefont {Zhang}, \citenamefont {Dial},
  \citenamefont {Chow},\ and\ \citenamefont
  {Gambetta}}]{jurcevicDemonstrationQuantumVolume2021}%
  \BibitemOpen
  \bibfield  {author} {\bibinfo {author} {\bibfnamefont {P.}~\bibnamefont
  {Jurcevic}}, \bibinfo {author} {\bibfnamefont {A.}~\bibnamefont
  {{Javadi-Abhari}}}, \bibinfo {author} {\bibfnamefont {L.~S.}\ \bibnamefont
  {Bishop}}, \bibinfo {author} {\bibfnamefont {I.}~\bibnamefont {Lauer}},
  \bibinfo {author} {\bibfnamefont {D.~F.}\ \bibnamefont {Bogorin}}, \bibinfo
  {author} {\bibfnamefont {M.}~\bibnamefont {Brink}}, \bibinfo {author}
  {\bibfnamefont {L.}~\bibnamefont {Capelluto}}, \bibinfo {author}
  {\bibfnamefont {O.}~\bibnamefont {G{\"u}nl{\"u}k}}, \bibinfo {author}
  {\bibfnamefont {T.}~\bibnamefont {Itoko}}, \bibinfo {author} {\bibfnamefont
  {N.}~\bibnamefont {Kanazawa}}, \bibinfo {author} {\bibfnamefont
  {A.}~\bibnamefont {Kandala}}, \bibinfo {author} {\bibfnamefont {G.~A.}\
  \bibnamefont {Keefe}}, \bibinfo {author} {\bibfnamefont {K.}~\bibnamefont
  {Krsulich}}, \bibinfo {author} {\bibfnamefont {W.}~\bibnamefont {Landers}},
  \bibinfo {author} {\bibfnamefont {E.~P.}\ \bibnamefont {Lewandowski}},
  \bibinfo {author} {\bibfnamefont {D.~T.}\ \bibnamefont {McClure}}, \bibinfo
  {author} {\bibfnamefont {G.}~\bibnamefont {Nannicini}}, \bibinfo {author}
  {\bibfnamefont {A.}~\bibnamefont {Narasgond}}, \bibinfo {author}
  {\bibfnamefont {H.~M.}\ \bibnamefont {Nayfeh}}, \bibinfo {author}
  {\bibfnamefont {E.}~\bibnamefont {Pritchett}}, \bibinfo {author}
  {\bibfnamefont {M.~B.}\ \bibnamefont {Rothwell}}, \bibinfo {author}
  {\bibfnamefont {S.}~\bibnamefont {Srinivasan}}, \bibinfo {author}
  {\bibfnamefont {N.}~\bibnamefont {Sundaresan}}, \bibinfo {author}
  {\bibfnamefont {C.}~\bibnamefont {Wang}}, \bibinfo {author} {\bibfnamefont
  {K.~X.}\ \bibnamefont {Wei}}, \bibinfo {author} {\bibfnamefont {C.~J.}\
  \bibnamefont {Wood}}, \bibinfo {author} {\bibfnamefont {J.-B.}\ \bibnamefont
  {Yau}}, \bibinfo {author} {\bibfnamefont {E.~J.}\ \bibnamefont {Zhang}},
  \bibinfo {author} {\bibfnamefont {O.~E.}\ \bibnamefont {Dial}}, \bibinfo
  {author} {\bibfnamefont {J.~M.}\ \bibnamefont {Chow}}, \ and\ \bibinfo
  {author} {\bibfnamefont {J.~M.}\ \bibnamefont {Gambetta}},\ }\href
  {https://iopscience.iop.org/article/10.1088/2058-9565/abe519} {\bibfield
  {journal} {\bibinfo  {journal} {Quantum Sci. Technol.}\ }\textbf {\bibinfo
  {volume} {6}},\ \bibinfo {pages} {025020} (\bibinfo {year}
  {2021})}\BibitemShut {NoStop}%
\bibitem [{\citenamefont {Ravi}\ \emph {et~al.}(2021)\citenamefont {Ravi},
  \citenamefont {Smith}, \citenamefont {Gokhale}, \citenamefont {Mari},
  \citenamefont {Earnest}, \citenamefont {Javadi-Abhari},\ and\ \citenamefont
  {Chong}}]{raviVAQEMVariationalApproach2021}%
  \BibitemOpen
  \bibfield  {author} {\bibinfo {author} {\bibfnamefont {G.~S.}\ \bibnamefont
  {Ravi}}, \bibinfo {author} {\bibfnamefont {K.~N.}\ \bibnamefont {Smith}},
  \bibinfo {author} {\bibfnamefont {P.}~\bibnamefont {Gokhale}}, \bibinfo
  {author} {\bibfnamefont {A.}~\bibnamefont {Mari}}, \bibinfo {author}
  {\bibfnamefont {N.}~\bibnamefont {Earnest}}, \bibinfo {author} {\bibfnamefont
  {A.}~\bibnamefont {Javadi-Abhari}}, \ and\ \bibinfo {author} {\bibfnamefont
  {F.~T.}\ \bibnamefont {Chong}},\ }\href {https://arxiv.org/abs/2112.05821}
  {\enquote {\bibinfo {title} {{VAQEM: A Variational Approach to Quantum Error
  Mitigation}},}\ } (\bibinfo {year} {2021})\BibitemShut {NoStop}%
\bibitem [{\citenamefont {Pokharel}\ and\ \citenamefont
  {Lidar}(2022)}]{pokharel2022demonstration}%
  \BibitemOpen
  \bibfield  {author} {\bibinfo {author} {\bibfnamefont {B.}~\bibnamefont
  {Pokharel}}\ and\ \bibinfo {author} {\bibfnamefont {D.~A.}\ \bibnamefont
  {Lidar}},\ }\href {https://arxiv.org/abs/2207.07647} {\enquote {\bibinfo
  {title} {Demonstration of algorithmic quantum speedup},}\ } (\bibinfo {year}
  {2022}),\ \Eprint {http://arxiv.org/abs/2207.07647} {arXiv:2207.07647
  [quant-ph]} \BibitemShut {NoStop}%
\bibitem [{\citenamefont {Maudsley}(1986)}]{Maudsley:1986ty}%
  \BibitemOpen
  \bibfield  {author} {\bibinfo {author} {\bibfnamefont {A.~A.}\ \bibnamefont
  {Maudsley}},\ }\href
  {https://www.sciencedirect.com/science/article/pii/0022236486901605}
  {\bibfield  {journal} {\bibinfo  {journal} {Journal of Magnetic Resonance
  (1969)}\ }\textbf {\bibinfo {volume} {69}},\ \bibinfo {pages} {488} (\bibinfo
  {year} {1986})}\BibitemShut {NoStop}%
\bibitem [{\citenamefont {Lidar}\ and\ \citenamefont
  {Brun}(2013)}]{Lidar-Brun:book}%
  \BibitemOpen
  \bibinfo {editor} {\bibfnamefont {D.}~\bibnamefont {Lidar}}\ and\ \bibinfo
  {editor} {\bibfnamefont {T.}~\bibnamefont {Brun}},\ eds.,\ \href
  {http://www.cambridge.org/9780521897877} {\emph {\bibinfo {title} {Quantum
  Error Correction}}}\ (\bibinfo  {publisher} {Cambridge University Press},\
  \bibinfo {address} {{Cambridge, UK}},\ \bibinfo {year} {2013})\BibitemShut
  {NoStop}%
\bibitem [{\citenamefont {Ezzell}\ \emph {et~al.}(2022)\citenamefont {Ezzell},
  \citenamefont {Pokharel}, \citenamefont {Tewala}, \citenamefont {Quiroz},\
  and\ \citenamefont {Lidar}}]{DD-survey}%
  \BibitemOpen
  \bibfield  {author} {\bibinfo {author} {\bibfnamefont {N.}~\bibnamefont
  {Ezzell}}, \bibinfo {author} {\bibfnamefont {B.}~\bibnamefont {Pokharel}},
  \bibinfo {author} {\bibfnamefont {L.}~\bibnamefont {Tewala}}, \bibinfo
  {author} {\bibfnamefont {G.}~\bibnamefont {Quiroz}}, \ and\ \bibinfo {author}
  {\bibfnamefont {D.~A.}\ \bibnamefont {Lidar}},\ }\href
  {https://arxiv.org/abs/2207.03670} {\enquote {\bibinfo {title} {{Dynamical
  decoupling for superconducting qubits: a performance survey}},}\ } (\bibinfo
  {year} {2022}),\ \Eprint {http://arxiv.org/abs/2207.03670} {arXiv:2207.03670
  [quant-ph]} \BibitemShut {NoStop}%
\bibitem [{\citenamefont {Uhrig}(2007)}]{Uhrig:2007qf}%
  \BibitemOpen
  \bibfield  {author} {\bibinfo {author} {\bibfnamefont {G.~S.}\ \bibnamefont
  {Uhrig}},\ }\href {http://link.aps.org/doi/10.1103/PhysRevLett.98.100504}
  {\bibfield  {journal} {\bibinfo  {journal} {Phys. Rev. Lett.}\ }\textbf
  {\bibinfo {volume} {98}},\ \bibinfo {pages} {100504} (\bibinfo {year}
  {2007})}\BibitemShut {NoStop}%
\bibitem [{\citenamefont {West}\ \emph
  {et~al.}(2010{\natexlab{b}})\citenamefont {West}, \citenamefont {Fong},\ and\
  \citenamefont {Lidar}}]{West:2010:130501}%
  \BibitemOpen
  \bibfield  {author} {\bibinfo {author} {\bibfnamefont {J.~R.}\ \bibnamefont
  {West}}, \bibinfo {author} {\bibfnamefont {B.~H.}\ \bibnamefont {Fong}}, \
  and\ \bibinfo {author} {\bibfnamefont {D.~A.}\ \bibnamefont {Lidar}},\ }\href
  {\doibase 10.1103/PhysRevLett.104.130501} {\bibfield  {journal} {\bibinfo
  {journal} {Phys. Rev. Lett.}\ }\textbf {\bibinfo {volume} {104}},\ \bibinfo
  {pages} {130501} (\bibinfo {year} {2010}{\natexlab{b}})}\BibitemShut
  {NoStop}%
\bibitem [{\citenamefont {Vuillot}()}]{vuillotErrorDetectionHelpful}%
  \BibitemOpen
  \bibfield  {author} {\bibinfo {author} {\bibfnamefont {C.}~\bibnamefont
  {Vuillot}},\ }\href {\doibase 10/gktsqq} {\bibfield  {journal} {\bibinfo
  {journal} {Quantum Information and Computation}\ }\textbf {\bibinfo {volume}
  {18}},\ 10/gktsqq},\ \Eprint {http://arxiv.org/abs/1705.08957}
  {arXiv:1705.08957} \BibitemShut {NoStop}%
\bibitem [{\citenamefont {Harper}\ and\ \citenamefont
  {Flammia}(2019)}]{Harper:2019aa}%
  \BibitemOpen
  \bibfield  {author} {\bibinfo {author} {\bibfnamefont {R.}~\bibnamefont
  {Harper}}\ and\ \bibinfo {author} {\bibfnamefont {S.~T.}\ \bibnamefont
  {Flammia}},\ }\href {\doibase 10.1103/PhysRevLett.122.080504} {\bibfield
  {journal} {\bibinfo  {journal} {Physical Review Letters}\ }\textbf {\bibinfo
  {volume} {122}},\ \bibinfo {pages} {080504} (\bibinfo {year}
  {2019})}\BibitemShut {NoStop}%
\bibitem [{\citenamefont {Urbanek}\ \emph {et~al.}(2020)\citenamefont
  {Urbanek}, \citenamefont {Nachman},\ and\ \citenamefont {{de
  Jong}}}]{urbanekErrorDetectionQuantum2020}%
  \BibitemOpen
  \bibfield  {author} {\bibinfo {author} {\bibfnamefont {M.}~\bibnamefont
  {Urbanek}}, \bibinfo {author} {\bibfnamefont {B.}~\bibnamefont {Nachman}}, \
  and\ \bibinfo {author} {\bibfnamefont {W.~A.}\ \bibnamefont {{de Jong}}},\
  }\href {\doibase 10.1103/PhysRevA.102.022427} {\bibfield  {journal} {\bibinfo
   {journal} {Physical Review A}\ }\textbf {\bibinfo {volume} {102}},\ \bibinfo
  {pages} {022427} (\bibinfo {year} {2020})},\ \Eprint
  {http://arxiv.org/abs/1910.00129} {arXiv:1910.00129} \BibitemShut {NoStop}%
\bibitem [{\citenamefont {Srinivasan}\ \emph {et~al.}(2022)\citenamefont
  {Srinivasan}, \citenamefont {Pokharel}, \citenamefont {Quiroz},\ and\
  \citenamefont {Boots}}]{srinivasanScalableMeasurementError2022}%
  \BibitemOpen
  \bibfield  {author} {\bibinfo {author} {\bibfnamefont {S.}~\bibnamefont
  {Srinivasan}}, \bibinfo {author} {\bibfnamefont {B.}~\bibnamefont
  {Pokharel}}, \bibinfo {author} {\bibfnamefont {G.}~\bibnamefont {Quiroz}}, \
  and\ \bibinfo {author} {\bibfnamefont {B.}~\bibnamefont {Boots}},\ }\href
  {http://arxiv.org/abs/2210.12284} {\enquote {\bibinfo {title} {Scalable
  {{Measurement Error Mitigation}} via {{Iterative Bayesian Unfolding}}},}\ }
  (\bibinfo {year} {2022}),\ \Eprint {http://arxiv.org/abs/2210.12284}
  {arXiv:2210.12284 [quant-ph]} \BibitemShut {NoStop}%
\bibitem [{\citenamefont {Krantz}\ \emph {et~al.}(2019)\citenamefont {Krantz},
  \citenamefont {Kjaergaard}, \citenamefont {Yan}, \citenamefont {Orlando},
  \citenamefont {Gustavsson},\ and\ \citenamefont
  {Oliver}}]{krantzQuantumEngineerGuide2019}%
  \BibitemOpen
  \bibfield  {author} {\bibinfo {author} {\bibfnamefont {P.}~\bibnamefont
  {Krantz}}, \bibinfo {author} {\bibfnamefont {M.}~\bibnamefont {Kjaergaard}},
  \bibinfo {author} {\bibfnamefont {F.}~\bibnamefont {Yan}}, \bibinfo {author}
  {\bibfnamefont {T.~P.}\ \bibnamefont {Orlando}}, \bibinfo {author}
  {\bibfnamefont {S.}~\bibnamefont {Gustavsson}}, \ and\ \bibinfo {author}
  {\bibfnamefont {W.~D.}\ \bibnamefont {Oliver}},\ }\href {\doibase
  10.1063/1.5089550} {\bibfield  {journal} {\bibinfo  {journal} {Applied
  Physics Reviews}\ }\textbf {\bibinfo {volume} {6}},\ \bibinfo {pages}
  {021318} (\bibinfo {year} {2019})}\BibitemShut {NoStop}%
\bibitem [{\citenamefont {Ofek}\ \emph {et~al.}(2016)\citenamefont {Ofek},
  \citenamefont {Petrenko}, \citenamefont {Heeres}, \citenamefont {Reinhold},
  \citenamefont {Leghtas}, \citenamefont {Vlastakis}, \citenamefont {Liu},
  \citenamefont {Frunzio}, \citenamefont {Girvin}, \citenamefont {Jiang},
  \citenamefont {Mirrahimi}, \citenamefont {Devoret},\ and\ \citenamefont
  {Schoelkopf}}]{Ofek:2016aa}%
  \BibitemOpen
  \bibfield  {author} {\bibinfo {author} {\bibfnamefont {N.}~\bibnamefont
  {Ofek}}, \bibinfo {author} {\bibfnamefont {A.}~\bibnamefont {Petrenko}},
  \bibinfo {author} {\bibfnamefont {R.}~\bibnamefont {Heeres}}, \bibinfo
  {author} {\bibfnamefont {P.}~\bibnamefont {Reinhold}}, \bibinfo {author}
  {\bibfnamefont {Z.}~\bibnamefont {Leghtas}}, \bibinfo {author} {\bibfnamefont
  {B.}~\bibnamefont {Vlastakis}}, \bibinfo {author} {\bibfnamefont
  {Y.}~\bibnamefont {Liu}}, \bibinfo {author} {\bibfnamefont {L.}~\bibnamefont
  {Frunzio}}, \bibinfo {author} {\bibfnamefont {S.~M.}\ \bibnamefont {Girvin}},
  \bibinfo {author} {\bibfnamefont {L.}~\bibnamefont {Jiang}}, \bibinfo
  {author} {\bibfnamefont {M.}~\bibnamefont {Mirrahimi}}, \bibinfo {author}
  {\bibfnamefont {M.~H.}\ \bibnamefont {Devoret}}, \ and\ \bibinfo {author}
  {\bibfnamefont {R.~J.}\ \bibnamefont {Schoelkopf}},\ }\href
  {http://dx.doi.org/10.1038/nature18949} {\bibfield  {journal} {\bibinfo
  {journal} {Nature}\ }\textbf {\bibinfo {volume} {536}},\ \bibinfo {pages}
  {441 EP } (\bibinfo {year} {2016})}\BibitemShut {NoStop}%
\bibitem [{\citenamefont {Ryan-Anderson}\ \emph {et~al.}(2022)\citenamefont
  {Ryan-Anderson}, \citenamefont {Brown}, \citenamefont {Allman}, \citenamefont
  {Arkin}, \citenamefont {Asa-Attuah}, \citenamefont {Baldwin}, \citenamefont
  {Berg}, \citenamefont {Bohnet}, \citenamefont {Braxton}, \citenamefont
  {Burdick}, \citenamefont {Campora}, \citenamefont {Chernoguzov},
  \citenamefont {Esposito}, \citenamefont {Evans}, \citenamefont {Francois},
  \citenamefont {Gaebler}, \citenamefont {Gatterman}, \citenamefont {Gerber},
  \citenamefont {Gilmore}, \citenamefont {Gresh}, \citenamefont {Hall},
  \citenamefont {Hankin}, \citenamefont {Hostetter}, \citenamefont {Lucchetti},
  \citenamefont {Mayer}, \citenamefont {Myers}, \citenamefont {Neyenhuis},
  \citenamefont {Santiago}, \citenamefont {Sedlacek}, \citenamefont {Skripka},
  \citenamefont {Slattery}, \citenamefont {Stutz}, \citenamefont {Tait},
  \citenamefont {Tobey}, \citenamefont {Vittorini}, \citenamefont {Walker},\
  and\ \citenamefont {Hayes}}]{Ryan-Anderson:22}%
  \BibitemOpen
  \bibfield  {author} {\bibinfo {author} {\bibfnamefont {C.}~\bibnamefont
  {Ryan-Anderson}}, \bibinfo {author} {\bibfnamefont {N.~C.}\ \bibnamefont
  {Brown}}, \bibinfo {author} {\bibfnamefont {M.~S.}\ \bibnamefont {Allman}},
  \bibinfo {author} {\bibfnamefont {B.}~\bibnamefont {Arkin}}, \bibinfo
  {author} {\bibfnamefont {G.}~\bibnamefont {Asa-Attuah}}, \bibinfo {author}
  {\bibfnamefont {C.}~\bibnamefont {Baldwin}}, \bibinfo {author} {\bibfnamefont
  {J.}~\bibnamefont {Berg}}, \bibinfo {author} {\bibfnamefont {J.~G.}\
  \bibnamefont {Bohnet}}, \bibinfo {author} {\bibfnamefont {S.}~\bibnamefont
  {Braxton}}, \bibinfo {author} {\bibfnamefont {N.}~\bibnamefont {Burdick}},
  \bibinfo {author} {\bibfnamefont {J.~P.}\ \bibnamefont {Campora}}, \bibinfo
  {author} {\bibfnamefont {A.}~\bibnamefont {Chernoguzov}}, \bibinfo {author}
  {\bibfnamefont {J.}~\bibnamefont {Esposito}}, \bibinfo {author}
  {\bibfnamefont {B.}~\bibnamefont {Evans}}, \bibinfo {author} {\bibfnamefont
  {D.}~\bibnamefont {Francois}}, \bibinfo {author} {\bibfnamefont {J.~P.}\
  \bibnamefont {Gaebler}}, \bibinfo {author} {\bibfnamefont {T.~M.}\
  \bibnamefont {Gatterman}}, \bibinfo {author} {\bibfnamefont {J.}~\bibnamefont
  {Gerber}}, \bibinfo {author} {\bibfnamefont {K.}~\bibnamefont {Gilmore}},
  \bibinfo {author} {\bibfnamefont {D.}~\bibnamefont {Gresh}}, \bibinfo
  {author} {\bibfnamefont {A.}~\bibnamefont {Hall}}, \bibinfo {author}
  {\bibfnamefont {A.}~\bibnamefont {Hankin}}, \bibinfo {author} {\bibfnamefont
  {J.}~\bibnamefont {Hostetter}}, \bibinfo {author} {\bibfnamefont
  {D.}~\bibnamefont {Lucchetti}}, \bibinfo {author} {\bibfnamefont
  {K.}~\bibnamefont {Mayer}}, \bibinfo {author} {\bibfnamefont
  {J.}~\bibnamefont {Myers}}, \bibinfo {author} {\bibfnamefont
  {B.}~\bibnamefont {Neyenhuis}}, \bibinfo {author} {\bibfnamefont
  {J.}~\bibnamefont {Santiago}}, \bibinfo {author} {\bibfnamefont
  {J.}~\bibnamefont {Sedlacek}}, \bibinfo {author} {\bibfnamefont
  {T.}~\bibnamefont {Skripka}}, \bibinfo {author} {\bibfnamefont
  {A.}~\bibnamefont {Slattery}}, \bibinfo {author} {\bibfnamefont {R.~P.}\
  \bibnamefont {Stutz}}, \bibinfo {author} {\bibfnamefont {J.}~\bibnamefont
  {Tait}}, \bibinfo {author} {\bibfnamefont {R.}~\bibnamefont {Tobey}},
  \bibinfo {author} {\bibfnamefont {G.}~\bibnamefont {Vittorini}}, \bibinfo
  {author} {\bibfnamefont {J.}~\bibnamefont {Walker}}, \ and\ \bibinfo {author}
  {\bibfnamefont {D.}~\bibnamefont {Hayes}},\ }\href
  {https://arxiv.org/abs/2208.01863} {\enquote {\bibinfo {title} {Implementing
  fault-tolerant entangling gates on the five-qubit code and the color code},}\
  } (\bibinfo {year} {2022}),\ \Eprint {http://arxiv.org/abs/2208.01863}
  {arXiv:2208.01863 [quant-ph]} \BibitemShut {NoStop}%
\bibitem [{\citenamefont {Marvian}\ and\ \citenamefont
  {Lidar}(2017)}]{Marvian-Lidar:16}%
  \BibitemOpen
  \bibfield  {author} {\bibinfo {author} {\bibfnamefont {M.}~\bibnamefont
  {Marvian}}\ and\ \bibinfo {author} {\bibfnamefont {D.~A.}\ \bibnamefont
  {Lidar}},\ }\href {https://link.aps.org/doi/10.1103/PhysRevLett.118.030504}
  {\bibfield  {journal} {\bibinfo  {journal} {Phys. Rev. Lett.}\ }\textbf
  {\bibinfo {volume} {118}},\ \bibinfo {pages} {030504} (\bibinfo {year}
  {2017})}\BibitemShut {NoStop}%
\bibitem [{\citenamefont
  {Maslov}(2016)}]{maslovAdvantagesUsingRelativephase2016}%
  \BibitemOpen
  \bibfield  {author} {\bibinfo {author} {\bibfnamefont {D.}~\bibnamefont
  {Maslov}},\ }\href {\doibase 10.1103/PhysRevA.93.022311} {\bibfield
  {journal} {\bibinfo  {journal} {Physical Review A}\ }\textbf {\bibinfo
  {volume} {93}},\ \bibinfo {pages} {022311} (\bibinfo {year}
  {2016})}\BibitemShut {NoStop}%
\bibitem [{\citenamefont {Arute}\ \emph {et~al.}(2019)\citenamefont {Arute},
  \citenamefont {Arya}, \citenamefont {Babbush}, \citenamefont {Bacon},
  \citenamefont {Bardin}, \citenamefont {Barends}, \citenamefont {Biswas},
  \citenamefont {Boixo}, \citenamefont {Brandao}, \citenamefont {Buell},
  \citenamefont {Burkett}, \citenamefont {Chen}, \citenamefont {Chen},
  \citenamefont {Chiaro}, \citenamefont {Collins}, \citenamefont {Courtney},
  \citenamefont {Dunsworth}, \citenamefont {Farhi}, \citenamefont {Foxen},
  \citenamefont {Fowler}, \citenamefont {Gidney}, \citenamefont {Giustina},
  \citenamefont {Graff}, \citenamefont {Guerin}, \citenamefont {Habegger},
  \citenamefont {Harrigan}, \citenamefont {Hartmann}, \citenamefont {Ho},
  \citenamefont {Hoffmann}, \citenamefont {Huang}, \citenamefont {Humble},
  \citenamefont {Isakov}, \citenamefont {Jeffrey}, \citenamefont {Jiang},
  \citenamefont {Kafri}, \citenamefont {Kechedzhi}, \citenamefont {Kelly},
  \citenamefont {Klimov}, \citenamefont {Knysh}, \citenamefont {Korotkov},
  \citenamefont {Kostritsa}, \citenamefont {Landhuis}, \citenamefont
  {Lindmark}, \citenamefont {Lucero}, \citenamefont {Lyakh}, \citenamefont
  {Mandr{\`a}}, \citenamefont {McClean}, \citenamefont {McEwen}, \citenamefont
  {Megrant}, \citenamefont {Mi}, \citenamefont {Michielsen}, \citenamefont
  {Mohseni}, \citenamefont {Mutus}, \citenamefont {Naaman}, \citenamefont
  {Neeley}, \citenamefont {Neill}, \citenamefont {Niu}, \citenamefont {Ostby},
  \citenamefont {Petukhov}, \citenamefont {Platt}, \citenamefont {Quintana},
  \citenamefont {Rieffel}, \citenamefont {Roushan}, \citenamefont {Rubin},
  \citenamefont {Sank}, \citenamefont {Satzinger}, \citenamefont {Smelyanskiy},
  \citenamefont {Sung}, \citenamefont {Trevithick}, \citenamefont
  {Vainsencher}, \citenamefont {Villalonga}, \citenamefont {White},
  \citenamefont {Yao}, \citenamefont {Yeh}, \citenamefont {Zalcman},
  \citenamefont {Neven},\ and\ \citenamefont {Martinis}}]{Arute:2019aa}%
  \BibitemOpen
  \bibfield  {author} {\bibinfo {author} {\bibfnamefont {F.}~\bibnamefont
  {Arute}}, \bibinfo {author} {\bibfnamefont {K.}~\bibnamefont {Arya}},
  \bibinfo {author} {\bibfnamefont {R.}~\bibnamefont {Babbush}}, \bibinfo
  {author} {\bibfnamefont {D.}~\bibnamefont {Bacon}}, \bibinfo {author}
  {\bibfnamefont {J.~C.}\ \bibnamefont {Bardin}}, \bibinfo {author}
  {\bibfnamefont {R.}~\bibnamefont {Barends}}, \bibinfo {author} {\bibfnamefont
  {R.}~\bibnamefont {Biswas}}, \bibinfo {author} {\bibfnamefont
  {S.}~\bibnamefont {Boixo}}, \bibinfo {author} {\bibfnamefont {F.~G. S.~L.}\
  \bibnamefont {Brandao}}, \bibinfo {author} {\bibfnamefont {D.~A.}\
  \bibnamefont {Buell}}, \bibinfo {author} {\bibfnamefont {B.}~\bibnamefont
  {Burkett}}, \bibinfo {author} {\bibfnamefont {Y.}~\bibnamefont {Chen}},
  \bibinfo {author} {\bibfnamefont {Z.}~\bibnamefont {Chen}}, \bibinfo {author}
  {\bibfnamefont {B.}~\bibnamefont {Chiaro}}, \bibinfo {author} {\bibfnamefont
  {R.}~\bibnamefont {Collins}}, \bibinfo {author} {\bibfnamefont
  {W.}~\bibnamefont {Courtney}}, \bibinfo {author} {\bibfnamefont
  {A.}~\bibnamefont {Dunsworth}}, \bibinfo {author} {\bibfnamefont
  {E.}~\bibnamefont {Farhi}}, \bibinfo {author} {\bibfnamefont
  {B.}~\bibnamefont {Foxen}}, \bibinfo {author} {\bibfnamefont
  {A.}~\bibnamefont {Fowler}}, \bibinfo {author} {\bibfnamefont
  {C.}~\bibnamefont {Gidney}}, \bibinfo {author} {\bibfnamefont
  {M.}~\bibnamefont {Giustina}}, \bibinfo {author} {\bibfnamefont
  {R.}~\bibnamefont {Graff}}, \bibinfo {author} {\bibfnamefont
  {K.}~\bibnamefont {Guerin}}, \bibinfo {author} {\bibfnamefont
  {S.}~\bibnamefont {Habegger}}, \bibinfo {author} {\bibfnamefont {M.~P.}\
  \bibnamefont {Harrigan}}, \bibinfo {author} {\bibfnamefont {M.~J.}\
  \bibnamefont {Hartmann}}, \bibinfo {author} {\bibfnamefont {A.}~\bibnamefont
  {Ho}}, \bibinfo {author} {\bibfnamefont {M.}~\bibnamefont {Hoffmann}},
  \bibinfo {author} {\bibfnamefont {T.}~\bibnamefont {Huang}}, \bibinfo
  {author} {\bibfnamefont {T.~S.}\ \bibnamefont {Humble}}, \bibinfo {author}
  {\bibfnamefont {S.~V.}\ \bibnamefont {Isakov}}, \bibinfo {author}
  {\bibfnamefont {E.}~\bibnamefont {Jeffrey}}, \bibinfo {author} {\bibfnamefont
  {Z.}~\bibnamefont {Jiang}}, \bibinfo {author} {\bibfnamefont
  {D.}~\bibnamefont {Kafri}}, \bibinfo {author} {\bibfnamefont
  {K.}~\bibnamefont {Kechedzhi}}, \bibinfo {author} {\bibfnamefont
  {J.}~\bibnamefont {Kelly}}, \bibinfo {author} {\bibfnamefont {P.~V.}\
  \bibnamefont {Klimov}}, \bibinfo {author} {\bibfnamefont {S.}~\bibnamefont
  {Knysh}}, \bibinfo {author} {\bibfnamefont {A.}~\bibnamefont {Korotkov}},
  \bibinfo {author} {\bibfnamefont {F.}~\bibnamefont {Kostritsa}}, \bibinfo
  {author} {\bibfnamefont {D.}~\bibnamefont {Landhuis}}, \bibinfo {author}
  {\bibfnamefont {M.}~\bibnamefont {Lindmark}}, \bibinfo {author}
  {\bibfnamefont {E.}~\bibnamefont {Lucero}}, \bibinfo {author} {\bibfnamefont
  {D.}~\bibnamefont {Lyakh}}, \bibinfo {author} {\bibfnamefont
  {S.}~\bibnamefont {Mandr{\`a}}}, \bibinfo {author} {\bibfnamefont {J.~R.}\
  \bibnamefont {McClean}}, \bibinfo {author} {\bibfnamefont {M.}~\bibnamefont
  {McEwen}}, \bibinfo {author} {\bibfnamefont {A.}~\bibnamefont {Megrant}},
  \bibinfo {author} {\bibfnamefont {X.}~\bibnamefont {Mi}}, \bibinfo {author}
  {\bibfnamefont {K.}~\bibnamefont {Michielsen}}, \bibinfo {author}
  {\bibfnamefont {M.}~\bibnamefont {Mohseni}}, \bibinfo {author} {\bibfnamefont
  {J.}~\bibnamefont {Mutus}}, \bibinfo {author} {\bibfnamefont
  {O.}~\bibnamefont {Naaman}}, \bibinfo {author} {\bibfnamefont
  {M.}~\bibnamefont {Neeley}}, \bibinfo {author} {\bibfnamefont
  {C.}~\bibnamefont {Neill}}, \bibinfo {author} {\bibfnamefont {M.~Y.}\
  \bibnamefont {Niu}}, \bibinfo {author} {\bibfnamefont {E.}~\bibnamefont
  {Ostby}}, \bibinfo {author} {\bibfnamefont {A.}~\bibnamefont {Petukhov}},
  \bibinfo {author} {\bibfnamefont {J.~C.}\ \bibnamefont {Platt}}, \bibinfo
  {author} {\bibfnamefont {C.}~\bibnamefont {Quintana}}, \bibinfo {author}
  {\bibfnamefont {E.~G.}\ \bibnamefont {Rieffel}}, \bibinfo {author}
  {\bibfnamefont {P.}~\bibnamefont {Roushan}}, \bibinfo {author} {\bibfnamefont
  {N.~C.}\ \bibnamefont {Rubin}}, \bibinfo {author} {\bibfnamefont
  {D.}~\bibnamefont {Sank}}, \bibinfo {author} {\bibfnamefont {K.~J.}\
  \bibnamefont {Satzinger}}, \bibinfo {author} {\bibfnamefont {V.}~\bibnamefont
  {Smelyanskiy}}, \bibinfo {author} {\bibfnamefont {K.~J.}\ \bibnamefont
  {Sung}}, \bibinfo {author} {\bibfnamefont {M.~D.}\ \bibnamefont
  {Trevithick}}, \bibinfo {author} {\bibfnamefont {A.}~\bibnamefont
  {Vainsencher}}, \bibinfo {author} {\bibfnamefont {B.}~\bibnamefont
  {Villalonga}}, \bibinfo {author} {\bibfnamefont {T.}~\bibnamefont {White}},
  \bibinfo {author} {\bibfnamefont {Z.~J.}\ \bibnamefont {Yao}}, \bibinfo
  {author} {\bibfnamefont {P.}~\bibnamefont {Yeh}}, \bibinfo {author}
  {\bibfnamefont {A.}~\bibnamefont {Zalcman}}, \bibinfo {author} {\bibfnamefont
  {H.}~\bibnamefont {Neven}}, \ and\ \bibinfo {author} {\bibfnamefont {J.~M.}\
  \bibnamefont {Martinis}},\ }\href {https://doi.org/10.1038/s41586-019-1666-5}
  {\bibfield  {journal} {\bibinfo  {journal} {Nature}\ }\textbf {\bibinfo
  {volume} {574}},\ \bibinfo {pages} {505} (\bibinfo {year}
  {2019})}\BibitemShut {NoStop}%
\bibitem [{\citenamefont {Khodjasteh}\ and\ \citenamefont
  {Viola}(2009)}]{khodjasteh:080501}%
  \BibitemOpen
  \bibfield  {author} {\bibinfo {author} {\bibfnamefont {K.}~\bibnamefont
  {Khodjasteh}}\ and\ \bibinfo {author} {\bibfnamefont {L.}~\bibnamefont
  {Viola}},\ }\href {https://link.aps.org/doi/10.1103/PhysRevLett.102.080501}
  {\bibfield  {journal} {\bibinfo  {journal} {Phys. Rev. Lett.}\ }\textbf
  {\bibinfo {volume} {102}},\ \bibinfo {pages} {080501} (\bibinfo {year}
  {2009})}\BibitemShut {NoStop}%
\bibitem [{\citenamefont {Khodjasteh}\ \emph {et~al.}(2010)\citenamefont
  {Khodjasteh}, \citenamefont {Lidar},\ and\ \citenamefont
  {Viola}}]{Khodjasteh:2010qd}%
  \BibitemOpen
  \bibfield  {author} {\bibinfo {author} {\bibfnamefont {K.}~\bibnamefont
  {Khodjasteh}}, \bibinfo {author} {\bibfnamefont {D.~A.}\ \bibnamefont
  {Lidar}}, \ and\ \bibinfo {author} {\bibfnamefont {L.}~\bibnamefont
  {Viola}},\ }\href {http://link.aps.org/doi/10.1103/PhysRevLett.104.090501}
  {\bibfield  {journal} {\bibinfo  {journal} {Phys. Rev. Lett.}\ }\textbf
  {\bibinfo {volume} {104}},\ \bibinfo {pages} {090501} (\bibinfo {year}
  {2010})}\BibitemShut {NoStop}%
\bibitem [{\citenamefont {Maciejewski}\ \emph {et~al.}(2020)\citenamefont
  {Maciejewski}, \citenamefont {Zimbor{\'a}s},\ and\ \citenamefont
  {Oszmaniec}}]{maciejewskiMitigationReadoutNoise2020}%
  \BibitemOpen
  \bibfield  {author} {\bibinfo {author} {\bibfnamefont {F.~B.}\ \bibnamefont
  {Maciejewski}}, \bibinfo {author} {\bibfnamefont {Z.}~\bibnamefont
  {Zimbor{\'a}s}}, \ and\ \bibinfo {author} {\bibfnamefont {M.}~\bibnamefont
  {Oszmaniec}},\ }\href {\doibase 10.22331/q-2020-04-24-257} {\bibfield
  {journal} {\bibinfo  {journal} {Quantum}\ }\textbf {\bibinfo {volume} {4}},\
  \bibinfo {pages} {257} (\bibinfo {year} {2020})}\BibitemShut {NoStop}%
\bibitem [{\citenamefont {Pokharel}\ \emph {et~al.}(2018)\citenamefont
  {Pokharel}, \citenamefont {Anand}, \citenamefont {Fortman},\ and\
  \citenamefont {Lidar}}]{Pokharel2018}%
  \BibitemOpen
  \bibfield  {author} {\bibinfo {author} {\bibfnamefont {B.}~\bibnamefont
  {Pokharel}}, \bibinfo {author} {\bibfnamefont {N.}~\bibnamefont {Anand}},
  \bibinfo {author} {\bibfnamefont {B.}~\bibnamefont {Fortman}}, \ and\
  \bibinfo {author} {\bibfnamefont {D.~A.}\ \bibnamefont {Lidar}},\ }\href
  {https://link.aps.org/doi/10.1103/PhysRevLett.121.220502} {\bibfield
  {journal} {\bibinfo  {journal} {Phys. Rev. Lett.}\ }\textbf {\bibinfo
  {volume} {121}},\ \bibinfo {pages} {220502} (\bibinfo {year}
  {2018})}\BibitemShut {NoStop}%
\bibitem [{\citenamefont {Ronnow}\ \emph {et~al.}(2014)\citenamefont {Ronnow},
  \citenamefont {Wang}, \citenamefont {Job}, \citenamefont {Boixo},
  \citenamefont {Isakov}, \citenamefont {Wecker}, \citenamefont {Martinis},
  \citenamefont {Lidar},\ and\ \citenamefont {Troyer}}]{speedup}%
  \BibitemOpen
  \bibfield  {author} {\bibinfo {author} {\bibfnamefont {T.~F.}\ \bibnamefont
  {Ronnow}}, \bibinfo {author} {\bibfnamefont {Z.}~\bibnamefont {Wang}},
  \bibinfo {author} {\bibfnamefont {J.}~\bibnamefont {Job}}, \bibinfo {author}
  {\bibfnamefont {S.}~\bibnamefont {Boixo}}, \bibinfo {author} {\bibfnamefont
  {S.~V.}\ \bibnamefont {Isakov}}, \bibinfo {author} {\bibfnamefont
  {D.}~\bibnamefont {Wecker}}, \bibinfo {author} {\bibfnamefont {J.~M.}\
  \bibnamefont {Martinis}}, \bibinfo {author} {\bibfnamefont {D.~A.}\
  \bibnamefont {Lidar}}, \ and\ \bibinfo {author} {\bibfnamefont
  {M.}~\bibnamefont {Troyer}},\ }\href
  {http://science.sciencemag.org/content/345/6195/420} {\bibfield  {journal}
  {\bibinfo  {journal} {Science}\ }\textbf {\bibinfo {volume} {345}},\ \bibinfo
  {pages} {420} (\bibinfo {year} {2014})}\BibitemShut {NoStop}%
\bibitem [{\citenamefont {Campbell}\ \emph {et~al.}(2019)\citenamefont
  {Campbell}, \citenamefont {Khurana},\ and\ \citenamefont
  {Montanaro}}]{campbell2019applyingquantum}%
  \BibitemOpen
  \bibfield  {author} {\bibinfo {author} {\bibfnamefont {E.}~\bibnamefont
  {Campbell}}, \bibinfo {author} {\bibfnamefont {A.}~\bibnamefont {Khurana}}, \
  and\ \bibinfo {author} {\bibfnamefont {A.}~\bibnamefont {Montanaro}},\ }\href
  {\doibase 10.22331/q-2019-07-18-167} {\bibfield  {journal} {\bibinfo
  {journal} {{Quantum}}\ }\textbf {\bibinfo {volume} {3}},\ \bibinfo {pages}
  {167} (\bibinfo {year} {2019})}\BibitemShut {NoStop}%
\bibitem [{\citenamefont {Sanders}\ \emph {et~al.}(2020)\citenamefont
  {Sanders}, \citenamefont {Berry}, \citenamefont {Costa}, \citenamefont
  {Tessler}, \citenamefont {Wiebe}, \citenamefont {Gidney}, \citenamefont
  {Neven},\ and\ \citenamefont {Babbush}}]{s2020compilation}%
  \BibitemOpen
  \bibfield  {author} {\bibinfo {author} {\bibfnamefont {Y.~R.}\ \bibnamefont
  {Sanders}}, \bibinfo {author} {\bibfnamefont {D.~W.}\ \bibnamefont {Berry}},
  \bibinfo {author} {\bibfnamefont {P.~C.~S.}\ \bibnamefont {Costa}}, \bibinfo
  {author} {\bibfnamefont {L.~W.}\ \bibnamefont {Tessler}}, \bibinfo {author}
  {\bibfnamefont {N.}~\bibnamefont {Wiebe}}, \bibinfo {author} {\bibfnamefont
  {C.}~\bibnamefont {Gidney}}, \bibinfo {author} {\bibfnamefont
  {H.}~\bibnamefont {Neven}}, \ and\ \bibinfo {author} {\bibfnamefont
  {R.}~\bibnamefont {Babbush}},\ }\href {\doibase 10.1103/PRXQuantum.1.020312}
  {\bibfield  {journal} {\bibinfo  {journal} {PRX Quantum}\ }\textbf {\bibinfo
  {volume} {1}},\ \bibinfo {pages} {020312} (\bibinfo {year}
  {2020})}\BibitemShut {NoStop}%
\bibitem [{\citenamefont {Pednault}\ \emph {et~al.}(2019)\citenamefont
  {Pednault}, \citenamefont {Gunnels}, \citenamefont {Nannicini}, \citenamefont
  {Horesh},\ and\ \citenamefont {Wisnieff}}]{ibm}%
  \BibitemOpen
  \bibfield  {author} {\bibinfo {author} {\bibfnamefont {E.}~\bibnamefont
  {Pednault}}, \bibinfo {author} {\bibfnamefont {J.~A.}\ \bibnamefont
  {Gunnels}}, \bibinfo {author} {\bibfnamefont {G.}~\bibnamefont {Nannicini}},
  \bibinfo {author} {\bibfnamefont {L.}~\bibnamefont {Horesh}}, \ and\ \bibinfo
  {author} {\bibfnamefont {R.}~\bibnamefont {Wisnieff}},\ }\href
  {https://arxiv.org/abs/1910.09534} {\bibfield  {journal} {\bibinfo  {journal}
  {arXiv preprint arXiv:1910.09534}\ } (\bibinfo {year} {2019})}\BibitemShut
  {NoStop}%
\bibitem [{\citenamefont {Mooney}\ \emph
  {et~al.}(2021{\natexlab{a}})\citenamefont {Mooney}, \citenamefont {White},
  \citenamefont {Hill},\ and\ \citenamefont
  {Hollenberg}}]{mooneyGenerationVerification27qubit2021}%
  \BibitemOpen
  \bibfield  {author} {\bibinfo {author} {\bibfnamefont {G.~J.}\ \bibnamefont
  {Mooney}}, \bibinfo {author} {\bibfnamefont {G.~A.~L.}\ \bibnamefont
  {White}}, \bibinfo {author} {\bibfnamefont {C.~D.}\ \bibnamefont {Hill}}, \
  and\ \bibinfo {author} {\bibfnamefont {L.~C.~L.}\ \bibnamefont
  {Hollenberg}},\ }\href {\doibase 10.1088/2399-6528/ac1df7} {\bibfield
  {journal} {\bibinfo  {journal} {Journal of Physics Communications}\ }\textbf
  {\bibinfo {volume} {5}},\ \bibinfo {pages} {095004} (\bibinfo {year}
  {2021}{\natexlab{a}})}\BibitemShut {NoStop}%
\bibitem [{\citenamefont {Mooney}\ \emph
  {et~al.}(2021{\natexlab{b}})\citenamefont {Mooney}, \citenamefont {White},
  \citenamefont {Hill},\ and\ \citenamefont
  {Hollenberg}}]{mooneyWholeDeviceEntanglement2021}%
  \BibitemOpen
  \bibfield  {author} {\bibinfo {author} {\bibfnamefont {G.~J.}\ \bibnamefont
  {Mooney}}, \bibinfo {author} {\bibfnamefont {G.~A.~L.}\ \bibnamefont
  {White}}, \bibinfo {author} {\bibfnamefont {C.~D.}\ \bibnamefont {Hill}}, \
  and\ \bibinfo {author} {\bibfnamefont {L.~C.~L.}\ \bibnamefont
  {Hollenberg}},\ }\href {\doibase 10.1002/qute.202100061} {\bibfield
  {journal} {\bibinfo  {journal} {Advanced Quantum Technologies}\ }\textbf
  {\bibinfo {volume} {4}},\ \bibinfo {pages} {2100061} (\bibinfo {year}
  {2021}{\natexlab{b}})}\BibitemShut {NoStop}%
\bibitem [{\citenamefont {Barron}\ and\ \citenamefont
  {Wood}(2020)}]{barronMeasurementErrorMitigation2020}%
  \BibitemOpen
  \bibfield  {author} {\bibinfo {author} {\bibfnamefont {G.~S.}\ \bibnamefont
  {Barron}}\ and\ \bibinfo {author} {\bibfnamefont {C.~J.}\ \bibnamefont
  {Wood}},\ }\href {http://arxiv.org/abs/2010.08520} {\enquote {\bibinfo
  {title} {Measurement {{Error Mitigation}} for {{Variational Quantum
  Algorithms}}},}\ } (\bibinfo {year} {2020}),\ \Eprint
  {http://arxiv.org/abs/2010.08520} {arXiv:2010.08520 [quant-ph]} \BibitemShut
  {NoStop}%
\bibitem [{\citenamefont {Bravyi}\ \emph {et~al.}(2021)\citenamefont {Bravyi},
  \citenamefont {Sheldon}, \citenamefont {Kandala}, \citenamefont {Mckay},\
  and\ \citenamefont {Gambetta}}]{bravyiMitigatingMeasurementErrors2021}%
  \BibitemOpen
  \bibfield  {author} {\bibinfo {author} {\bibfnamefont {S.}~\bibnamefont
  {Bravyi}}, \bibinfo {author} {\bibfnamefont {S.}~\bibnamefont {Sheldon}},
  \bibinfo {author} {\bibfnamefont {A.}~\bibnamefont {Kandala}}, \bibinfo
  {author} {\bibfnamefont {D.~C.}\ \bibnamefont {Mckay}}, \ and\ \bibinfo
  {author} {\bibfnamefont {J.~M.}\ \bibnamefont {Gambetta}},\ }\href {\doibase
  10/gjtqvd} {\bibfield  {journal} {\bibinfo  {journal} {Physical Review A}\
  }\textbf {\bibinfo {volume} {103}},\ \bibinfo {pages} {042605} (\bibinfo
  {year} {2021})}\BibitemShut {NoStop}%
\bibitem [{\citenamefont {Quek}\ \emph {et~al.}(2022)\citenamefont {Quek},
  \citenamefont {Franca}, \citenamefont {Khatri}, \citenamefont {Meyer},\ and\
  \citenamefont {Eisert}}]{Quek:2022}%
  \BibitemOpen
  \bibfield  {author} {\bibinfo {author} {\bibfnamefont {Y.}~\bibnamefont
  {Quek}}, \bibinfo {author} {\bibfnamefont {D.~S.}\ \bibnamefont {Franca}},
  \bibinfo {author} {\bibfnamefont {S.}~\bibnamefont {Khatri}}, \bibinfo
  {author} {\bibfnamefont {J.~J.}\ \bibnamefont {Meyer}}, \ and\ \bibinfo
  {author} {\bibfnamefont {J.}~\bibnamefont {Eisert}},\ }\href
  {https://arxiv.org/abs/2210.11505} {\enquote {\bibinfo {title} {Exponentially
  tighter bounds on limitations of quantum error mitigation},}\ } (\bibinfo
  {year} {2022})\BibitemShut {NoStop}%
\bibitem [{\citenamefont {Nielsen}\ and\ \citenamefont
  {Chuang}(2010)}]{nielsen2010quantum}%
  \BibitemOpen
  \bibfield  {author} {\bibinfo {author} {\bibfnamefont {M.~A.}\ \bibnamefont
  {Nielsen}}\ and\ \bibinfo {author} {\bibfnamefont {I.~L.}\ \bibnamefont
  {Chuang}},\ }\href@noop {} {\emph {\bibinfo {title} {Quantum computation and
  quantum information}}}\ (\bibinfo  {publisher} {{Cambridge University
  Press}},\ \bibinfo {year} {2010})\BibitemShut {NoStop}%
\end{thebibliography}%
\bibliographystyle{apsrev4-1}

\end{document}